\RequirePackage{fix-cm}
\documentclass[twocolumn,epjc3]{svjour3}  

\usepackage[numbers]{natbib}
\usepackage{amsmath}
\usepackage{amsfonts}
\usepackage{amssymb}
\usepackage{mathtools}
\usepackage{graphicx}
\graphicspath{{./}}
\usepackage{colordvi}
\usepackage[dvipsnames]{xcolor}
\usepackage{xspace}
\usepackage{acronym}
\usepackage{booktabs}
\usepackage{adjustbox}
\usepackage[tight]{units}
\usepackage[normalem]{ulem}
\usepackage{hyperref}
\usepackage{comment}
\usepackage{pifont}
\usepackage[utf8]{inputenc}

\hypersetup{
  colorlinks=true,        % false: boxed links; true: colored links
  linkcolor=blue,         % color of internal links
  citecolor=cyan,         % color of links to bibliography
}

\def\ie{i.e.\@\xspace}
\def\eg{e.g.\@\xspace}
\def\cf{cf.\@\xspace}

\newacro{AMR}{adaptive mesh refinement}
\newacro{BH}{black hole}
\newacro{BNS}{binary neutron star}
\newacro{BHF}{Brueckner-Hartree-Fock}
\newacro{CCSN}{Core\- Collapse\- Supernova}
\newacro{CFL}{Courant-Friedrichs-Lewy}
\newacro{EOS}{equation of state}
\newacroplural{EOS}{equations of state}
\newacro{GRB}{gamma ray burst}
\newacro{sGRB}{short gamma ray burst}
\newacro{GR}{general relativity}
\newacro{GRLES}{general-\-relativ\-istic large-\-eddy simulation}
\newacro{GRHD}{general-relativistic hydrodynamics}
\newacro{GRMHD}{general-relativistic magneto\-hydro\-dynamics}
\newacro{MHD}{magneto\-hydro\-dynamics}
\newacro{GW}{gravitational wave}
\newacro{HR}{high resolution}
\newacro{HRSC}{high-resolution shock-capturing}
\newacro{MNS}{massive neutron star}
\newacro{HMNS}{hyper-massive neutron star}
\newacro{LES}{large-eddy simulation}
\newacro{LL}{long lived}
\newacro{LR}{low resolution}
\newacro{MoL}{method of lines}
\newacro{N2LO}{next-to-next-to-leading order}
\newacro{N3LO}{next-to-next-to-next-to-leading order}
\newacro{NR}{numerical relativity}
\newacro{NS}{neutron star}
\newacro{NLS}{neutrino leakage scheme}
\newacro{PC}{prompt collapse}
\newacro{DC}{delayed collapse}
\newacro{RMF}{relativistic mean field}
\newacro{RMS}{root-mean-square}
\newacro{SNA}{single nucleus approximation}
\newacro{SR}{standard resolution}
\newacro{LORENE}{Langage Objet pour la RElativité NumériquE}
\newacro{NSE}{nuclear statistical equilibrium}
\newacro{VSL}{very short lived}
\newacro{MRI}{magneto-rotational instability}
\newacro{FWHM}{full width at half maximum}

\newcommand{\secref}[1]{Sec.~\ref{sec:#1}}
\newcommand{\figref}[1]{Fig.~\ref{fig:#1}}
\newcommand{\tabref}[1]{Table~\ref{tab:#1}}

\newcommand{\nue}[0]{\nu_{\textnormal e}}
\newcommand{\nua}[0]{\bar{\nu}_{\textnormal e}}
\newcommand{\nux}[0]{\nu_{\textnormal x}}
\newcommand{\msun}[0]{M_{\odot}}
\newcommand{\ltilde}[0]{\tilde{\Lambda}}
\newcommand{\ye}[0]{Y_{\textnormal e}}
\newcommand{\cmark}{\ding{51}}%
\newcommand{\xmark}{\ding{55}}%
\newif\ifmathmode
\newcommand{\SA}[2]{\relax\ifmmode\mathmodetrue\else\mathmodefalse\fi\ifmathmode{\textcolor{gray}{#1}}{\textcolor{red}{#2}}\else{\textcolor{gray}{#1}}{\textcolor{orange}{---#2---}}\fi}

\begin{document}

\title{Neutrino emission from binary neutron star mergers: characterising
  light curves and mean energies}

\author{Marco Cusinato$^{1}$
  \and
  Federico Maria Guercilena$^{2,3}$
  \and
  Albino Perego$^{3,2}$
  \and 
  Domenico Logoteta$^{4,5}$
  \and
  David Radice$^{6,7,8}$
  \and
  Sebastiano Bernuzzi$^{9}$
  \and
  Stefano Ansoldi$^{10}$
}

\institute{Dipartimento di Fisica, Universit\`{a} di Trieste, Via A. Valerio 2, 34127 Trieste, Italy
  \and
  INFN-TIFPA, Trento Institute for Fundamental Physics and Applications,
  via Sommarive 14, I-38123 Trento, Italy
  \and
  Dipartimento di Fisica, Universit\`{a} di Trento, Via Sommarive 14,
  38123 Trento, Italy 
  \and
  Dipartimento di Fisica, Universit\`{a} di Pisa, Largo B. Pontecorvo, 3 I-56127 Pisa, Italy
  \and
  INFN, Sezione di Pisa, Largo B. Pontecorvo, 3 I-56127 Pisa, Italy
  \and
  Institute for Gravitation and the Cosmos, The Pennsylvania State University, University Park, PA 16802, USA
  \and
  Department of Physics, The Pennsylvania State University, University Park, PA 16802, USA
  \and
  Department of Astronomy \& Astrophysics, The Pennsyvlania State University, University Park, PA 16802, USA
  \and
  Theoretisch-Physikalisches Institut, Friedrich-Schiller Universit\"{a}t Jena, 07743, Jena, Germany  
  \and
  Dipartimento di Scienze Matematiche, Informatiche e Fisiche, Univerit\`{a} degli Studi di Udine, via delle Scienze 206, I-33100 Udine, Italy  
}

\date{\today}

% ------------------------------------------------------------------------------
%\begin{document}
% ------------------------------------------------------------------------------

\maketitle

\begin{abstract}
Neutrinos are copiously emitted by neutron star mergers, due to the high
temperatures reached by dense matter during the merger and its aftermath.
Neutrinos influence the merger dynamics and shape the properties of the ejecta,
including the resulting $r$-process nucleosynthesis and kilonova emission. In
this work, we analyse neutrino emission from a large sample of binary neutron
star merger simulations in Numerical Relativity, covering a broad range of
initial masses, nuclear equation of state and viscosity treatments. We extract
neutrino luminosities and mean energies, and compute quantities of interest such
as the peak values, peak broadnesses, time averages and decrease time scales. We
provide a systematic description of such quantities, including their dependence
on the initial parameters of the system. We find that for equal-mass systems the
total neutrino luminosity (several $\unit[10^{53}]{erg~s^{-1}}$) decreases as
the reduced tidal deformability increases, as a consequence of the less violent
merger dynamics. Similarly, tidal disruption in asymmetric mergers leads to
systematically smaller luminosities. Peak luminosities can be twice as large as
the average ones. Electron antineutrino luminosities dominate (initially by a
factor of 2-3) over electron neutrino ones, while electron neutrinos and heavy
flavour neutrinos have similar luminosities. Mean energies are nearly constant
in time and independent on the binary parameters. Their values reflect the
different decoupling temperature inside the merger remnant. Despite present
uncertainties in neutrino modelling, our results provide a broad and physically
grounded characterisation of neutrino emission, and they can serve as a
reference point to develop more sophisticated neutrino transport schemes.

\end{abstract}

% ------------------------------------------------------------------------------
\section{Introduction}
\label{sec:introduction}
% ------------------------------------------------------------------------------

\Ac{BNS} mergers represent one of the main research topics in modern
astrophysics. Due to the wide range of densities and temperatures required to
study the dynamics of these events \citep{Perego:2019adq}, their understanding
connects several branches of physics spanning from nuclear physics to
relativistic hydrodynamics in strong-field conditions. They can be considered
natural laboratories to investigate the behaviour of matter at extreme
densities, which cannot be produced in Earth-based facilities (see \eg{}
\citep{Shibata:2019wef,Radice:2020ddv,Baiotti:2016qnr} for recent reviews).

\Ac{BNS} mergers are prominent sources of \acp{GW}
\citep{LIGOScientific:2018mvr, LIGOScientific:2021qlt}, and a primary target for
ground-based \ac{GW} detector facilities such as LIGO
\citep{TheLIGOScientific:2014jea}, VIRGO\citep{TheVirgo:2014hva} and KAGRA
\citep{Aso:2013eba}. Furthermore, they have long been considered one of the most
likely progenitors of high-energy astronomical signals such as \acp{sGRB}
\citep{Eichler:1989ve, Narayan:1992iy, Lee:2007js,Nakar:2007yr} and kilonovae
\cite{Li:1998bw, Kulkarni:2005jw}, see also \cite{Fernandez:2015use,
  Metzger:2019zeh} for recent reviews. Kilonovae (sometimes also referred to as
macronovae) are powered by the decay of radioactive heavy elements that are
synthesised in the ejecta of \ac{BNS} mergers \citep[see \eg{}][and references
  therein]{Cowan:2019pkx, Perego:2021dpw}. This aspect links these systems to
open issues regarding the evolution of the chemical composition of the Galaxy
and of the Cosmos. \ac{BNS} mergers have indeed emerged as sites (perhaps the
main ones) of production of heavy elements in the Universe
\citep{Korobkin:2012uy, Rosswog:2017sdn, Kasen:2017sxr, Drout:2017ijr}.

All these hypotheses recently received a direct confirmation by the first
multimessenger detection of a \ac{BNS} merger. This event (hereafter referred to
as GW170817) was observed as a \ac{GW} signal \citep{TheLIGOScientific:2017qsa},
followed by a \ac{sGRB} (GRB170817A) and, finally, by a kilonova lasting from a
few hours to several days after the merger \citep{LIGOScientific:2017ync,
  2017ApJ...848L..33A, Chornock:2017sdf, Cowperthwaite:2017dyu, Coulter:2017wya,
  Drout:2017ijr, Evans:2017mmy, Goldstein:2017mmi,Hallinan:2017woc,
  Kasliwal:2017ngb, Murguia-Berthier:2017kkn, Nicholl:2017ahq, Rosswog:2017sdn,
  Smartt:2017fuw, Soares-Santos:2017lru, Savchenko:2017ffs, Tanvir:2017pws,
  Tanaka:2017qxj, Troja:2017nqp, Villar:2017wcc, Waxman:2017sqv,
  Kasliwal:2018fwk, Waxman:2019png}. This detection opened the era of
multimessenger astronomy from compact binary mergers. A second detection of a
\ac{GW} signal from a \ac{BNS} merger, GW190425 \citep{Abbott:2020uma}, was
observed a couple of years later, but without the firm identification of
associated electromagnetic counterparts.

\ac{BNS} mergers produce copious amounts of neutrinos, starting from the latest
moments of the inspiral until the merger remnant collapses or cools down. This
emission is a key element in the dynamics of the system. On one hand, neutrinos
are thought to play a significant role in the jet-launching mechanism that
powers \acp{sGRB}
\citep[e.g.][]{Eichler:1989ve,Ruffert:1998vp,Rosswog:2002rt,Rosswog:2003ts}.
Neutrino absorption and energy deposition in the funnel above the poles of the
merger remnant could contribute to clean this region, reducing its baryon
density and allowing the launch of a relativistic jet
\cite{Mosta:2020hlh,Sun:2022vri}. It has also been suggested that
neutrino/antineutrino pair annihilation could deposit an amount of energy
compatible with the one necessary to explain \acp{sGRB} \citep[see
  e.g.][]{Eichler:1989ve, Ruffert:1996by, Rosswog:2003tn, Zalamea:2010ax,
  Dessart:2008zd, Just:2015dba, Perego:2014fma, Fujibayashi:2017xsz}. Neutrino
absorption is also likely one of the mechanisms for matter ejection from
\ac{BNS} mergers, in association to the production of neutrino-driven winds on
time scales of $\unit[\sim100]{ms}$ after the merger
\citep[\eg][]{Rosswog:2002rt, Dessart:2008zd, Perego:2017xth,
  Fujibayashi:2017xsz, Fujibayashi:2017xsz}. Even more importantly,
neutrino-matter interactions affect the composition of the ejecta, by driving
the evolution of the relative abundance of neutrons and protons, starting from
the decompression of beta-equilibrated, cold \ac{NS} matter. The neutron
richness in the ejecta directly impacts the outcome of the $r$-process
nucleosynthesis and of the resulting kilonova signal \citep{Metzger:2014ila,
  Martin:2015hxa, Miller:2019dpt}. It was shown that the neutrino transport used
in the simulations influences essential ejecta properties like the radial speed,
the electron fraction and the entropy \citep{Foucart:2016rxm, Perego:2017wtu,
  Nedora:2020qtd}. To reliably model these phenomena it is therefore of the
utmost importance to characterise the properties of neutrino emission in
\ac{BNS} mergers.

\Ac{BNS} mergers are intrinsically multi-\-dimensional events. Moreover, their
thermodynamic conditions are such that the neutrino optical depth decreases by
several orders of magnitude from the optically thick central remnant to the
optically thin accretion disc \cite{Endrizzi:2019trv}. The quantitative
modelling of neutrino production and diffusion in \ac{BNS} mergers is, thus, a
non-trivial task that has only been made possible by the advent of sophisticated
numerical simulations in three spatial dimensions. The employed transport
methods range from light bulb models in Newtonian spacetime, to moment schemes,
and even to Monte Carlo schemes in full \ac{GR} \citep[\eg{}][]{Ruffert:1996by,
  Rosswog:2003rv, Shibata:2011kx, Galeazzi:2013mia, 
  Foucart:2017mbt, Ardevol-Pulpillo:2018btx, Foucart:2018gis, Gizzi:2019awu,
  Weih:2020qyh,Foucart:2020qjb, Gizzi:2021ssk,Radice.etal:2022}. Our
understanding of neutrino physics and transport in \ac{BNS} mergers largely
benefits from \ac{CCSN} modelling \citep[][and references
  therein]{Janka:2017vlw, Mezzacappa:2020oyq}. However, compared to the wealth
of literature regarding neutrinos in \ac{CCSN}e, only few studies in the past
have examined neutrino luminosities and mean energies in \ac{BNS} mergers
\citep{Ruffert:1996by, Rosswog:2003rv, Sekiguchi:2015dma, Palenzuela:2015dqa,
  Foucart:2015gaa, Foucart:2016rxm, Wu:2017drk, George:2020veu,
  Kullmann:2021gvo,Radice.etal:2022}. From these seminal studies, a few robust
features emerged. Due to the initial neutron richness, electron antineutrinos
dominate over the other flavours. Moreover, heavy flavour neutrinos are more
energetic, since they decouple deeper inside the remnant. Additionally, more
compact \acp{BNS} produce more violent mergers, resulting in larger neutrino
luminosities. Despite the general consensus about these features, quantitative
differences have emerged, such that both the absolute and the relative
importance of the different neutrino species, as well as their temporal
evolution during the transition between the merger and the remnant cooling phase
still remain largely unexplored. One of the main reasons behind these
limitations is that neutrino luminosities are only studied for a few
milliseconds, while neutrino cooling is relevant during the entire cooling
phase, lasting up to tens of seconds.

In this work, we consider \ac{BNS} simulations spanning a wide range in total
mass, mass ratio, and dense matter \ac{EOS}. Moreover, we consider some of the
longest \ac{BNS} merger simulations in 3+1 \ac{NR}. We also consider the effects
of the inclusion of physical viscosity of magnetic origin in our simulations.
Based on this ample trove of data, we endeavour to find patterns, trends and
commonalities in the temporal evolution of the neutrino luminosities and mean
energies. We strive to identify in neutrino data universal relations, \ie,
relations between parameters describing neutrino emission and quantities
characterising \ac{BNS} models that are \ac{EOS} independent. Similar relations
have been found in the context of \ac{NS} structure and \ac{GW} emission
\citep{Yagi:2016bkt, Carson:2019rjx, Paschalidis:2017qmb, Pani:2015nua,
  Rezzolla:2016nxn, Godzieba:2020bbz, Bernuzzi:2014kca}. The broad scope of our
data sample, which allows us to avoid as much as possible being biased towards a
too specialised subset of \ac{BNS} merger configurations, represents a major
innovation of this work.

All the simulations considered in this work, in addition to being homogeneous
with respect to the general numerical setup, share the same neutrino physics
input and transport scheme. In particular, the minimal set of necessary neutrino
reactions has been included (see the main text and \tabref{reactions} for
details). Moreover, neutrino transport is taken in account using the combination
of a leakage scheme and a so-called M0 scheme. These schemes attempt to strike a
balance between computational cost and physical realism. In our setup, neutrinos
are assumed to be massless and we neglect neutrino oscillations.

The paper is organised as follows: in \secref{methods} we summarise the
numerical methods employed to perform the simulations, which we base our
analysis on; \secref{sim_method_oview} describes our simulation sample, the
overall properties of neutrino emission, and the analysis strategy that we
follow; \secref{results} contains the main results of this work, in the form of
a detailed analysis of the properties of neutrino emission in \ac{BNS} mergers
and their likely explanation in terms of the system dynamics. We discuss our
results in the context of multimessenger astrophysics in \secref{discussion}. We
finally summarise our findings and discuss their implications in
\secref{conclusion}. Several appendices provide additional details on our
analysis, including information about each simulation in our sample.

% ------------------------------------------------------------------------------
\section{Methods}
\label{sec:methods}
% ------------------------------------------------------------------------------

% ------------------------------------------------------------------------------
\subsection{Numerical setup}
\label{sec:NR}
% ------------------------------------------------------------------------------

We base our analysis on results collected from a large sample of \ac{BNS}
mergers simulations in \ac{NR}. All simulations share the same numerical setup
and evolution scheme. In the following, we summarise them and we briefly
introduce the codes used to produce our data. More details can be found in
Ref.~\citep{Radice:2018pdn}.

The \ac{BNS} initial data are evolved with the
infrastructure provided by the \verb|Einstein Toolkit| \citep{Loffler:2011ay,
  Schnetter:2003rb,Reisswig:2012nc}. The hyperbolic sector of Einstein's field
equations is evolved with the Z4c formalism \citep{Bernuzzi:2009ex}, implemented
in the \verb|CTGamma| solver \citep{Pollney:2009yz,Reisswig:2013sqa}. Moreover,
general relativistic hydrodynamics is handled by the \verb|WhiskyTHC| code
\citep{Radice:2012cu,Radice:2013hxh, Radice:2013xpa, Radice:2016dwd,
  Radice:2018pdn}. The code solves Euler's equations for the balance of energy
and momentum:
\begin{equation}
	\label{eq:euler}
	\nabla_\nu T^{\mu\nu} = Q u^{\mu}\,,
\end{equation}
where $T^{\mu\nu}$ is the stress-energy tensor and $Q$ is the net energy
deposition rate due to the absorption and emission of neutrinos and
antineutrinos (see \secref{nu-physics}). \verb|WhiskyTHC| evolves neutron and
proton number densities separately as:
\begin{equation}
	\label{eq:conservation}
	\nabla_\mu\left(n_{\textnormal{p,n}} u^\mu\right) = R_{\textnormal{p,n}},
\end{equation}
where $n_{\textnormal{p,n}}$ are the proton and neutron number densities,
respectively, $u^\mu$ is the fluid four-velocity and $R_{\textnormal{p,n}}$ is
the net lepton number exchange rate due to the absorption and emission of
electron flavour neutrinos and antineutrinos. Due to charge neutrality the
electron fraction is directly related to the proton number density, \ie
$Y_{\textnormal{e}} \equiv n_{\textnormal{e}}/(n_p+n_n) =
n_{\textnormal{p}}/(n_p+n_n)$. Neutrino emission and cooling are handled with a
leakage scheme, while neutrino absorption and heating in optically thin
conditions are treated with the so-called M0 scheme (see \secref{nu-physics}).
Eqs.~\eqref{eq:euler} and~\eqref{eq:conservation} are closed by a
finite-temperature, composition dependent, nuclear \ac{EOS} (see \secref{EOS}).
The code also implements the \ac{GRLES} method to account for turbulent
viscosity of magnetic origin (see \secref{viscosity}).

The computational domain of the simulations is a cube of side $\unit[\sim
  3024]{km}$ centred on the binary’s centre of mass. The code uses a box-in-box
Berger-Oliger \ac{AMR} scheme with refluxing \citep{Berger:1984zza,
  Berger:1989a} provided by the \verb|Carpet| module of the \verb|Einstein|
\verb|Toolkit|, and composed of seven refinement levels. The finest refinement
level covers both \acp{NS} during the inspiral and the remnant after the merger,
and it has a resolution of $h\approx\unit[246]{m}$ (for grid setup named here
low resolution; LR \acused{LR}), $h\approx\unit[185]{m}$ (standard resolution;
SR \acused{SR}) or $h\approx\unit[123]{m}$ (high-resolution; HR\acused{HR})
\citep[see also][]{Bernuzzi:2020txg}.

% ------------------------------------------------------------------------------
\subsection{Relevant simulation parameters}
\label{sec:sim_parameters}
% ------------------------------------------------------------------------------

Each \ac{BNS} is characterised by the gravitational masses of the two \acp{NS}
at infinity, $M_{A,B}$\footnote{Here and in the following the subscripts $A$ and
$B$ refer to the most and least massive star of a \ac{BNS} system,
respectively.}. The total gravitational mass and mass ratio are defined as
$M_{\textnormal{tot}}=M_A+M_B$ and $q=M_A/M_B$, respectively. A further
characterisation system
is provided by the dimensionless reduced tidal deformability $\ltilde$, since it
also depends on the stars' \ac{EOS}. It is a weighted average of the
dimensionless tidal deformabilities $\Lambda_i$, $i\in{A,B}$, of the two
\acp{NS}, defined as
\citep{Favata:2004wz}:
\begin{equation}
\label{eq:tilde_lambda}
  \ltilde =
  \frac{16}{3}\frac{\left(M_A+12M_B\right)M_A^4\Lambda_A}{M_\textnormal{tot}^5} +
  (A\leftrightarrow B)\,.
\end{equation}
In Eq.~\eqref{eq:tilde_lambda} the notation $(A\leftrightarrow B)$ indicates a
second term identical to the first except that the indices $A$ and $B$ are
exchanged. The dimensionless tidal deformabilities in turn are related to the
quadrupolar Love number, $k_2$, describing the static quadrupolar deformation of
a star in the gravitoelectric field of the companion
\citep{Hinderer:2007mb}, by:
\begin{equation}
  \Lambda_i=\frac{2}{3} k_2 C_i^{-5}\,,
\end{equation}
where $C_i=GM_i/c^2R_i$ is the \ac{NS} compactness and $R_i$ is the areal radius
prior to deformation.

The initial data for all the selected simulations are constructed by solving for
irrotational stars of varying masses and different \acp{EOS}, using the spectral
elliptic solver \verb|LORENE| \citep{Gourgoulhon:2001ec}. The binaries are set
to quasi-circular orbits at an initial separation which, in most cases, is
\unit[45]{km}. This orbital separation corresponds to an inspiral phase of $2-3$
orbits before merger. Note that our results do not depend sensitively on the
initial separation or the number of orbits before merger, since neutrino
emission is linked to the dynamics of the system in the post-merger phase. The
\ac{EOS} used in solving for the initial data are the minimum temperature slice
of the \ac{EOS} table used for the evolution composition fixed assuming
neutrinoless beta-equilibrium.

In the following, we use the term \textit{model} to describe a \ac{BNS} system
with a given combination of initial masses and \ac{EOS}. For each model, we can
have multiple realisations of it, \ie \textit{simulations}, which differ from
one another by having been run at different resolution, or by including or not
a model of the magnetic viscosity of turbulent origin.

% ------------------------------------------------------------------------------
\subsection{Input physics}
\label{sec:in-physics}
% ------------------------------------------------------------------------------

% ------------------------------------------------------------------------------
\subsubsection{Neutrino transport}
\label{sec:nu-physics}
% ------------------------------------------------------------------------------

Since the focus of the present work are the properties of neutrino emission, we
provide here a brief, yet fairly detailed, description of the methods of
neutrino transport implemented in the simulations that we use. These methods (a
leakage scheme and the so-called M0 scheme) are described in detail in
Refs.~\citep{Galeazzi:2013mia, Radice:2016dwd} and references therein. They are
both ``grey'' schemes, \ie schemes in which the dependence of various quantities
on the energy of the neutrinos is not explicitly taken into account: instead,
energy-averaged quantities are considered. They account for three
distinct neutrino species: electron neutrinos, $\nue$; electron antineutrinos,
$\nua$; and a collective species for heavy neutrinos, $\nux$. The last one
models muonic and tauonic neutrinos and antineutrinos as a single species
of statistical weight 4.

%-------------------------------------------------------------------------------
\paragraph{Neutrino emission.}
\label{sec:nu_leakage}
%-------------------------------------------------------------------------------

The emission of neutrinos from the fluid and the subsequent loss of energy
is described by a \ac{NLS}. It is based on the method
outlined in Ref.~\citep{Ruffert:1996qu}, where the local thermodynamical
equilibrium chemical potential is used everywhere for all neutrino species while
computing opacities as in Ref.~\citep{Rosswog:2003tn}. \tabref{reactions} lists
the reactions taken into account by this scheme to compute the neutrino
production free rates, $R_\nu^{\textnormal{free}}$, $\nu\in\{\nue,\nua,\nux\}$,
the free energy release, $Q_\nu^{\textnormal{free}}$, and the neutrino
absorption, $\kappa_{\nu,\textnormal{a}}$, and scattering,
$\kappa_{\nu,\textnormal{s}}$, opacities. These reactions include charged
current absorption reactions on free nucleons, namely electron neutrino and
antineutrino absorption on free neutrons and protons, respectively; and their
inverse reactions. The direct ones are the main responsible for the absorption
of $\nue$ and $\nua$ both in optically thick and thin conditions, and they
provide a relevant contribution to neutrino opacity. The inverse ones are the
main processes responsible for the production of electron neutrinos and
antineutrinos in hot and dense matter. Additionally, we consider the production
of neutrino pairs of all flavours through electron-positron annihilation,
nucleon-nucleon bremsstrahlung and plasmon decay. The first one is expected to
be the most relevant source of $\nux$'s in mildly and non-degenerate matter
conditions, while the second one at very high density
\citep{Thompson:2002mw,Burrows:2004vq}. We neglect their explicit contribution
to the absorption opacity, since we expect it to be subdominant due to the pair
nature of the inverse reactions, while their thermalisation effect is implicitly
taken into account inside a \ac{NLS}. Neutrino scattering off free nucleons is
included as a major source of scattering opacity for neutrinos of all flavours
and it is treated in the elastic approximation. In the case of $\nue$'s and
$\nua$'s, this opacity contribution is comparable to the one of absorption
reactions, while in the case of $\nux$ this is the dominant one \citep[see
  e.g.][]{Endrizzi:2019trv}. Coherent scattering off nuclei is also included,
even if the paucity of nuclei makes its impact negligible in the context of
\ac{BNS} mergers. It is important to recall that, at leading order, both the
absorption and the scattering opacity off free nucleons depends quadratically on
the energy of the incoming neutrinos. This quadratic dependence is taken into
account when computing absorption opacities for the M0 scheme.

The scheme distinguishes number density weighted opacities,
$\kappa^0_{\nu,\textnormal{a}}$ and $\kappa^0_{\nu,\textnormal{s}}$, that
determine the rate at which neutrinos diffuse out of the material, from energy
density weighted opacities, $\kappa^1_{\nu,\textnormal{a}}$ and
$\kappa^1_{\nu,\textnormal{s}}$, that determine the rate at which energy is
released due to the loss of neutrinos. The neutrino optical depth $\tau_\nu$ is
evolved in time following the scheme presented in \citep{Neilsen:2014hha}, which
allows the optical depth profile to adapt to the complex geometry of the system.
In particular, the optical depth evolves as:
\begin{equation}
  \tau_\nu^{n+1}=\max(
  (k_{\nu,\textnormal{s}} +
  k_{\nu,\textnormal{a}})\textnormal{d}l + \tau_\nu^n)\,,
\end{equation}
where $\textnormal{d}l$ is a local displacement of one grid point and the
maximum is taken over all spatial directions.

The optical depth is used to define the effective emission rates:
\begin{equation}
  \label{eq:R_eff}
  R_\nu^\textnormal{eff}
  =\frac{R_\nu^{\textnormal{free}}}
  {1+t^0_\textnormal{diff}(t^0_\textnormal{loss})^{-1}}\,,
\end{equation}
where $
t_\textnormal{diff}$ is the effective diffusion time
\begin{equation}
  \label{eq:diffusion_time}
  t_\textnormal{diff}^0 = \mathcal{D}
  \frac{(\tau^0_\nu)^2}{\kappa^0_{\nu,\textnormal{a}}+\kappa^0_{\nu,\textnormal{s}}}\,,
\end{equation}
and $t_\textnormal{loss}$ is the neutrino emission time scale
\begin{equation}
  \label{eq:nu_emission_time}
  t_\textnormal{loss}^0 = \frac{R_\nu^{\textnormal{free}}}{n_\nu}\,.
\end{equation}
In Eq.~\eqref{eq:diffusion_time}, $\mathcal{D}$ is a (dimensionless) tuning
parameter set to 6\footnote{The value of this parameter was suggested as 3 by
\citep{Ruffert:1995fs, Rosswog:2003rv} by random walk arguments. Calibration
against more sophisticated transport methods led to a larger value
\citep{OConnor:2009iuz}. The obtained luminosities are also consistent with
similar approximate schemes employed in BNS merger simulations, e.g.
\cite{Sekiguchi:2011zd, Palenzuela:2015dqa,Lehner:2016lxy}}, and $n_\nu$ in
Eq.~\eqref{eq:nu_emission_time} is the neutrino number density computed assuming
thermal and weak equilibrium. The effective energy emission rates
$Q_\nu^\textnormal{eff}$ are computed with the same procedure as
$R_\nu^\textnormal{eff}$, but using the appropriate opacities and optical
depths. This method of computing effective rates provides a smooth interpolation
between an estimate of the diffusion rate in optically thick condition and the
local production rate in optically thin conditions, based on the optical depth.

\begin{table*}
  \centering
  \begin{tabular}{llc}
    \toprule
    Name & Reaction & Reference\\
    \midrule
    Electron neutrino capture on free neutron &
    $\nue+\textnormal{n}\leftrightarrow\textnormal{p}+\textnormal{e}^-$ &
    \citep{Bruenn:1985en}\\
    Electron antineutrino capture on free protons &
    $\nua+\textnormal{p}\leftrightarrow\textnormal{n}+\textnormal{e}^+$ &
    \citep{Bruenn:1985en}\\
    Electron-positron annihilation &
    $\textnormal{e}^++\textnormal{e}^-\rightarrow\nu+\overline{\nu}$ &
    \citep{Ruffert:1996qu}\\
    Plasmon decay &
    $\gamma+\gamma\rightarrow\nu+\overline{\nu}$ &
    \citep{Ruffert:1996qu}\\
    Nucleon-nucleon Bremsstrahlung &
    $\textnormal{N}+\textnormal{N}\rightarrow\textnormal{N}+
    \textnormal{N}+\nu+\overline{\nu}$ &
    \citep{Burrows:2004vq}\\
    Scattering off nucleons &
    $\nu+\textnormal{N}\rightarrow\nu+\textnormal{N}$ &
    \citep{Ruffert:1996qu}\\
    Scattering off nuclei &
    $\nu+\textnormal{A}\rightarrow\nu+\textnormal{A} $ &
    \citep{Shapiro:1983du}\\
    \bottomrule
  \end{tabular}
  \caption{Weak reactions accounted for in the neutrino transport schemes. The
    following notation is used: $\textnormal{N}\in\{\textnormal{n},
    \textnormal{p}\}$ denotes a free nucleon, $\textnormal{A}$ a nucleus,
    $\nu\in\{\nue,\nua,\nux\}$ a neutrino. The "Reference" column accounts for
    the corresponding rate implementation.}
  \label{tab:reactions}
\end{table*}

%-------------------------------------------------------------------------------
\paragraph{Neutrino transport and absorption in optically thin conditions.}
\label{se:nu_M0}
%-------------------------------------------------------------------------------

Neutrino transport and absorption in optically thin conditions is accounted for
by the moment scheme introduced in \citep{Radice:2016dwd}, called M0 scheme.
Neutrinos are split into two components: a free-streaming one,
$n_\nu^\textnormal{fs}$, and a trapped one, $n_\nu^\textnormal{trap}$, which is
treated with the \ac{NLS} previously described. The M0 scheme evolves the zeroth
moment of the distribution function of free streaming neutrinos, and allows to
compute their number densities and average energies on a polar grid.

This scheme assumes that neutrinos propagate radially at the speed of light
along four-vectors:
\begin{equation}
  \label{eq:nu_propagation}
  k^\alpha = u^\alpha + r^\alpha\,,
\end{equation}
where $r^\alpha$ represents the spatial direction of propagation orthogonal to
the fluid four-velocity $u^\alpha$. This assumption implies that the neutrino
number current $J^\alpha$ equals $n_\nu^\textnormal{fs}k^\alpha$. Under these
assumptions it is possible to show that the free-streaming neutrino number
density, $n_\nu^{\rm fs}$, satisfies:
\begin{equation}
  \label{eq:Boltzmann_first_moment}
  \nabla_\alpha(n_\nu^\textnormal{fs}k^\alpha) = R^\textnormal{eff}_\nu -
  \kappa_{\nu,\textnormal{a}} n_\nu^\textnormal{fs} \,,
\end{equation}
where $\kappa_{\nu,\textnormal{a}}$ is the absorption opacity. This finally
results in an evolution equation for the neutrino number density, namely:
\begin{align}
  \label{eq:nu_number_balance_fs}
  \partial_t(\sqrt{-g}n^\textnormal{fs}_\nu k^t) +
  &\partial_r(\sqrt{-g}n^\textnormal{fs}_\nu
  k^r)=\nonumber\\
  &=\sqrt{-g}(R^\textnormal{eff}_\nu-\kappa^\textnormal{eff}_\nu
  n^\textnormal{fs}_\nu)\,,
\end{align}
where $g$ is the four-metric determinant in spherical coordinates. This equation
is solved on a series of independent radial rays using a first order,
fully-implicit, finite volume method.

Free-streaming neutrino mean energies are estimated under the additional
assumption of a stationary spacetime. Accordingly,
$t^\alpha:=(\partial_t)^\alpha$ is assumed to be a Killing vector so that
$p^\alpha_\nu (\partial_t)_\alpha$, with $p^\alpha$ being the neutrino
four-momentum, is conserved. Therefore the quantity $\varepsilon_\nu = -p^\alpha
t_\alpha$ represents the energy of neutrinos as seen by the ``coordinate
observer'' (a non-physical observer with four-velocity $t^\alpha$), and can be
rewritten as $\varepsilon_\nu = E_\nu\chi$, with $E_\nu$ the neutrino energy as
measured by an observer comoving with the fluid and $\chi=-k_\alpha t^\alpha$.
Within this approximation, the evolution equation for the average neutrino
energy is written as:
\begin{equation}
  \label{eq:energy_evolution}
  n_\nu^\textnormal{fs}k^t\partial_t \varepsilon_\nu +
  n_\nu^\textnormal{fs}\partial_r k^r \varepsilon_\nu=(\chi
  Q_\nu^\textnormal{eff}-\varepsilon_\nu R^\textnormal{eff}_\nu)\,,
\end{equation}
where $Q_\nu^\textnormal{eff}$ and $R_\nu^\textnormal{eff}$ are the effective
neutrino energy and number emission rates
taken from the \ac{NLS}. This equation is solved using a fully-implicit upwind
1st order finite-difference method.

The coupling with hydrodynamics is handled by interpolating quantities from/to
the standard Cartesian \ac{AMR} grid at every timestep, by means of trilinear
interpolation. In the setup of our sample of simulations, the M0 grid consists
of $2048$ rays uniformly spaced in latitude and longitude with a radial
resolution $\Delta r\approx\unit[244]{m}$.

The neutrino number and energy rates computed by the combined leakage and M0
schemes appear as sources in the Euler equations for the \ac{NS} matter, see
\secref{NR}. The coupling in this case is handled, at every timestep, by first
advancing the hydrodynamic quantities in time disregarding neutrino
contributions; neutrino sources are then added to the Euler equations with a
semi-implicit first-order method, in an operator split approach.

% ------------------------------------------------------------------------------
\subsubsection{Equations of state}
\label{sec:EOS}
% ------------------------------------------------------------------------------

In our simulation sample, we consider six finite temperature, composition
dependent \acp{EOS}, namely: LS220 \citep{Lattimer:1991nc}, SLy4
\citep{Douchin:2001sv, daSilvaSchneider:2017jpg}, DD2 \citep{Typel:2009sy,
  Hempel:2009mc}, SFHo \citep{Steiner:2012xt}, BHB$\Lambda\phi$
\citep{Banik:2014qja}, and BLh \citep{Logoteta:2020yxf}. They are widely used in
the literature on \ac{BNS} mergers and are broadly consistent with current
constraints, including astrophysical constraints derived from \ac{GW}
observations \citep{TheLIGOScientific:2017qsa, Abbott:2018exr,
  LIGOScientific:2019fpa, Abbott:2020uma, De:2018uhw}. The above \acp{EOS}
satisfy properties of symmetric nuclear matter at saturation density. They also
provide values for the symmetry energy and its slope in agreement with recent
experimental estimates \citep{Tsang:2012se, Lattimer:2012xj}, with the possible
exception of PREX II results \cite{PREX:2021umo} that reported a quite large
value of the slope of the symmetry energy at saturation density. The matter
modelled by these \acp{EOS} is composed of neutron, protons, electrons,
positrons and photons. One of them, namely BHB$\Lambda\phi$, also includes
$\Lambda$-hyperons. In all our \acp{EOS} we do not take into account the
presence of muons. They would lead to a slight softening of the \ac{EOS} and
their correct inclusion in the \ac{EOS} may be important to describe the
emission spectrum of neutrinos in a more accurate way. This task is left for
future work.

The LS220 and SLy4 \acp{EOS} are based on a non-relativistic liquid drop model
with a Skyrme-like interaction. This model includes surfaces effects and
considers in the low density region an ideal classical gas formed by $\alpha$
particles and heavy nuclei. The latter are treated using the \ac{SNA}. The SLy4
\ac{EOS} employed in this work is constructed on the original Skyrme
parametrisation proposed in Ref.~\citep{Douchin:2001sv} for cold nuclear matter.
It is extended to finite temperature \citep{daSilvaSchneider:2017jpg}, employing
an improved version of the LS220 framework that includes non-local isospin
asymmetric terms, a better treatment of nuclear surface properties, and a more
consistent treatment of heavy nuclei sizes.

The DD2, SFHo, and BHB$\Lambda\phi$ \acp{EOS} are based on \ac{RMF} models.
Besides single nucleons, their composition includes light nuclei (such as
deuterium, tritium, helium) as well as heavy nuclei in \ac{NSE}.
%BHB$\Lambda\phi$ takes into account also the presence of the $\Lambda$-hyperon.
The Lagrangian that models the mean-field nuclear interaction is parametrised
differently for the three \acp{EOS}. While DD2 and BHB$\Lambda\phi$ use density
dependent coupling constants, the SFHo parametrisation employs constant
couplings adjusted to reproduce \ac{NS} radius measurements from low-mass X-ray
binaries. In all three cases, the resulting \ac{RMF} equations are solved in
Hartree’s approximation.

The BLh \ac{EOS} is a microscopical, finite temperature \ac{EOS} obtained as an
extension to the zero-temperature BL \ac{EOS} \citep{Bombaci:2018ksa}. At
densities larger than $0.05$ {\rm fm}$^{-3}$, the latter was derived in the
framework of the non-relativistic many-body \ac{BHF} approach. The
nucleon-\-nucleon interactions are described through a potential derived in the
context of chiral effective theory \citep{Machleidt:2011zz}. They include
two-body interactions \citep{Piarulli:2016vel} calculated up to \ac{N3LO}, and
an effective treatment of three-body interaction up to \acf{N2LO}
\citep{Logoteta:2016nzc}. Both interactions include contributions from
$\Delta$-excitation in the intermediate states of the nucleon-nucleon and
three-nucleon interactions. Finite temperature and arbitrary nuclear composition
effects are calculated using the ﬁnite temperature extension of the
Brueckner–\-Bethe–\-Goldstone quantum many-body theory in the \ac{BHF}
approximation. At low densities the BLh \ac{EOS} is smoothly connected to the
SFHo \ac{EOS}.

The \acp{EOS} employed in this work have been chosen in order to cover a broad
range of stiffness. The stiffest \ac{EOS} is the DD2 \ac{EOS}, while the softest
is the SLy4 \ac{EOS}. These two \acp{EOS} support cold, non-rotating \acp{NS}
maximum masses of $\unit[2.42]{M_\odot}$ and $\unit[2.06]{M_\odot}$,
respectively. Operating on a broad stiffness range is important on one hand to
avoid as much as possible any bias in our analysis, and on the other to allow us
to look for universal relations in our data.

% ------------------------------------------------------------------------------
\subsubsection{Viscosity}
\label{sec:viscosity}
% ------------------------------------------------------------------------------

Slightly more than one third of the models analysed in this work employs the
\ac{GRLES} method of Ref.~\citep{Radice:2017zta} to investigate the impact of
turbulent viscosity on the merger dynamics \citep[see also][for an alternative
  version of this formalism]{Duez:2020lgq}.

In essence, the \ac{GRLES} method consists in taking into account that, due to
finite resolution, any simulation deals only with a coarse-grained version of
the hydrodynamics equations. Formally, this means introducing a linear filtering
operator on the hydrodynamics variables that removes features at small scales
(in our case this is simply the cell-averaging of the finite-volume
discretization of the equations).

In turn, this implies that applying the filtering to the hydrodynamics equations
requires the introduction of closure terms. In the resulting equations, the
turbulent viscosity, $\nu_\textnormal{T}$, is expressed in terms of the sound
speed, $c_\textnormal{s}$, and a free parameter, $\ell_\textnormal{mix}$, that
sets the characteristic length at which the turbulence operates, as
$\nu_\textnormal{T}=\ell_\textnormal{mix} c_\textnormal{s}$. In the simulations
that we consider, $\ell_\textnormal{mix}$ is estimated as a function of the rest
mass density by fitting the results of very high resolution magnetohydrodynamics
\ac{BNS} merger simulations \citep{Radice:2020ids, Kiuchi:2017zzg}.

% ------------------------------------------------------------------------------
\section{Overview of simulations and analysis methods}
\label{sec:sim_method_oview}
% ------------------------------------------------------------------------------

% ------------------------------------------------------------------------------
\subsection{Simulations sample}
\label{sec:sample}
% ------------------------------------------------------------------------------

For our analysis, we consider a subset of the simulations presented in
Ref.~\citep{Perego:2019adq}, whose setup is generic and not targeted to model a
specific \ac{BNS} configuration. In addition, we consider a subset of the
simulations targeted to GW170817 and extensively discussed in
Refs.~\citep{Bernuzzi:2020txg, Nedora:2019jhl, Nedora:2020pak}, and data
extracted from more recent simulations targeted to GW190425
\citep{Camilletti:2022jms}. Finally, we include also eight simulations which
have not been published in earlier works but are presented for the first time in
this article. In summary, we work on a sample of 66 simulations of 51 models of
\ac{BNS} mergers. The range of total gravitational mass $M_{\textnormal{tot}}$
spanned by these models is $\left[2.600,\,3.438\right]\msun$ and the range of
mass ratio $q$ is $\left[1.0,\,1.82\right]$.

The reduced dimensionless tidal deformability of our set of models spans the
wide range $\ltilde \in \left[90,\,1108\right]$. By comparison, data from the
only two detected \ac{GW} signals compatible with \ac{BNS} mergers, namely
GW170817 and GW190425, suggest that for those systems $\ltilde<700$ at the 90\%
confidence level \citep{Chatziioannou:2020pqz}. However, we remind the reader that
$\ltilde$ depends on the masses and mass ratio of the stars, so future events
could also have larger $\ltilde$.

Regarding resolution, the sample consists of 15 \ac{LR} simulations, 49 \ac{SR}
simulations, and 2 \ac{HR} simulations. Where possible, we have decided to work
with \ac{SR} simulations because these tend to offer a better balance between
accuracy and time extent of the post-merger data. Finally, 25 simulations out of
the 66 employ the \ac{GRLES} method described in \secref{viscosity} to account
for viscous effects.

For each simulation we consider the neutrino energy
luminosities and mean energies as extracted at the edge of the M0 computational
domain and integrated over the outermost coordinate sphere. The luminosities and
the mean energies are given in retarded time with respect to the time of merger
(computed as the instant where the amplitude of the strain of the GW is
maximum). The main properties of our sample of \ac{BNS} simulations are
summarised in Tables \ref{tab:sim_info_PC_VSL} and \ref{tab:sim_info_DC_LL} of
\ref{sec:sim_table}.

% ------------------------------------------------------------------------------
\subsection{Neutrino emission: a qualitative overview}
\label{sec:qual_overview}
% ------------------------------------------------------------------------------

\begin{figure*}
  \centering
  \includegraphics[width=\textwidth]{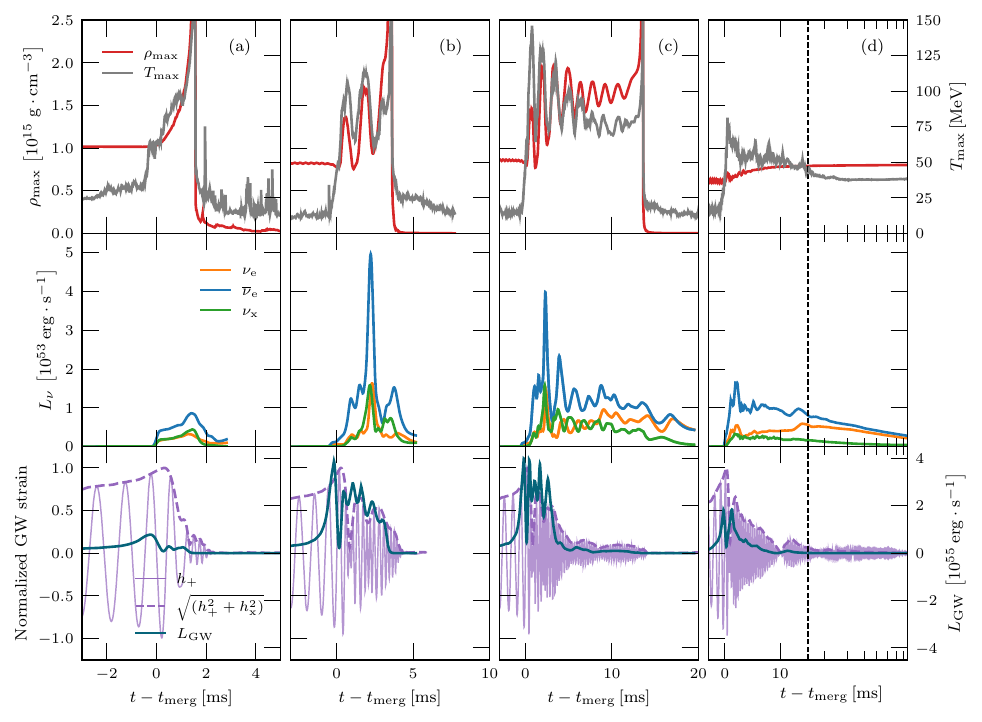}
  \caption{Evolution of the maximum density and temperature (top panels),
    normalised \ac{GW} strain and \ac{GW} luminosity (middle panels), and
    neutrino luminosity for the three neutrino species (bottom panels) for four
    models representative of the considered simulation categories. (a): \ac{PC}
    simulation with LS220 \ac{EOS}, \ac{NS} masses of $1.772\msun$ and
    $1.065\msun$, and no viscosity; (b): \ac{VSL} simulation of an equal mass
    binary ($1.364\msun$) with SFHo \ac{EOS} and no viscosity; (c): \ac{DC}
    simulation of an equal mass binary ($1.364\msun$) with SLy4 \ac{EOS} and no
    viscosity; (d): \ac{LL} simulation with DD2 \ac{EOS}, \ac{NS} masses of
    $1.509\msun$ and $1.235\msun$, with viscosity. The dashed vertical line in
    panel (d) indicates the change from linear to logarithmic scale in the time
    axis.}
  \label{fig:qualitative_overview}
\end{figure*}

We first present an overview of the observed properties of neutrino emission
that are common to large subsets of models and simulations in our sample.
According to the remnant fate, we distinguish our simulations into four
categories: \acf{PC}, \acf{VSL}, \acf{DC} and \acf{LL}. We define \ac{PC}
simulations as the ones for which, at the time of merger, the minimum of the
lapse function over the computational domain decreases monotonically. These
conditions provide a proxy for detecting the collapse of the central object to a
\ac{BH}, \ie for all these simulations the remnant is too heavy to sustain the
formation of an \ac{MNS}. In \ac{VSL} simulations, the merger remnant does not
collapse promptly, but within \unit[5]{ms} from the merger. \ac{DC} simulations
are those for which the collapse happens between \unit[5]{ms} and the end of the
simulation. Finally, in \ac{LL} simulations no \ac{BH} are observed until the
simulation end.

Before discussing the main qualitative features, it is useful to summarise the
origin of the neutrino emission. Neutrinos are emitted mostly from three sites:
1) from matter expanding from the contact interface between the two stars at
merger and soon after it; 2) from the merger remnant, before collapse; 3) from
the innermost, hot part of the post-merger accretion disc. The relative
importance of these three sites varies depending on the dynamics of the system.
For example, in \ac{PC} simulations the remnant collapses immediately and the
accretion disc has very low mass, making the contact interface during merger the
only significant source of neutrinos. In \ac{LL} models with high-mass ratio,
the disc can be rather massive and contribute more to the neutrino emission than
in near equal-mass models. The definitions of the above simulation classes
therefore are also motivated by the mechanics of neutrino emission, since every
simulation within one of these groups has similar properties and behaviour
regarding neutrino luminosities and mean energies.

With reference to \figref{qualitative_overview}, we observe that in all cases
the neutrino luminosity increases just before the merger. During the inspiral,
tidal interaction heats up the two \acp{NS}, however this effect is expected to
be small, $T \unit[\lesssim 1]{MeV}$ \citep[see e.g.][]{Lai:1993di}, and not
accompanied by an intense neutrino emission. However, a non-negligible
luminosity is observed in our simulations also during the inspiral. This is due
to a spurious numeric increase in temperature ($T \unit[\lesssim 10]{MeV}$) at
the \ac{NS} surfaces resulting from the fast \ac{NS} motion inside a much more
dilute atmosphere. Note that this has a negligible effect on the ejecta
composition, since the emitting matter represents a small fraction both of the
total mass and of the ejecta. A significant increase is observed around merger,
due to the direct contact between the \ac{NS} surfaces. This process continues
during the merger and its immediate aftermath, causing the neutrino luminosity
to peak at this time to typical values around
$\unit[10^{53}]{erg~s^{-1}}=\unit[100]{Bethe}$. This is primarily due to the
rapid increase in matter temperature (up to several tens of MeV) due to the
\ac{NS} collision and core fusion, two processes in which kinetic bulk energy is
efficiently converted into thermal energy available to be radiated in neutrinos.

\Ac{PC} simulations present a single, relatively low peak generally between the
merger and \unit[1]{ms} after it. This is due to the main source of neutrinos,
the merger remnant, being cut off by its collapse. In \ac{VSL} simulations this
peak is also present, but typically a few times higher than for \ac{PC} ones. By
contrast, simulations have, typically, between 3 to 4 well defined luminosity
peaks in the first \unit[10-15]{ms} after the merger for each neutrino flavour.
We notice that what we consider as the ``first peak'' is always the highest one,
and we disregard smaller, secondary peaks in the luminosity before it. While
these secondary peaks are likely physical in origin, they cannot be modelled
robustly in our simulations. In particular, their number, position and width
vary with resolution (see \ref{sec:resolution_dep}). While
this statement can apply to the highest peak as well, it can still be
unambiguously defined. Therefore we focus on its analysis. These luminosity
peaks are likely related to the oscillations of the \ac{MNS} in the early post
merger. In this phase, the contractions and expansions of the merger remnant as
it evolves towards a more stable configuration drive shock waves outwards
through the remnant itself and the surrounding matter, raising its temperature
via shock heating and therefore enhancing neutrino emission. Additionally,
matter compressed at the \ac{NS} collision interface and between the two merging
cores is heated up and expelled from the centre of the remnants, expanding and
decreasing its density inside the forming accretion disc. It is however
non-trivial to link neutrino luminosity peaks to, \eg, features in the density
evolution of the \ac{MNS} or in the \ac{GW} signal, as
\figref{qualitative_overview} illustrates. This is due to the fact that
neutrinos can escape the system only when produced or transported outside the
neutrinosphere, which is located $\unit[\sim20]{km}$ from the remnant and is
itself evolving and growing in radius \citep{Endrizzi:2019trv}, making it very
difficult to look for time coincidences.

Most \ac{VSL} remnants approach a \ac{BH}-torus configuration shortly after
merger. We observe that after this point the neutrino luminosity decreases very
rapidly, even if it does not drop to zero, as the inner, hot parts of the
remaining disc are still neutrino sources. A similar behaviour can be seen in
the \ac{DC} case, but the drop in luminosity is not as steep as in \ac{VSL}
simulations and the post collapse luminosity is $\lesssim 50\%$ of the one
before merger. This is due to the fact that the accretion disc mass is usually
larger if the system is less massive (i.e. less prone to a fast collapse) or
asymmetric (i.e. more prone to a tidal deformation of the secondary). Indeed,
since the disc formation process lasts for several milliseconds after merger
\citep{Radice:2018pdn,Nedora:2020pak}, a faster collapse of the central \ac{MNS}
prevents the formation of a massive disc that can sustain a significant
luminosity also after the \ac{MNS} collapse. We recall, in this respect, that
the collapse of the \ac{MNS} drags inside the apparent horizon roughly half of
the disc mass, corresponding to the innermost, hotter portion of the disc.

In \ac{LL} simulations, after the first oscillatory phase, the neutrino luminosity decreases
exponentially in time at a much smaller and steady rate, remaining comparable to
the luminosity observed in the first milliseconds after merger on time scales
even of hundreds of milliseconds, \ie{} comparable to the \ac{MNS} lifetime. In
a \ac{MNS}+disc configuration, both the central object and the disc
significantly contribute to the neutrino emission. The cooling of the central
object and the release of gravitational energy inside the accretion disc are
both active mechanisms in sustaining the neutrino emission over the longer
cooling and accretion time scales. In particular, the optical depth for the most
relevant neutrino energies inside the disc is of the order of a few, while it is
two to three orders of magnitude larger inside the central \ac{MNS}. As a
consequence, the cooling time scale of the disc is a few ms and its luminosity
is sustained until accretion takes place, while the cooling time scale of the
\ac{MNS} is of several seconds and the corresponding luminosity lasts until the
central object is hot enough \citep[see \eg{}][]{Perego:2014fma}. In
Ref.~\cite{Radice:2018xqa}, it was estimated that a LL remnant should liberate
$\sim 0.08 M_{\odot} c^2$ in its cooling phase. This corresponds to $\sim 1.4
\times 10^{53} {\rm erg}$. This is compatible with a total neutrino luminosity
of the order of $10^{53}$-$10^{52}{\rm erg~s^{-1}}$, lasting for a few seconds.

Regarding the relative abundance of neutrino species, during and after the
merger positron captures on free neutrons are favoured since matter is initially
extremely neutron rich ($Y_{\textnormal{e}}\sim0.1$) and hot
$(T\sim\unit[10-50]{MeV})$. Therefore the electron antineutrino luminosity is
dominant in every model. For electron neutrinos the most relevant production
reaction is the capture of electrons on free protons. Due to the relative
paucity of protons, $\nue$ are emitted in a subdominant fashion with respect to
$\nua$. Moreover $\nue$ are also more easily absorbed in typically thin
conditions in their way out from the remnant. Around the time of merger heavy
flavour neutrinos are emitted with a luminosity comparable to that of electron
neutrinos. These heavy flavour neutrinos are produced by very hot matter ($T
\sim $ tens of MeV) initially expelled from the bouncing remnant and rapidly
expanding in optically thin conditions. Electron-positron annihilation and
plasmon decay (which are the most relevant reactions producing $\nux$'s) have an
extreme dependence on temperature (with production rates $Q_{\nux}\propto
T^9$, \citep[see e.g.][]{Galeazzi:2013mia}). Once the remnant has settled on a
quasi-stationary configuration, $\nux$'s emission mostly reduces to the thermal
diffusion from the optically thick central remnant. On the other hand electron
(anti-)neutrinos are mostly produced via electron/positron captures on nucleons,
reactions with a milder dependence on the temperature and happening also inside
the accretion disc. Therefore, as the system stabilises and cools, the heavy
neutrino production is significantly reduced with respect to the other neutrino
flavours. In the case of LL simulations, we also note that with time the
difference in luminosity between $\nue$ and $\nua$ tends to decrease, such that
for all long lasting simulations we observe that $L_{\nua} \sim L_{\nue}$. This
is due to the matter being leptonised, reducing the dominance of the $\nua$'s
emission mechanisms.

The neutrino mean energies present a different pattern with respect to the
neutrino luminosities. In the first few milliseconds after merger, we observe
that they oscillate wildly and rapidly. However, this might be an artefact due
to the approximate character of the neutrino transport schemes we rely on. We
therefore do not attempt to characterise this phase any further. After this
oscillatory phase the neutrino mean energies show a much more stable behaviour,
in fact they are nearly constant until the end of simulation or the collapse to
\ac{BH} of the merger remnant. Clearly this second phase is only present in
\ac{DC} and \ac{LL} simulations.
This behaviour is related to the thermodynamic conditions of matter around the
surfaces of neutrino decoupling. Neutrinos leave the system if emitted outside
the neutrinosphere, and their energy distribution is strongly influenced by the
temperature of the emitting medium at the density where thermal and weak
decoupling between neutrinos and matter occurs. In the aftermath of
\ac{BNS} mergers, the neutrinospheres for each flavour and neutrino energy are
mostly determined by the density profile inside the disc
\citep{Endrizzi:2019trv}, and the latter changes very slowly, only over the
accretion time scale. This in turn implies that the neutrinos are emitted by
matter whose thermodynamic conditions do not significantly vary within the
analysed time.

% ------------------------------------------------------------------------------
\subsection{Analysis strategy}
\label{sec:strategy}
% ------------------------------------------------------------------------------

Based on the general features summarised in \secref{qual_overview}, we focus our
analysis on neutrino luminosities $L_\nu$ and mean energies $E_\nu$ for all
three flavours, \ie for $\nu\in\{\nue,\nua,\nux\}$.

For all simulations we consider the peak luminosity $L_{\textnormal{peak},\nu}$,
which is simply the highest peak for a given simulation. We also examine the \ac{FWHM} $\Gamma$
of the peak by fitting the neutrino luminosity in a window of width \unit[1]{ms}
centred on the peak luminosity time $t_{\textnormal{peak}}$. As a fitting
function, we employ a Gaussian function:
\begin{equation}
  \label{eq:Gauss_fit}
  L=L_\textnormal{peak}\exp{\left(-\frac{\left(t-t_\textnormal{peak}
      \right)^2}{2\sigma^2}\right)}\,,
\end{equation}
where the amplitude and peak centre position are fixed as the peak luminosity
and time, respectively, while the peak width $\sigma$ is the fitting parameter.
Finally the \ac{FWHM} is related to $\sigma$ as:
\begin{equation}
  \label{eq:FWHM}
  \Gamma = 2\sqrt{2\ln{2}}\sigma\,.
\end{equation}

For \ac{DC} and \ac{LL} simulations, we also analyse the values of the
time-averaged luminosity $\langle L_\nu \rangle$ and the time-averaged neutrino
mean energy $\langle E_\nu \rangle$. Explicitly, the time average of a quantity
$X_\nu$ is computed as:
\begin{equation}
    \label{eq:time_average}
    \langle X_\nu \rangle =
    \frac{1}{t_{\textnormal{stop}}-t_{\textnormal{merg}}}
    \int^{t_{\textnormal{stop}}}_{t_\textnormal{merg}}X_\nu(t)dt\,,
\end{equation}
where $t_\textnormal{merg}$ is the time of merger and $t_\textnormal{stop}$ is a
suitable final time. To these time-averages we associate their standard
deviations, computed as:
\begin{equation}
  \label{eq:luminosity_std}
  \sigma_{X_\nu} = \sqrt{\langle X_\nu^2 \rangle-\langle X_\nu \rangle^2}\,,
\end{equation} 
where $\langle X_\nu^2 \rangle$ is the average of $X^2$. For the luminosity, the
time average is computed using a window starting at the time of merger and
extending either to \unit[10]{ms} after merger or until \ac{BH} formation. This
window has been chosen to be long enough so that computing the average is
meaningful, but not so long that in \ac{LL} simulations the final value is
influenced by the late time decrease. For the time-averaged neutrino mean
energies we select a different time window, extending from the point at which
the neutrino mean energies begin to stabilise (typically \unit[2-5]{ms} after
the merger), to either the end of the simulation or \ac{BH} formation. In this
case too the window has been chosen to be long enough to get a meaningful
average energy. Differently from the case of the time-averaged luminosity
however, the final computed value is not sensitive to the end point of the
window, because as noted above mean energies are essentially constant until
collapse or the end of the simulation.

% ------------------------------------------------------------------------------
\section{Results}
\label{sec:results}
% ------------------------------------------------------------------------------

% ------------------------------------------------------------------------------
\subsection{Luminosity peak and peak broadness}
\label{sec:peakL}
% ------------------------------------------------------------------------------

\begin{figure}
  \centering
  \includegraphics[width=.5\textwidth]{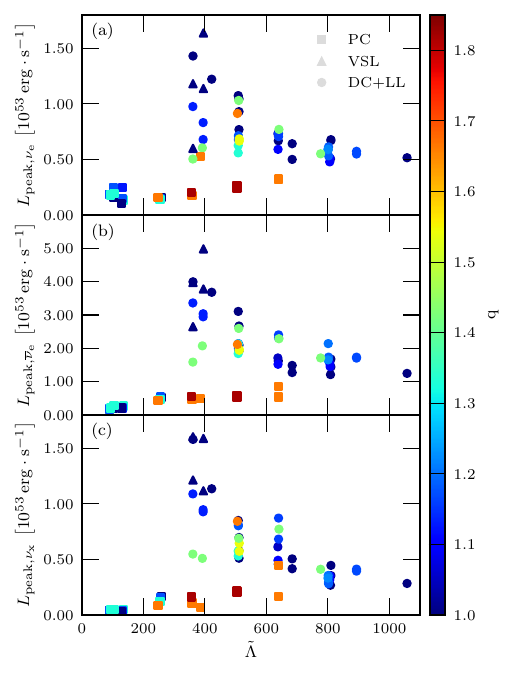}
  \caption{Peak luminosity $L_\textnormal{peak}$ plotted against the reduced
    dimensionless tidal deformability $\ltilde$ for electron neutrinos (panel
    (a)), electron antineutrinos (b) and heavy lepton neutrinos (c). Colour
    indicates the \ac{BNS} mass ratio. Note the different abscissa scales in the
    three panels.}
  \label{fig:peak_lum}
\end{figure}

We start by exploring the peak luminosities for the different neutrino species.
\figref{peak_lum} displays their dependency on the tidal deformability and mass
ratio for our \ac{BNS} models. The peak luminosities approximately span the
range $\unit[1\cdot10^{52}-5.5\cdot10^{53}]{erg~s^{-1}}$ for electron
antineutrinos, while the other two flavours do not go beyond
$\unit[\sim1.7\cdot10^{53}]{erg~s^{-1}}$ even in the most extreme cases. The
extreme neutron richness and high temperatures of the \ac{MNS} matter enhance
the production of electron antineutrinos, hence the differences in the peak
strengths. Within the observed ranges we notice that the peak luminosity values
follow very similar trends in different neutrino species. There is a roughly
constant factor of $\sim3$ between $\nue$ and $\nux$ neutrinos with respect to
$\nua$ ones. This similarity can be understood by noting that the qualitative
behaviour of neutrino emission in this phase is influenced more by the bulk
dynamics of matter than the specifics of neutrino interactions.

\ac{PC} simulations have very low peak luminosities, up to six
times lower than other models. For symmetric systems, this is due to two related
phenomena. The merger remnant collapses right after merger and a massive disc
cannot form since most of the matter is caught in the collapse. 
While equal-mass \ac{PC} simulations cluster at low values of $\ltilde$ (bottom lower
part of \figref{peak_lum}), high-$q$ models with higher $\ltilde$ can also
result in a prompt collapse. With respect to $q$ (and thus to $\ltilde$) we
observe a slightly upward trend, which can be understood by noting that the
lighter object is more easily tidally disrupted as $q$ increases, allowing for a
more massive disc that contributes to neutrino emission.

The remaining simulation categories show a different and much stronger
dependence on $\ltilde$. Equal- or nearly equal-mass \ac{DC} and \ac{LL} models
generally have higher peak luminosity than their asymmetric counterparts and the
luminosity peak values present a downward trend with respect to $\ltilde$.
Systems characterised by a higher tidal deformability contain less compact
stars, which collide in a less violent fashion. Under these conditions, shock
heating is less prominent and neutrino emission is correspondingly smaller. We
observe the largest peak luminosities for $380 \lesssim \ltilde \lesssim 420$;
note however that the limits of this interval depend on the sample of \acp{EOS}
and masses that we consider, and might change by considering a wider sample.
Systems with higher mass ratio fall on a second branch, because the tidal
disruption of one of the two stars leads to less violent coalescences. Because
an increased tidal disruption also tends to increase the disc mass and its
contribution to the neutrino emission, this trend is not strictly monotonic with
respect to the mass ratio. This behaviour can be contrasted to the analogous one
of the time-averaged neutrino luminosity, where both branches (the equal- and
unequal-mass ones) show a much more well delineated trend with respect to
$\ltilde$ (see \secref{averageL}).

Note finally that the \ac{VSL} simulations provide a sort of transition between
the $q \approx 1$ maximum of the \ac{DC}+\ac{LL} sample and the \ac{PC} $q
\approx 1$ branch.

\begin{figure}
  \centering
  \includegraphics[width=.5\textwidth]{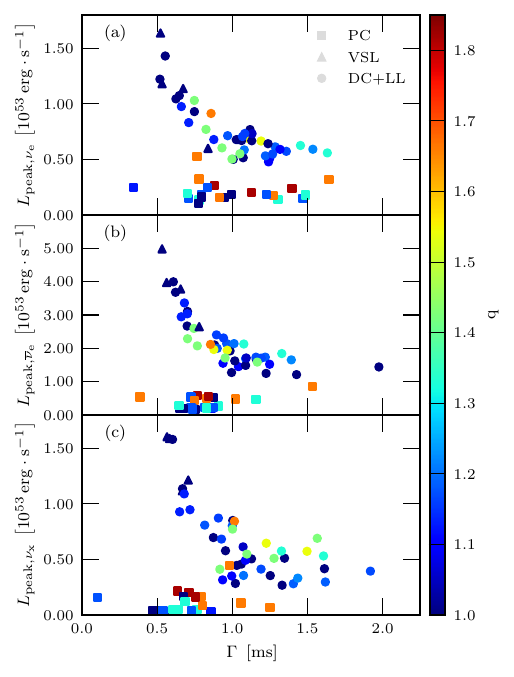}
  \caption{Peak luminosity, $L_\textnormal{peak}$, as a function of the
    \ac{FWHM} of the peak, $\Gamma$, for electron neutrinos (panel (a)),
    electron antineutrinos (b) and heavy lepton neutrinos (c). Colour indicates
    the \ac{BNS} mass ratio. Note the different abscissa scales in the three
    panels.}
  \label{fig:peak_lum_FWHM}
\end{figure}
\begin{figure}
  \centering
  \includegraphics[width=.5\textwidth]{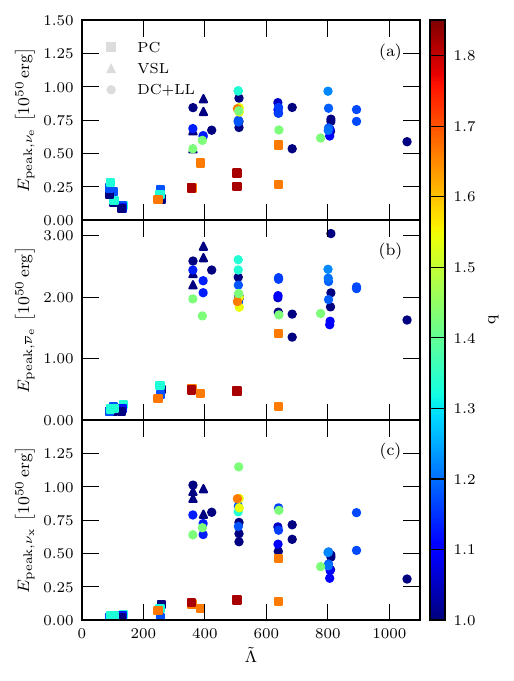}
  \caption{Energy emitted by the peak $E_{\textnormal{peak}}$ as a function of the
    reduced dimensionless tidal parameter $\ltilde$ for electron neutrinos
    (panel (a)), electron antineutrinos (b) and heavy lepton neutrinos (c).
    Colour indicates the \ac{BNS} mass ratio.}
  \label{fig:Ene_peak_lambda}
\end{figure}

We then consider the broadness of the first peak of the neutrino luminosity,
computed as detailed in \secref{strategy}. To measure the goodness of the fit,
we consider the relative residuals between the data and the fit at fixed times.
We observe that, for every flavour and every simulation, they do not exceed 5\%
at any point in the fit interval. Furthermore the coefficient of determination
of the fit $R^2$ is $\sim0.99$ in all cases. To further test the goodness of the
fit we also compare the fitted $\Gamma$ values with the ones calculated directly
from the light curves as the FWHM (when the peak shapes allow this calculation,
\ie for \ac{PC} and \ac{VSL} simulations). The differences in the results of
these two procedures does not exceed the $20\%$ level in most cases, barring two
\ac{PC} outliers, with very low and broad peaks and a relative difference of
$30\%$. \figref{peak_lum_FWHM} presents the dependence of $L_\textnormal{peak}$
to $\Gamma$. Here too we observe two trends, one for the \ac{PC} simulations and
one for the other three categories. In the \ac{VSL}, \ac{DC} and \ac{LL}
simulations, as the peak luminosity decreases with increasing $\ltilde$, the
peak broadness increases instead. The peaks of the neutrino luminosity arise as
the results of shock waves generated by the oscillations of the merger remnant.
The time scale of these oscillations is $\unit[\sim1]{ms}$, which is indeed the
typical value of $\Gamma$. The time scale of the oscillations is related to the
free-fall time scale of the remnant, which scales as
$t_{\textnormal{ff}}\propto\langle{\rho}\rangle^{-1/2}$, where
$\langle{\rho}\rangle$ is the mean density of the \ac{MNS}. Since stars with
higher deformability have generally lower $\langle\rho \rangle$, their
oscillations time scale is longer, and $\Gamma$ is broader. This observation can
be recast in a way which is physically more meaningful. We note that for the
\ac{VSL}+\ac{DC}+\ac{LL} branch, $L_{\textnormal{peak}}$ and $\Gamma$ are
loosely inversely proportional to each other, ans their product roughly constant
(see \figref{peak_lum_FWHM}). The time integral of the Gaussian we employed as
fitting function,
\begin{equation}
  E_{\textnormal{peak}}=\frac{1}{2}\sqrt{\frac{\pi}{\ln2}}
  ~L_{\textnormal{peak}}\Gamma\,,
\end{equation}
represents an estimate of the energy released by the first neutrino peak. We
plot this quantity in \figref{Ene_peak_lambda}. Clearly $E_{\textnormal{peak}}$
is broadly constant, with typical values of $\unit[0.75\cdot10^{50}]{erg}$ for
$\nue$'s and $\nux$'s, and of $\unit[2.25\cdot10^{50}]{erg}$ for
$\nua$'s
with a maximum deviation of a factor of 2. This allows us to provide a very concise
characterisation of the first neutrino luminosity peak: as long as the remnant
does not collapse promptly after merger, the first luminosity peak releases a
roughly constant amount of energy of $\unit[\approx 6 \times 10^{50}]{erg}$.

Finally, it is also clear that the argument outlined above does not apply to
\ac{PC} simulations, which due to immediate collapse have not only very low
$L_\textnormal{peak}$, but also very low $E_{\textnormal{peak}}$. Furthermore no
time scale argument can apply to a collapsed remnant since it does not emit
neutrinos. Indeed while the typical values of $\Gamma$ are the same for \ac{PC}
simulations too, they do not follow any particular trend with respect to either
$\ltilde$ or $q$.

% ------------------------------------------------------------------------------
\subsection{Time-averaged luminosities}
\label{sec:averageL}
% ------------------------------------------------------------------------------

\begin{figure}
  \centering
  \includegraphics[width=.5\textwidth]{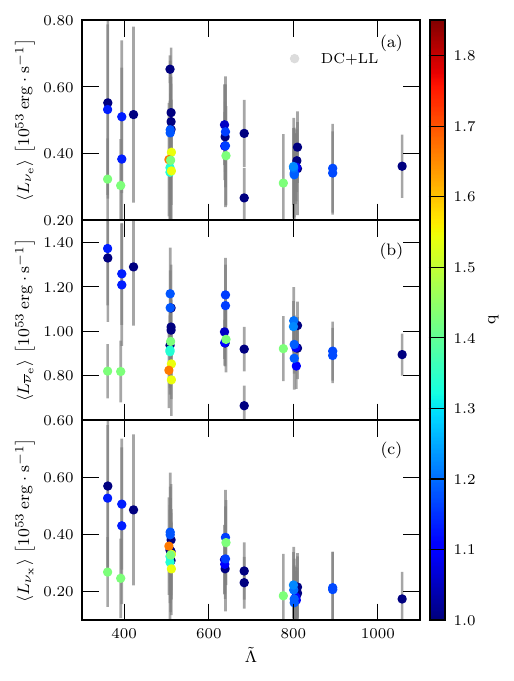}
  \caption{Average luminosity $\langle L_\nu\rangle$ as a function of
    the reduced dimensionless tidal deformability $\ltilde$ for electron
    neutrinos (panel (a)), electron antineutrinos (b) and heavy lepton neutrinos
    (c) for \ac{DC} and \ac{LL} simulations. Colour indicates the \ac{BNS} mass
    ratio while grey bars the standard deviation of the values.}
  \label{fig:average_lum}
\end{figure}

Next we examine the average neutrino luminosity for different neutrino species,
showed in \figref{average_lum}. We limit the analysis to the \ac{DC} and \ac{LL}
classes, since for the other two the average luminosity is not well defined. The
values span the range $\unit[0.6\cdot10^{53}-1.4\cdot10^{53}]{erg~s^{-1}}$ for
electron antineutrinos. The other two flavours mostly vary in the range
$\unit[0.2\cdot10^{53}-0.6\cdot10^{53}]{erg~s^{-1}}$. Similarly to the peak
luminosities, different neutrino species follow very similar trends, differing
in this case 
by a roughly constant factor between 2 and 2.5\,.
The physical explanation of this trend outlined in the previous section applies
here too. The reduction of the scaling factor with respect to 
$L_{\rm peak}$ is due to the unbalanced $\nua$ emission,
which leptonises the remnant, partially suppressing its own emission mechanism.

\begin{figure}
  \centering
  \includegraphics[width=.5\textwidth]{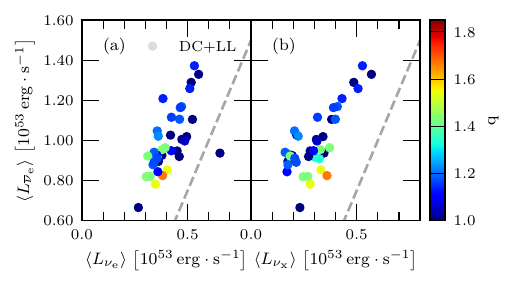}
  \caption{Average luminosity $\langle L_\nu\rangle$ of two neutrino flavours
    plotted against each other for \ac{DC} and \ac{LL} simulations. Panel (a):
    $\langle L_{\nua}\rangle$ vs. $\langle L_{\nue}\rangle$; panel (b): $\langle
    L_{\nua}\rangle$ vs. $\langle L_{\nue}\rangle$. Colour indicates the
    \ac{BNS} mass ratio and the dashed lines have slopes of 2.5, for
    comparison.}
  \label{fig:lum_vs_lum}
\end{figure}

To corroborate these observations we consider the dependence of the average
luminosities for a flavour on the other two, as shown in \figref{lum_vs_lum}.
Clearly there is a linear correlation between the average luminosities of any
two flavours, with a proportionality factor of $\sim2.5$ between electron
neutrinos and antineutrinos, and a slightly smaller factor between electron
antineutrinos and heavy neutrinos (however we refrain from fitting a straight
line trough our data points, judging their quality too poor to warrant such an
analysis).

In \figref{average_lum} we also see that equal-mass models values decrease with
increasing tidal deformability, and in this case too the explanation outlined in
\secref{peakL} holds true. Also in this case varying the mass ratio creates a
second branch, with generally smaller average luminosities than equal-mass
binaries. It is however much more prominent in the case of average luminosities
and our data suggests it is monotonically increasing with respect to $\ltilde$,
at least for $\ltilde \lesssim 700$. 

The explanation of the differences between the peak and the average luminosities
is in the act of taking a time average. Peak luminosities are associated to a
transient and quite violent phase, whose properties cannot be satisfactorily
described with a single parameter such as $\ltilde$ or $q$. Therefore it is to
be expected for the peak luminosities to show a larger variability. On the other
hand taking an average value can help to better isolate a trend present in the
data, as shown in \figref{average_lum}. We find further support for this line of
reasoning by looking at the grey bars in \figref{average_lum}, representing the
time variability of the data around the average values. The bars are quite wide,
spanning a range that in some cases is as wide as the value of the average value
to which they are associated: stated differently, the neutrino luminosities
oscillate rather widely as a function of time. Note that the origin of this
variability is physical, being linked to \eg the oscillations of the central
object. Moreover their width also shows a trend with $\ltilde$: \ac{BNS} mergers
characterised by smaller $\ltilde$ and $q \approx 1$ present a more significant
variability between the peaks and the valleys in the luminosity behaviour,
reflecting the more violent dynamics of the merger.

Different resolutions and/or the inclusion of physical viscosity in the
simulations do not seem to have a significant impact on the major results
concerning the peak and average luminosities. A more detailed discussion about
these points is documented in \ref{sec:viscosity_dep} and
\ref{sec:resolution_dep}.

% ------------------------------------------------------------------------------
\subsection{Long term behaviour of the luminosity}
\label{sec:longL}
% ------------------------------------------------------------------------------

\begin{figure}
  \centering
  \includegraphics[width=.5\textwidth]{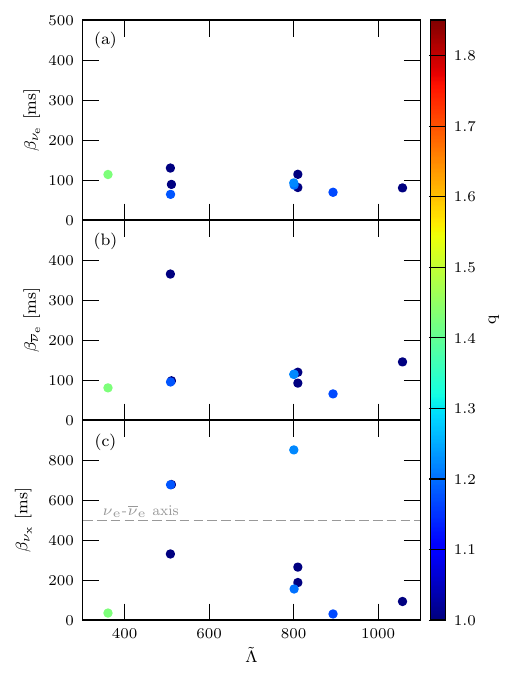}
  \caption{Neutrino luminosity decay time scale $\beta_\nu$ as a function of the
    reduced dimensionless tidal deformability $\ltilde$ for electron neutrinos
    (panel (a)), electron antineutrinos (b) and heavy lepton neutrinos (c) for
    the longest \ac{LL} simulations. Colour indicates the \ac{BNS} mass ratio
    and the dashed line in panel (c) indicates the upper limit of the other two
    panels for ease of comparison.}
  \label{fig:coeff_beta}
\end{figure}

In order to better characterise the time evolution of the neutrino luminosity
over longer time scales, 
we focus on \ac{LL} remnants and
only select simulations that extend further than \unit[20]{ms} after merger, for
a total of 10 simulations that last between 40 and \unit[110]{ms}. We choose the
simple model
\begin{equation}
  L_{\nu}(t)=L_{0,\nu}\exp \left(\frac{\alpha_{\nu}}{t}-\frac{t}{\beta_{\nu}}
  \right) \text{ for }\nu\in{(\nue,\nua,\nux)}\,,
    \label{eq:long_term_exp}
\end{equation}
and fit $L_{0,\nu}$, $\alpha_{\nu}\geq0$ and $\beta_{\nu}>0$ to the neutrino
luminosity curves, starting from the time at which all neutrino flavours
monotonically decrease until the end of the simulation.
Eq.~\eqref{eq:long_term_exp} is an exponential decay, augmented by a term which
allows for deviations from a purely decaying exponential at early times.

The relative residuals between the data and the fit 
vary by up
to 15\% for heavy lepton neutrinos, and up to 10\% for other flavours. We also
observe that the largest residuals are observed at early times, when the
luminosity is still characterised by residual oscillations. We also compute the
coefficient of determination $R^2$, which equals $\sim0.93$ for heavy lepton
neutrinos and $\sim0.99$ for the other two flavours. We conclude that
Eq.~\eqref{eq:long_term_exp} is a good description of the long term evolution of
neutrino luminosities.
  
We focus on the coefficients $\beta_\nu$, \ie the time scale over which the
luminosity drops, shown in \figref{coeff_beta}. This quantity does not seem to
correlate with either $\ltilde$ or $q$, but a few interesting observations are
possible. Typical values of $\beta_{\nu}$ for electron neutrinos and
antineutrinos are of the order of \unit[100]{ms}. Barring a few outlying points,
the corresponding value for heavy neutrinos is between 100 and \unit[400]{ms}.
These are rather long time scales, compared to the dynamical time scales
associated with the \ac{MNS} ($\unit[\sim1]{ms}$). Clearly the decline of
neutrino emission reactions is a steady and relatively slow process, associated
with the cooling of matter in the remnant,
and indeed a
time scale of several hundreds of milliseconds is more in line with both the
cooling time scale of the \ac{MNS} and with the accretion time scale of the disc
\citep[see \eg][and \ref{sec:viscosity_dep}]{Perego:2014fma,Fernandez:2013tya}.

The difference in the decrease rate between $\nue$/$\nua$'s and $\nux$'s is
related to two causes. First, the neutrino origin: $\nue$'s and $\nua$'s are
both emitted by the accretion disc and the central \ac{MNS}, while $\nux$'s
mostly by the latter. Second, the different mean energies at the decoupling
surfaces: $\nux$'s decouple deeper inside the remnant and their spectrum is
significantly harder (see next section). These hotter neutrinos still diffuse
between the equilibrium decoupling surface and the last scattering surface, due
to the opacity provided by quasi-elastic scattering off free baryons. Since the
cross section for this process depends quadratically on the neutrino energy, the
opacity for $\nux$'s (and consequently also its cooling time scale) is
significantly larger and the cooling of the deepest layers proceeds at a slower
pace.

Extrapolating Eq.~\eqref{eq:long_term_exp} to
late times, the total emitted energy
would be a few times $\unit[10^{52}]{erg}$, \ie{}
  almost one order of magnitude smaller than expected. We speculate that
 the exponential decrease we
 observe for $\nue$'s and $\nua$'s is mostly due to the evolution of the
  accretion luminosity. However, once a significant portion of
  the disc has been consumed, the luminosity coming from the cooling of the
  central object will take over and it will likely decrease with a different
  time scale, which our fit over a limited time window cannot account for.

% ------------------------------------------------------------------------------
\subsection{Time-averaged mean energies}
\label{sec:averageE}
% ------------------------------------------------------------------------------

\begin{figure*}
  \centering
  \includegraphics[width=\textwidth]{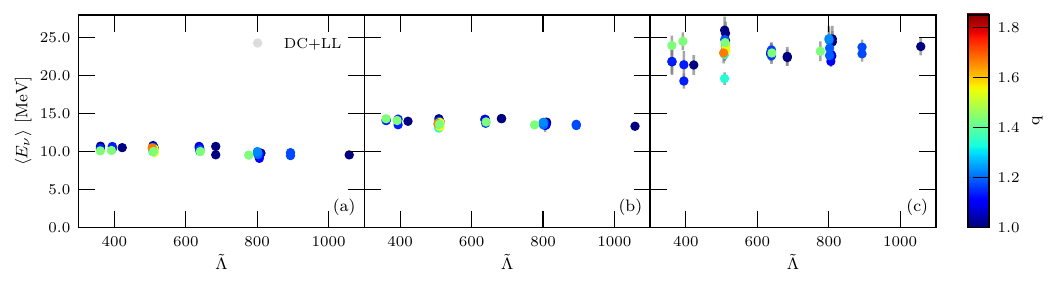}
  \caption{Neutrino average mean energy $\langle E_\nu\rangle$ plotted against
    the reduced tidal deformability $\ltilde$ for electron neutrinos (panel
    (a)), electron antineutrinos (b) and heavy lepton neutrinos (c) for \ac{DC}
    and \ac{LL} simulations. Grey bars indicate the standard deviation in time
    from the average value.}
  \label{fig:mean_ene_av}
\end{figure*}

The neutrino mean energies of \ac{DC} and \ac{LL} simulations plotted in
\figref{mean_ene_av} present a radically different behaviour compared to the
luminosities. The typical energy values are $\unit[\sim10]{MeV}$,
$\unit[\sim14]{MeV}$ and $\unit[\sim23]{MeV}$ for electron neutrinos, electron
antineutrinos and heavy neutrinos, respectively (note that these are the same
values reported in Ref.~\citep{Endrizzi:2019trv} and references therein). This
hierarchy can be explained in relation to the properties of the neutrino
decoupling regions. Of relevance here are the equilibrium surfaces, where
neutrinos decouple from the fluid but are not yet free-streaming. It has been
shown (for long-lived remnants) that these surfaces lie at increasing radii
further away from the remnant for heavy neutrinos, electron antineutrinos and
electron neutrinos, in this order. As temperature also decreases further away
from the remnant, this explains the energy hierarchy between neutrino flavours.
Furthermore the grey bars, representing the time variability of the mean
energies (\cf \secref{strategy}), are extremely small, not being even visible in
the leftmost two panels. Stated differently the neutrino mean energies are
constant in the early post merger phase. This can be explained by noting that
the thermodynamic conditions of matter at the surface of neutrino decoupling are
constant in time in the later part of the evolution, since the location of the
neutrinosphere itself does not evolve significantly at this point
\citep{Endrizzi:2019trv}.

A similar observation also explains why the neutrino mean do not depend on the
masses or \ac{EOS}. The thermodynamics condition at the neutrinosphere are not
only constant in time, but being located at rather large radii
($\unit[\sim20]{km}$), they are also rather insensitive to variations in the
bulk dynamics of the system. We speculate that this could result from two
reasons. On one hand, since the location of the equilibrium decoupling surface
depends at leading order on the matter density, it is likely that the properties
of the accretion disc (and in particular of the density-temperature profile) are
rather independent from the specific binary system, especially once the disc has
reached a quasi-stationary state and a high degree of axisymmetry. On the other
hand, matter temperature also influences the neutrino opacity, mostly through
the energy of the diffusion of thermal neutrinos. If a disc is hotter, the
larger temperatures increase the opacity inside the disc, moving the decoupling
surfaces at larger radii and, thus, lower temperatures. Clearly, these two
effects tends to compensate each other, providing similar decoupling
temperatures in all cases.

% ------------------------------------------------------------------------------
\section{Discussion}
\label{sec:discussion}
% ------------------------------------------------------------------------------

\subsection{Comparison with \ac{GW} luminosities}

\begin{figure}
  \centering
  \includegraphics[width=0.5\textwidth]{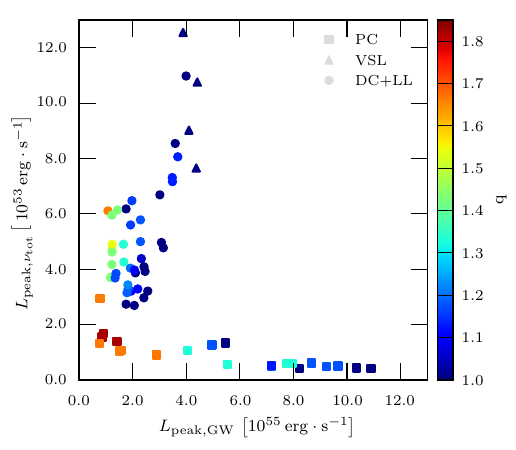}
  \caption{Neutrino peak luminosity $L_{\nu\textnormal{,peak}}$ as a function
    \ac{GW} peak luminosity $L_{\textnormal{GW,peak}}$. Colour indicates the
    \ac{BNS} mass ratio.}
  \label{fig:LGW_VS_Lnu}
\end{figure}

Neutrinos provide the most relevant radiation loss from merging \acp{BNS} on the
cooling time scale of the remnant, but the inspiral and the early post-merger
($\unit[\lesssim20]{ms}$) are \ac{GW}-dominated \citep{Bernuzzi:2015opx}. In
\figref{LGW_VS_Lnu} we present a comparison between the \ac{GW} and total
neutrino peak luminosities, the former being computed as the first peak that the
\ac{GW} luminosity reaches during the merger. We recognise three different
regimes. For \ac{VSL}, \ac{DC} and \ac{LL} near-equal mass ($q \lesssim 1.25$)
systems, there is a correlation between the luminosity in \acp{GW} and $\nu$'s.
This is due to the fact that neutrino radiation is emitted by the same matter
that produces also the \ac{GW} emission. Since the binary properties that boost
the \ac{GW} emission \citep[see, e.g.][]{Zappa:2017xba} are the same ones that
increase the remnant temperature, the two luminosities increase together. If the
mass ratio becomes significantly higher than 1, $\ltilde$ decreases and both
$L_{\rm peak,GW}$ and $L_{{\rm peak},\nu}$ decrease, but the reduction in
\acp{GW} is less significant. This is due to the fact that the strong-field
behaviour for $L_{\rm GW,peak}$ is not precisely captured by $\ltilde$, but by
the so-called $\kappa_2^L$ parameter \cite{Zappa:2017xba}. In particular
$\kappa_2^L$ is the perturbative parameter that enters the 5th order
post-Newtonian term related to the tidal effects in the binary evolution.
Finally, in the \ac{PC} cases the two luminosities follow opposite trends:
\ac{GW} emission is the brightest for \acp{PC} resulting from symmetric
\acp{BNS} merger, but these are the systems for which $L_{\rm peak,\nu}$ is the
lowest (see \secref{peakL}). This effect is partially mitigated by the tidal
disruption of the secondary happening in the high-$q$ cases.

\subsection{Influence on the electron fraction and kilonova colour}

Neutrino interactions change the electron fraction, $\ye$, of matter through
charged current reactions, including electron, positron, $\nue$ and $\nua$
captures on free neutrons and protons. All these reactions are relevant inside
the neutrino surfaces to change $\ye$ from cold, neutrino-less,
$\beta$-equilibrium conditions ($\ye \sim 0.05$ for the relevant densities) to
finite temperature, neutrino trapped equilibrium conditions. Additionally,
neutrino emission and irradiation can further change $\ye$ also outside the
neutrino surface in out-of-equilibrium conditions. Simulations including
neutrino transport can follow in detail the evolution of the ejecta properties.
The contribution of the different processes and their outcome can also be
analysed in post-processing \citep[see e.g.][]{Martin:2017dhc}. Here we want to
focus on a simpler question: how does the variation in the neutrino luminosity
observed in our simulation sample possibly translate in a variation of $\ye$ for
a representative expanding fluid elements?

To answer this question, we assume a simplified model for the evolution of
the $\nu_i$ (with $i=\nue,\nua$) luminosities:
\begin{equation}
L_{\nu_i}(t) =
    \begin{cases}
    \langle L_{\nu_i} \rangle & 0 < t < t_{\nu_i} \, , \\
    \langle L_{\nu_i} \rangle \exp \left( -\frac{t-t_{\nu_i}}{\beta_{\nu}}
    \right) & t \geq t_{\nu_i} \,,
    \end{cases}
\end{equation}
where $\langle L_{\nu_i} \rangle$ are the average luminosity presented in
\secref{averageL} and $t$ the time after the escape of the fluid element from
the neutrino surface. We set $\beta_{\nua}=\beta_{\nue}=\unit[100]{ms}$, based
on \figref{coeff_beta}. For $\nua$'s we assume $t_{\nua}=\unit[10]{ms}$ while
$t_{\nue}$ is fixed by the condition
\begin{equation}
  L_{\nue}(t_{\nue}) = L_{\nua}(t_{\nue})\,,
\end{equation}
meaning that the two luminosities are the same on the time scale set by
$\beta_{\nu}$, as visible in the long term evolution of our \ac{LL} models. 

We further consider constant mean energies, equal to the average ones extracted
from the simulations and presented in \secref{averageE}. We compute the
evolution of $\ye$ based on the equation
\begin{equation}
    \frac{\textnormal{d}Y_e}{\textnormal{d}t} = \lambda_{\nue} (1-\ye) -
    \lambda_{\nua} \ye\,,
    \label{eq: Ye evolution}
\end{equation}
where $\lambda_{\nue}$ and $\lambda_{\nua}$ are the $\nue$ and $\nua$ capture
rates, respectively. The expressions of $\lambda_{\nue}$ and $\lambda_{\nua}$
are taken from equations (C.4)-(C.10) and (3) of \citep{Martin:2017dhc}. To
better focus on the role of luminosities, we neglect the impact of electron and
positron captures outside the neutrino surfaces. This approximation is valid as
long as the temperature in the ejecta expanding outside the neutrino surfaces is
below a few MeV \cite{Qian:1996xt}. According to Ref.\cite{Endrizzi:2019trv},
typical temperatures at the relevant outermost $\nu_e$ surfaces are 3-5 MeV,
depending on the \ac{EOS} stiffness. According to Ref.\cite{Martin:2017dhc},
$e^{\pm}$ captures alone combine in such a way that they do not change
significantly $\ye$ (see their ``capture'' case in figure 6 or 7; however, see
\citep{Sekiguchi:2015dma} for different conclusions, possibly due to hotter
ejecta). From equation (3) the neutrino flux depends on the radial distance and
its evolution. We consider $R(t) = v t + R_0$ where $R_0$ is the typical radial
distance of the neutrino surface and $v$ the ejecta speed. We further know that
neutrino emission is not isotropic, due to the shadow effect provided by dense
matter in the disc along the equatorial plane. We then consider two possible
directions identified by the polar angle $\theta$, namely $\theta=0$ (polar
direction) and $\theta=\pi/2$ (orbital plane), and the angular dependence
implied by equation (3) in \citep{Martin:2017dhc}, assuming $\alpha = 2$, which
corresponds to a polar flux three times larger than the equatorial one.

\begin{figure*}
  \centering
  \includegraphics[width=\textwidth]{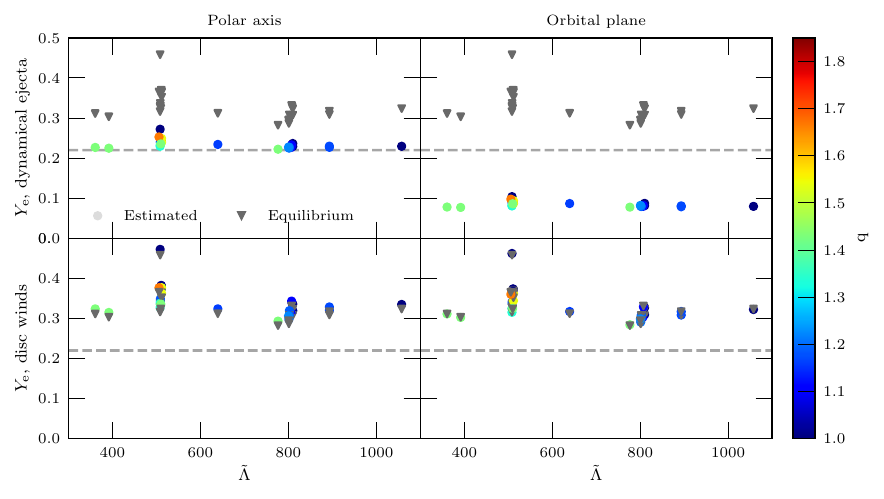}
  \caption{Estimated electron fraction $Y_{\textnormal{e}}$ of an ejected fluid
    element as a function of the reduced tidal deformability $\ltilde$ for
    dynamical ejecta along the polar axis (top left panel, $\ye(t=0)=0.2$) and
    orbital plane (top right, $\ye(t=0)=0.05$), and for a disc wind along the
    polar axis (bottom left, $\ye(t=0)=0.2$) and orbital plane (bottom right,
    $\ye(t=0)=0.2$) for \ac{LL} simulations. Colours indicate the \ac{BNS} mass
    ratio while dashed lines indicate the threshold below which the mass
    fraction of lanthanides and actinides produced in the $r$-process
    nucleosynthesis increases above 10\% for typical ejecta conditions
    ($\ye=0.22$). The grey triangles represent, for each simulation, the
    corresponding equilibrium $Y_e$. See the text for more details.}
  \label{fig:Ye_ejecta}
\end{figure*}

We consider two kinds of ejecta: the dynamical and the disc wind ejecta. The
dynamical ejecta \citep[see \eg{}][]{Ruffert:1995fs,Rosswog:1998hy,
  Rosswog:2001fh,Rosswog:2003tn, Oechslin:2006uk,Sekiguchi:2011zd,
  Rosswog:2012wb,Bauswein:2013yna, Sekiguchi:2015dma,Radice:2016dwd,
  Lehner:2016lxy,Sekiguchi:2016bjd, Foucart:2016rxm,Bovard:2017mvn,
  Radice:2018pdn,Vincent:2019kor,
  Perego:2020evn,Nedora:2020pak,Kullmann:2021gvo} are the matter expelled within
a few dynamical time scales after merger ($\unit[\lesssim5]{ms}$), with typical
average speeds ranging between 0.1-0.3$c$, by tidal torques and shock waves
propagating inside the remnant. We compute the speed of the ejecta as a function
of $\ltilde$ and $q$, based on the fitting formula equation (6) presented in
\citep{Nedora:2020qtd}, using in particular results from the \texttt{M0RefSet}
dataset. This fit is a second order polynomial that predicts the largest speeds
either for $q \lesssim 1.2$ and $\tilde{\Lambda} \lesssim 400$ (corresponding to
very violent PC mergers) or for $\tilde{\Lambda} \gtrsim 1000$, in which tidal
ejection is very effective. Disc winds \citep[see \eg{}][]{Dessart:2008zd,
  Metzger:2008av, Metzger:2008jt, Fernandez:2013tya, Siegel:2014ita,
  Just:2014fka, Metzger:2014ila, Perego:2014fma, Martin:2015hxa,
  Fujibayashi:2017puw, Siegel:2017jug, Metzger:2018uni, Fernandez:2018kax,
  Nedora:2019jhl, Miller:2019dpt, Fujibayashi:2020dvr, Ciolfi:2020wfx,
  Mosta:2020hlh, Just:2021cls, Shibata:2021bbj} are possibly expelled on the
disc evolution time scale ($\sim$ 10ms - 1s) by a variety of mechanisms,
including neutrino absorption itself, nuclear recombination following viscous
spreading of the disc, spiral wave triggered by long-standing $m=1$ bar modes in
the remnant, magnetic processes. In this case, the ejection speed is expected to
be $\sim 0.05-0.1~c$. In our calculation, we consider a representative value of
$0.08c$\footnote{Magnetically-driven and spiral wave winds could be
characterised by larger speeds, closer to the ones of dynamical ejecta, see
\eg{} \cite{Fernandez:2018kax,Nedora:2019jhl,Nedora:2020pak}.}. However, in
order to account for the fact that disc winds are produced on rather long
time scales, $\mathcal{O}(\unit[100]{ms})$, we actually consider a lower
velocity value of $0.008c$ for the first \unit[100]{ms} of evolution, necessary for a fluid element to reach the typical nuclear recombination radius (\unit[250-300]{km}) inside the disc within \unit[100]{ms} \citep[see e.g.][]{Metzger:2008jt,Fernandez:2013tya,Perego:2014fma,Radice:2018xqa,Fujibayashi:2017puw}. Finally, we smoothly connect the two values. Therefore for the wind velocity, we impose:
%\begin{equation}
$v_{\textnormal{wind}} = 0.008\cdot(1 - f) + 0.08\cdot f$,
%\end{equation}
where
%\begin{equation}
$f = ( 1 + \exp{\left( -(t - \unit[100]{ms})/\unit[2]{ms} \right)})^{-1}$.
%\end{equation}
For the ejecta expelled in the orbital plane we assume $R_0=\unit[20]{km}$,
corresponding to the typical radius of the most relevant $\nue$ and $\nua$
neutrino surfaces inside the disc \citep{Endrizzi:2019trv}. For the ejecta
expelled along the polar axis, we consider $R_0=\unit[15]{km}$, corresponding to
the radius of the \ac{MNS}.

This analysis critically relies on the initial $\ye$. A distributions of $\ye$
at the neutrino surface is expected. However, here we rely on representative
values just to focus on the impact of the luminosity variation. For the
dynamical ejecta of tidal origin moving across the equator, we take $\ye(t=0) =
0.05$. This matter is indeed expected not to be significantly reprocessed by
weak processes \citep[e.g.][]{Korobkin:2012uy,Bernuzzi:2020txg}. Dynamical
ejecta expanding close to the poles are more significantly re-processed by
shock-heating and we take $\ye(t=0) = 0.2$ \cite{Sekiguchi:2015dma}. In the case
of the disc wind ejecta, since the ejection happens on the longer viscous time
scale, $e^{\pm}$-captures have time to act and we consider again $\ye(t=0) =
0.2$, irrespectively of the direction \citep{Wu:2016pnw}.

In \figref{Ye_ejecta} we present the final results of our $\ye$ calculations (at
1 second after merger), for \ac{LL} simulations only. In the top (bottom)
panels, we collect results for the dynamical (wind) ejecta, while in the left
(right) panels, along the polar axis (equatorial plane). The grey triangles mark
the equilibrium $\ye$, $Y_{\textnormal{e,eq}}$, defined as the value of $\ye$
obtained by assuming $\textnormal{d}Y_e/\textnormal{d}t=0$ in Eq.~\eqref{eq: Ye
  evolution} and no evolution of the radius \citep[see e.g.][]{Qian:1996xt}. In
practice, it is the value of the electron fraction that the fluid element would
reach if the neutrino absorption time scales were significantly smaller than the
expansion time scales. We observe that $0.28 \lesssim Y_{\textnormal{e,eq}}
\lesssim 0.38$, without any clear trend with $\ltilde$. 
Due to the fast expansion, the final $\ye$ computed by integrating 
Eq.~\eqref{eq: Ye evolution} is smaller than $Y_{\textnormal{e,eq}}$ for the dynamical 
ejecta, but comparable for disc winds, which expand more slowly.
Additionally, the different flux intensities produce an appreciable difference
between the polar and the equatorial directions, that it is more pronounced
in the case of the dynamical ejecta. 
Finally, in all cases there is a weak trend both with
respect to $\ltilde$ and $q$: the change in $\ye$ is smaller for \acp{BNS} with
higher tidal deformability and for more asymmetric binaries: this is consistent
with the variation of the luminosities observed in \secref{results}. To asses
possible systematics, we repeat our calculations using $\ye(t=0)=0.15$ for all
ejecta types and directions. Despite the fact that the differences in the final
$\ye$ decreases among the different cases, we found qualitatively similar
results.

While appropriate to study general and robust trends, we stress that a detailed
evolution requires to extract $Y_e$ from the simulations. We notice, however,
that our results are in good qualitative agreement with simulations results. In
particular, the polar irradiation is effective in increasing $\ye$ in all
possible configurations, due to the larger radiation flux. By comparing the
calculated values of $\ye$ in the different cases with the equilibrium ones, we
can however conclude that the dependence on the final $\ye$ on $\ltilde$ and $q$
is rather weak.

Additionally, in \figref{Ye_ejecta} we highlight $\ye = 0.22$, corresponding to
the value of $\ye$ above which the mass fraction of synthesised lanthanides and
actinides drops below 10\% \citep[see e.g.][]{Lippuner:2015gwa}. Ejecta with
$\ye$ above or around this value is more prone to power a blue kilonova, while
for the ejecta whose electron fraction is below that value the production of
lanthanides and actinides provides larger opacities to photons, resulting in a
redder kilonova peaking at later times. Our results confirm previous findings:
equatorial ejecta tend to produce red kilonovae in all configurations, while
polar dynamical ejecta produced in equal mass and more compact mergers blue
ones.

\subsection{Comparison with previous results}

The major outcomes of our work are in good qualitative agreement with previous
works. For example, a comparison between the neutrino luminosities produced by
\ac{BNS} mergers with different masses and/or different \acp{EOS} in Numerical
Relativity was carried out in
Refs.~\citep{Sekiguchi:2015dma,Sekiguchi:2016bjd,Palenzuela:2015dqa}. The
reported qualitative behaviours are similar to what we find, with peak
luminosities of the order of several $\unit[10^{53}]{erg~s^{-1}}$, dominant
$\nua$ emission, and an oscillatory phase lasting \unit[10-15]{ms} post merger
followed by a slow decay. As in our analysis, the softer SFHo \ac{EOS}
(resulting in smaller $\ltilde$'s) provides systematically larger luminosities.
A relevant difference is the relative importance between $\nue$'s and $\nux$'s,
whose luminosities are comparable in our simulations and in simulations from
Ref.\citep{Palenzuela:2015dqa}, while $\nue$ luminosities from
Ref.~\citep{Sekiguchi:2015dma,Sekiguchi:2016bjd} are smaller but closer to the
$\nua$ ones. This difference is likely related to the different implementation
details of the neutrino treatment.

Compared with Newtonian simulations, as for example the ones presented in
Ref.~\citep{Ruffert:1996by, Rosswog:2003rv, Rosswog:2012wb}, we see again a
qualitative agreement, but some quantitative differences. In these cases, the
lower neutrino luminosities were probably a consequence of the lower remnant
temperature observed in the less violent merger dynamics that characterise
Newtonian gravity simulations employing stiff \acp{EOS}. It is interesting to
note that values of the luminosities intermediate between ours and the ones
obtained in Newtonian simulations were obtained in Ref.~\citep{George:2020veu},
using a Smoothed Particle Hydrodynamics code with conformally flat spacetime
approximation coupled with a leakage scheme. The duration of the oscillatory
phase were in these models also shorter than ours, probably resulting from a
possibly different post merger dynamics of the remnant. However, the hierarchy
and the numerical values of the mean energies were very compatible with ours and
rather independent on the \ac{BNS} properties. Also the relative importance of
$\nue$ and $\nux$ is closer to our results. Finally, we compare our results with
the ones reported in \citep{Foucart:2016rxm} for a light \ac{BNS} merger
(1.2-1.2 $\msun$) employing the LS220 \ac{EOS}. The rather low luminosities
obtained in this case are in overall agreement with the fact that such a system
is characterised by a relatively large value of $\ltilde$. $\nux$ luminosities
are however more relevant here than in our results. This difference is partially
explained by the larger $\langle E_{\nux} \rangle$ obtained in that analysis. A
more important difference is represented by the different evolution of the
luminosities with time. While also in this case one can see fast oscillations in
all neutrino luminosities on the dynamical time scale, the presence of a strong
peak in the very first post-merger phase is not present in these simulations. On
the contrary, all luminosities tend to increase up to the end of the simulation.
The relatively short duration of the simulation and the need of considering the
neutrino time of flight make the comparison harder in this case.

\subsection{Limitations of the present analysis}
\label{sec:limitations}

It is important to note the several limitations that affect the present
analysis. First of all, since the data we work on has been generated by
numerical simulations, the usual caveats that apply in this context apply in our
case as well, namely the loss of accuracy due to finite resolution and the
difficulty of obtaining proper convergence in the post-merger phase. In
addition, a more serious limitation concerns the algorithms for neutrino
transport that we rely on. They attempt to strike a balance between
computational cost and physical realism, but in doing so neglect some of the
finer details of neutrino dynamics. One such example is the assumption of purely
radial propagation of neutrinos in the M0 scheme, which only approximately
reflects the complex geometry of \ac{BNS} systems. Moreover, a grey \ac{NLS} is
not a proper transport scheme, since it only approximates the diffusion regime
through time scale arguments.

Detailed comparisons between different neutrino treatments in the context of
\acp{CCSN} and \ac{BNS} mergers recently addressed the problem of the accuracy
of approximate neutrino transport schemes in astrophysical environments. The
general outcome is that, while well gauged leakage schemes can still provide a
qualitatively correct picture, the comparison with more sophisticated transport
scheme revels possibly relevant differences at a quantitative level
\citep[\eg{}][]{Dessart:2008zd, Perego:2015agy, Richers:2017awc,
  OConnor:2018sti, Pan:2018vkx, Cabezon:2018lpr, Gizzi:2021ssk}. In the context
of \acp{CCSN} (for which more detailed models are available and the geometry of
the problem is simpler), the accuracy in the neutrino luminosity and mean energy
provided by leakage schemes can be even of the order of $20-30\%$, once directly
compared with moment schemes or even with Boltzmann transport
\citep[e.g.][]{Perego:2015agy,Pan:2018vkx, Cabezon:2018lpr}. In the more complex
and less studied case of \ac{BNS} mergers, the discrepancy possibly increases up
to a factor of a few. A recent direct comparison between the leakage+M0 scheme
(used in this work) and a new M1 scheme \citep{Radice.etal:2022}, both
implemented within the \texttt{WhiskyTHC} code and using the same progenitors
and the same microphysics, revealed that the two schemes provide qualitatively
similar features, but the former tends to overestimate the luminosity by a
factor of $\sim$2. Additionally, the non-trivial angular dependence also
introduces additional uncertainties \citep[see, e.g.,][]{Gizzi:2021ssk}. Because
of these reasons, we have decided to focus mostly on peak and integrated
quantities, stressing in particular trends with respect to global binary
properties and to the neutrino flavours.

Furthermore the neutrino treatment employed in this work uses what we think is
the minimal set of neutrino-matter reactions necessary to account for in
\ac{BNS} merger scenarios, both in terms of reactions and reaction rate
implementations. However a detailed analysis of the role and impact of these and
other missing reactions is presently lacking. One of the main reason is that the
large uncertainties that still plague neutrino transport in \ac{BNS} merger
simulations do not allow to robustly address this problem. In parallel to the
improvement of transport schemes, it would be desirable also to improve the
level of microphysics in the simulations, for example by extending the set of
reactions and by implementing more detailed reactions rates and opacities, more
consistent with nuclear matter properties \citep[see
  \eg{}][]{Horowitz:2001xf,Roberts:2016mwj,Oertel:2020pcg_pub}.

Finally, neutrinos are expected to undergo flavour conversions due to their
small, but non-zero, masses. Neutrino oscillations will occur for the neutrinos
emitted during and after a \ac{BNS} merger. These oscillations will certainly
happen in vacuum and due to matter interaction, in both cases relatively far
from the merger remnant. However, collective and resonant neutrino oscillations
could also happen closer to the neutrino surfaces and above the remnant
\citep[see
  \eg{}][]{Malkus:2012ts,Malkus:2015mda,Zhu:2016mwa,Wu:2017qpc,Richers:2019grc}.
These effects are not included in our simulation setup, but since we are mainly
interested in characterising the energy loss from the remnant this should not be
a major limitation. The possible impact on the ejecta composition and on the
neutrino-antineutrino annihilation is possibly more relevant and requires
dedicated studies, that at the moment are usually done in a post-processing
fashion \citep{Frensel:2016fge,Wu:2017drk,George:2020veu}.

% ------------------------------------------------------------------------------
\section{Conclusions}
\label{sec:conclusion}
% ------------------------------------------------------------------------------

% The paper in one sentence
In this paper, we have analysed the behaviour of the neutrino luminosities and
mean energies produced in the merger of two \acp{NS} and during the first tens
of ms after it.
% Method
We have considered the outcome of 66 \ac{BNS} merger simulations in Numerical
Relativity, exploring 51 distinct models. The various models differ because of
the masses of the binary constituents and the employed \ac{NS} \ac{EOS}. Each
model can correspond to multiple simulations due to the inclusion of viscosity
of physical origin and because of the numerical grid's resolution. The
simulation sample is homogeneous in terms of numerical setup and all simulations
include neutrino emission through a grey neutrino leakage scheme
\citep{Galeazzi:2013mia} coupled to an M0 scheme for the propagation of neutrino
radiation in optically thin conditions \citep{Radice:2016dwd}.

% Summarize the major outcomes
Despite the large variety of conditions, we found that the behaviour of the
neutrino luminosities presents qualitative similarities, mostly depending on the
fate of the remnant.
% Overview of the emission, starting from the LL cases
Assuming that the central remnant does not collapse to a \ac{BH} within the
first \unit[20]{ms} after merger (\ie \ac{DC} and \ac{LL} cases in our
classification), the early post merger phase is characterised by an intense
neutrino emission (with total luminosities in excess of several times
$\unit[10^{53}]{erg~s^{-1}}$), showing a more pronounced first peak (usually
occurring within the first \unit[2-3]{ms} after merger), followed by ample
oscillations whose period is comparable to the dynamical time scale of the
merger remnant. After \unit[10-15]{ms}, the oscillations subside and the
luminosities enter an exponentially decreasing phase.

% Origin of the oscillations
The bulk properties of the remnant, and in particular the matter temperature,
determine the intensity of the emission. More symmetric and compact \acp{BNS},
resulting in more violent mergers and hotter remnants, have larger luminosities.
The formation of shock waves produced by the bouncing central \ac{MNS} and their
propagation through the remnant up to the neutrino surfaces produce this
characteristic peak structure. We additionally find that, unless the merger
results in a \ac{PC}, the neutrino luminosity correlates with the \ac{GW}
luminosities, since they both are enhanced by the same \ac{BNS} properties.

% BH formation
Neutrinos come both from the cooling of the optically thick central \ac{MNS} and
from the innermost part of the accretion disc. The formation of a \ac{BH} in the
centre removes not only the \ac{MNS}, but also a significant fraction of the
disc. Then neutrino luminosities are significantly reduced when a \ac{BH} forms.
If the gravitational collapse happens within the first \unit[5]{ms} (\ac{VSL}
simulations), only the first peak is present. If it happens promptly, \ie
without the formation of a \ac{MNS} (\ac{PC} simulations), only a weak and broad
peak is observed.

% Dependence of Lambda
We then studied the dependence of both the peak and average luminosity (where
the latter is computed over the oscillatory phase) on the reduced tidal
deformability parameter $\ltilde$. We found that for equal or nearly equal
\ac{BNS} mergers that do not collapse too quickly to a \ac{BH} (\ie \ac{LL} and
\ac{DC} cases), the luminosity significantly increases (up to a factor of 3) as
$\ltilde$ decreases, down to $\ltilde \approx 380$. \ac{BNS} mergers
characterised by similar $\ltilde$, but with mass ratios significantly different
from 1 produce a less intense neutrino emission, due to the less violent nature
of the tidally dominated merger dynamics. \Ac{PC} cases populate the
low-$\ltilde$, low-$L_{\nu}$ portion of the result space, with an increasing
trend in both quantities for more asymmetric binaries.

% Dependence of Gamma
We additionally focused on the main luminosity peak. We found that, once the
\ac{PC} cases are excluded, in all cases the peak intensity anti-correlates with
the peak width $\Gamma$: stronger peaks last less than weaker ones. Indeed, the
energy emitted by this peak, $E_{\rm peak} \sim L_{\nu,{\rm peak}} \Gamma$ does
not show any clear trend for non-\ac{PC} models.

% Flavour dependence
All the trends described above apply to all modelled neutrino species, \ie
$\nue$, $\nua$, and $\nux$, the latter being a collective species for heavy
flavour (anti)neutrinos. Due to the neutron richness of the system and to the
tendency of neutron rich matter to leptonise when decompressed and heated up,
$\nua$ emission dominates over $\nue$ and $\nux$, at least during the early
post-merger phase. In particular,
$L_{\textnormal{peak},\nua}\sim3L_{\textnormal{peak},\nue}$ and
$L_{\textnormal{peak},\nue}\sim L_{\textnormal{peak},\nux}$. Similar relations
hold for the luminosity averaged over the first \unit[10]{ms}, even if $\langle
L_{\nua} \rangle / \langle L_{\nue} \rangle \sim 2-2.5$. The reduction of the
difference between the $\nue$ and $\nua$ luminosities becomes more evident at
later times, during the steadily declining phase, as a consequence of the early
remnant leptonisation, driving its neutron-to-proton content towards a new
equilibrium.

% Mean energies
Finally, we investigated the value of the neutrino mean energies and their
dependence on the \ac{BNS} parameters. We found that $\langle E_{\nu} \rangle$
is the least sensitive quantity, for all neutrino flavours, with $\langle
E_{\nue} \rangle \approx \unit[10]{MeV}$, $\langle E_{\nua} \rangle \approx
\unit[14-15]{MeV}$ and $\langle E_{\nux} \rangle \approx \unit[20-25]{MeV}$.
This hierarchy can be easily explained in terms of the different location of the
neutrino surfaces \citep{Endrizzi:2019trv}.

% Impact on Ye
Building on a simplified, yet physically motivated model for the neutrino
luminosities based on our results, we have studied the potential impact of
neutrino irradiation on the electron fraction of the matter expelled from a
\ac{BNS} merger. Our results agree with previous findings: for example, $Y_e$
increases more significantly due to $\nue$ absorption in the polar direction and
for slower disc winds. We further found that the range in luminosities has a
weaker effect than the expansion time scale and the remnant geometry.

The major strengths of this work are the wide sample of models we employed and
their relative homogeneity in terms of numerical setup. They qualify it as the
first systematic study of the properties of the neutrino emission over a wide
sample of \ac{BNS} models available at present. As mentioned in
\secref{limitations}, there are several areas in which our approach could be
improved. Yet we believe that the results presented in this work are relevant
and possibly very useful. This stems chiefly from two considerations. First of
all, while obtaining more precise, accurate and realistic data is indeed
desirable, it is important to start building a phenomenological and theoretical
picture from the data as they are available at present. Secondly, while more
realistic neutrino treatments and overall improvements in simulation machinery
will undoubtedly provide quantitative corrections to the data we collected and
presented here, we believe that our approach captures the fundamental aspects of
neutrino emission in \ac{BNS} mergers. Moreover, our characterisation of
neutrino emission will likely work and find usefulness also as a reference
point, to gauge the accuracy, performance and overall behaviour of the
aforementioned advanced schemes.

Our analysis could also serve as input to study the detectability of neutrinos
produced in a \ac{BNS} merger \citep[see \eg{}][]{Lehner:2016lxy}. Due to their
small cross sections, it will be impossible to detect thermal MeV-neutrinos
produced by a merger at the typical distance of several tens of Mpc (or even
more) we usually expect to observe them. However, in the very unlikely case of a
Galactic \ac{BNS} merger, Hyper-Kamiokande \citep{2018arXiv180504163H} will be
able to detect several tens of thousands neutrinos, similar to the case of a
\ac{CCSN} or even larger due to the larger neutrino luminosities, especially for
$\nua$'s. A \ac{BNS} merger occurring in the outskirt of our Galaxy (where it is
more plausible to happen rather than inside the Galactic disc) will still result
in a few thousands events. A handful of neutrinos could possibly be detected
also if the merger happens in a nearby galaxy, up to a distance of a few times
$10^3 {\rm kpc}$. Our analysis could also be expanded towards the study of the
spatial dependence of neutrino emission, as well as the its late post-merger
properties. This information will be key to study, for example, the role of
neutrino flavour conversions. However we leave these topics for future works.

% ------------------------------------------------------------------------------
\begin{acknowledgements}

We thank the European COST Action CA16214 PHAROS ``The multi- messenger physics
and astrophysics of neutron stars'' for the useful and stimulating discussions
that have inspired this work, and the \texttt{CoRe} collaboration for providing
simulation data. MC, AP and SB acknowledge the INFN for the usage of computing
and storage resources through the \texttt{tullio} cluster in Turin. AP
acknowledges the usage of computer resources under a CINECA-INFN agreement
(allocation INF20\_teongrav and INF21\_teongrav). He also acknowledge PRACE for
awarding him access to Joliot-Curie at GENCI@CEA. SB acknowledges funding from
the EU H2020 under ERC Starting Grant, no.BinGraSp-714626, and from the Deutsche
Forschungsgemeinschaft, DFG, project MEMI number BE 6301/2-1. DR acknowledges
funding from the U.S. Department of Energy, Office of Science, Division of
Nuclear Physics under Award Number(s) DE-SC0021177 and from the National Science
Foundation under Grants No. PHY-2011725, PHY-2020275, PHY-2116686, and
AST-2108467. FMG acknowledges funding from the Fondazione CARITRO, program Bando
post-doc 2021, project number 11745. \Ac{NR} simulations were performed on
SuperMUC-LRZ (Gauss project pn56zo), Marconi-CINECA (ISCRA-B project HP10BMHFQQ,
INF20\_teongrav and INF21\_teongrav allocation); Bridges, Comet, Stampede2 (NSF
XSEDE allocation TG-PHY160025), NSF/NCSA Blue Waters (NSF AWD-1811236),
Joliot-Curie at GENCI@CEA (PRACE-ra5202) supercomputers. This research used
resources of the National Energy Research Scientific Computing Center, a DOE
Office of Science User Facility supported by the Office of Science of the
U.S.~Department of Energy under Contract No.~DE-AC02-05CH11231.
\end{acknowledgements}

\appendix

% ------------------------------------------------------------------------------
\section{Influence of viscosity treatment}
\label{sec:viscosity_dep}
% ------------------------------------------------------------------------------

\begin{figure}
  \centering
  \includegraphics[width=.5\textwidth]{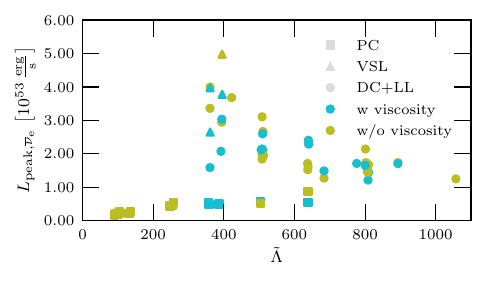}
  \caption{Same as the middle panel of \figref{peak_lum}, but distinguishing
    simulations which take into account the fluid viscosity by means of the
    \ac{GRLES} method from those that do not.}
  \label{fig:Peak_visc}
\end{figure}

\begin{figure}
  \centering
  \includegraphics[width=.5\textwidth]{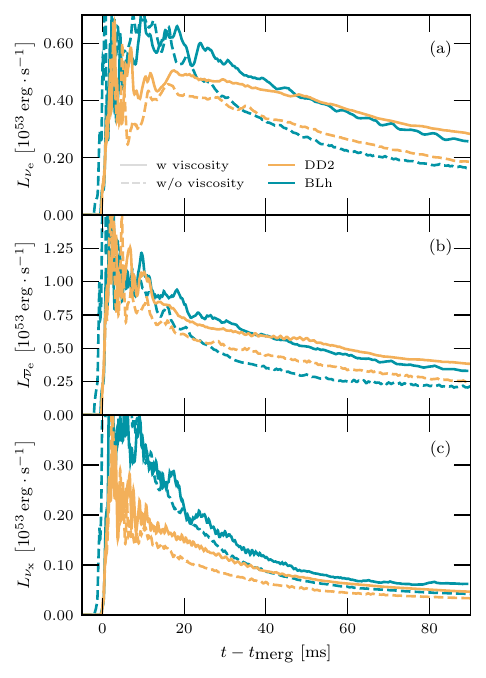}
  \caption{Comparison of the long term behaviour of the luminosities for the
    three different neutrino flavours (top: $\nue$, middle: $\nua$, bottom:
    $\nux$) for two equal mass simulations with $M_A=M_B=1.364 \msun$ employing
    the DD2 (orange lines) and the BLh (green lines) \ac{EOS}. Solid (dashed)
    lines refer to the simulation with (without) GRLES viscosity.}
  \label{fig:Comparison_visc}
\end{figure}

Part of our simulation sample employs an implementation of the \ac{GRLES} method
to effectively model the viscosity that results from the amplification of
magnetic fields and the \ac{MRI} in the post-merger. In this appendix, we
explore the impact that the inclusion or lack of viscosity has on the properties
of neutrino emission. To this end Fig.~\ref{fig:Peak_visc} presents the same
data that has been analysed in Secs.~\ref{sec:peakL}, but separating the
simulations employing the \ac{GRLES} technique from those that do not employ it.

Rather interestingly, the figure highlights how the viscosity has essentially no
impact on the neutrino emission, at least during the first \unit[10]{ms} after
merger A similar behaviour is observed also for the time average luminosities
and mean energies. The explanation is related to the fact that many properties
of neutrino emission are influenced primarily by the bulk dynamics of \ac{BNS}
systems and /or by the thermodynamic conditions of matter at the surface of
neutrino decoupling. Small scale effects due to viscosity can happen on a time
scale comparable to the disc dynamical time scale
\begin{eqnarray}
  t_{\rm dyn} & \sim & \frac{2 \pi}{\Omega_{\rm K}} \approx
  0.011~{\rm s} \left( \frac{M_{\rm rem}}{2.7 \msun} \right)^{-1/2}
  \left( \frac{R_{\rm disc}}{100 {\rm km}} \right)^{3/2} \, ,
\end{eqnarray}
(where $\Omega_{\rm K}$ is the Keplerian angular speed, $M_{\rm rem}$ the
central remnant mass, and $R_{\rm disc}$ the typical disc extension) only on
length scales comparable or smaller than the mixing length. Since the \ac{GRLES}
model was calibrated on \ac{MRI} data for which the mixing length is $\ell_{\rm
  mix} \lesssim \unit[25]{m}$, and only in a narrow density band (for both
higher and lower densities, it decreases rapidly \citep{Radice:2020ids}), the
inclusion of viscous effects have almost no impact on the bulk motion inside the
remnant ($R_{\rm rem} \sim \unit[15]{km} \gg \ell_{\rm mix}$) and inside the
disc ($R_{\rm disc} \sim \unit[100]{km} \gg \ell_{\rm mix}$) during the first ms
after merger.

On the other hand, viscosity induces matter accretion on the longer viscous time
scale. The latter can be estimated as $t_{\rm vis} \sim \nu_{\rm T}/R_{\rm
  disc}$ where $\nu_{\rm T}$ is the viscosity coefficient. For a Keplerian,
Shakura-Sunyaev disc \citep{Shakura:1972te} whose viscosity coefficient is
parametrised in terms of a dimensionless $\alpha$ parameter,
\begin{eqnarray}
  t_{\rm vis} & \sim & \alpha^{-1} \left( \frac{H}{R} \right)^{-1} \Omega_{\rm
    K}^{-1} \approx 12.8~{\rm s} \left( \frac{\alpha}{5\cdot10^{-4}}
  \right)^{-1} \times \nonumber \\ & & \left( \frac{H/R}{1/2} \right)^{-1}
  \left( \frac{M_{\rm rem}}{2.7 \msun} \right)^{-1/2} \left( \frac{R_{\rm
      disc}}{\unit[100]{km}} \right)^{3/2} \,,
\end{eqnarray}
where $H/R$ is the disc aspect ratio. Note that in this formula we used the
estimate $\alpha = \ell_{\rm mix} H$ obtained by considering a Shakura-Sunyaev 
disc with $\ell_{\rm mix}=\unit[25]{m}$. However \ac{BNS} discs are not thin as in the
Shakura-Sunyaev model, so this is only a qualitative estimate. More realistic values
inside the disc are $\alpha \sim 0.01$, see \cite{Kiuchi:2017zzg}, and the resulting
accretion time scale are $\sim \mathcal{O}(1~{\rm s})$.
On such a time scale, simulations employing physical 
viscosity provide larger neutrino luminosities (especially for $\nue$'s and $\nua$'s) 
due to the enhanced accretion rate, as visible in Fig.~\ref{fig:Comparison_visc}.

% ------------------------------------------------------------------------------
\section{Resolution dependence}
\label{sec:resolution_dep}
% ------------------------------------------------------------------------------

\begin{figure*}
    \centering
    \includegraphics[width=\textwidth]{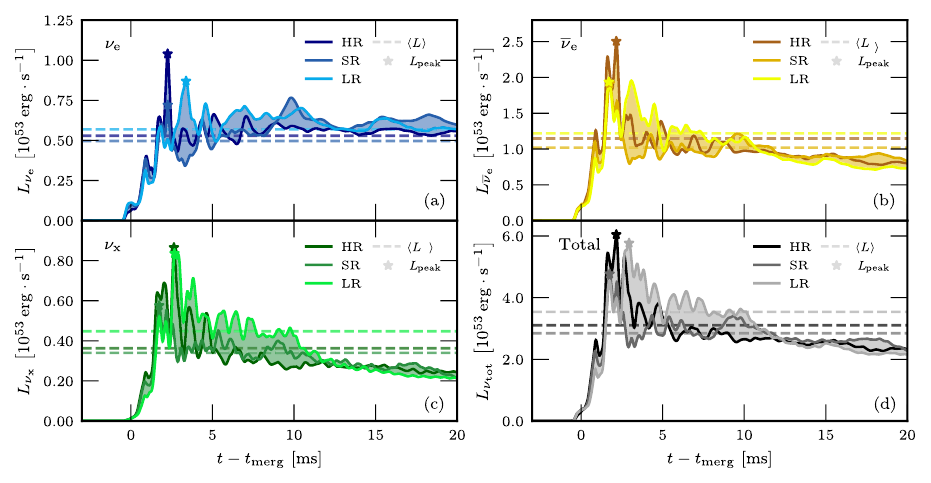}
    \caption{Resolution comparison for the neutrino luminosity evolution of the
      three neutrino flavours ($\nue$, $\nua$ and $\nux$ in panels (a), (b) and
      (c), respectively) and for the total neutrino luminosity (panel (d)). The
      model is a \ac{LL} simulation of an equal mass system
      ($M_A=M_B=\unit[1.364]{\msun}$) employing the BLh \ac{EOS} and \ac{GRLES}
      viscosity. Filled regions cover the range containing all three
      resolutions.}
    \label{fig:resolution_comparison}
\end{figure*}

The simulations used in
this study make use of a box-in-box \ac{AMR} grid with three possible
resolutions, namely: \ac{LR}, \ac{SR} and \ac{HR}. Here we study the effect of
different resolutions on the neutrino emission by considering one model for
which all the three resolutions are available as representative. In each panel
of \figref{resolution_comparison} we present the luminosities obtained by the
different simulations for each of the three neutrino species, alongside their
sum (bottom right panel). The shaded area represents the maximum variability
between resolutions.

On one hand, within the first $\sim\unit[10]{ms}$ after merger, the three
resolutions can differ up to a factor $\sim2$ at corresponding times. This is
due to the fact that the neutrino luminosity oscillates very rapidly and widely,
as a consequence of the complex remnant's dynamics. Clearly point-wise
differences become comparable to the oscillation amplitude as soon as the
remnant's dynamics (characterised by bounces and sound/shock waves, and strongly
dependent on the resolution) accumulates a difference comparable to the
luminosity oscillation periods. Enlarging our view on the whole \unit[0-10]{ms}
interval, we recognise that different resolutions produce a very similar global
behaviour, even if the \ac{HR} simulation tends to have higher maximum peaks and
lower secondary peaks with respect to the \ac{LR} one. On the other hand, in the
exponentially decreasing phase (after the $\unit[10-15]{ms}$ mark), the three
resolutions show a much closer behaviour.

Finally resolution mainly affects the luminosity peak, while the average
luminosities and mean energies are less influenced. In order to quantify their
variations, we average the values of $L_{\textnormal{peak}}$, $\langle L
\rangle$ and $\langle E\rangle$ over the three available resolutions and
consider the maximum relative deviations from these values. While peak
luminosities can vary by up to 20\% from their average value, for average
luminosities and average mean energies this figure is reduced to 15\% and 8\%,
respectively.

This analysis suggests that while the precise values we quote in our results
could of course be improved if we had access to higher-resolution simulations,
the trends we identify are robust and valid.

% ------------------------------------------------------------------------------
\section{Data tables}
\label{sec:sim_table}
% ------------------------------------------------------------------------------

In this section we collect detailed data pertaining to our simulation sample and
our results. \tabref{sim_info_PC_VSL} lists \ac{PC} and \ac{VSL} simulations,
providing details on the initial conditions, \ac{EOS} and the value of peak
luminosities. The same data is provided for \ac{DC} and \ac{LL} simulations in
\tabref{sim_info_DC_LL}. In \tabref{sim_av_info}, we list the values of average
luminosities and average mean energies, and their standard deviations for
\ac{DC} and \ac{LL} simulations.

\begin{table*}
        \centering
        \begin{tabular}{c| c| c| c| c| c| c| c| c| c| c| c}
            \hline\hline
            EOS	&	$M_A$	&	 $M_B$	&	 $q$	&	 $\tilde{\Lambda}$	&	Visc	&	Res	&	  $t_\textnormal{BH}\,\left(t_\textnormal{end}\right)$	&	\multicolumn{3}{c|}{$L_{\textnormal{Peak}}\,\left[10^{53}\, \textnormal{erg}\cdot\textnormal{s}^{-1}\right]$}					&	Reference	\\
	&	$\left[M_\odot\right]$	&	$\left[M_\odot\right]$	&		&		&		&		&	$\left[\textnormal{ms}\right]$	&	$\nue$	&	$\nua$	&	$\nux$	&		\\\hline\hline
\multicolumn{12}{c}{\Acl{PC}}		\\\hline
BLh	&	1.856	&	1.020	&	1.820	&	504	&	\cmark	&	SR	&	1.034 (9.489)	&	0.267	&	0.579	&	0.215	&	\cite{Bernuzzi:2020txg}	\\
BLh	&	1.856	&	1.020	&	1.820	&	504	&	\xmark	&	SR	&	1.128 (2.394)	&	0.237	&	0.515	&	0.205	&	\cite{Bernuzzi:2020txg}	\\
BLh	&	1.914	&	1.437	&	1.332	&	135	&	\xmark	&	SR	&	0.377 (17.495)	&	0.134	&	0.266	&	0.050	&	\cite{Camilletti:2022jms}	\\
BLh	&	1.795	&	1.527	&	1.176	&	131	&	\xmark	&	SR	&	0.400 (21.321)	&	0.150	&	0.219	&	0.041	&	\cite{Camilletti:2022jms}	\\
BLh	&	1.750	&	1.557	&	1.124	&	133	&	\xmark	&	SR	&	17.518 (25.888)	&	0.248	&	0.202	&	0.034	&	\cite{Camilletti:2022jms}	\\
BLh	&	1.654	&	1.654	&	1.000	&	130	&	\xmark	&	SR	&	0.399 (12.294)	&	0.107	&	0.212	&	0.032	&	\cite{Camilletti:2022jms}	\\
DD2	&	1.654	&	1.654	&	1.000	&	258	&	\xmark	&	SR	&	2.907 (2.883)	&	0.157	&	0.529	&	0.165	&	\cite{Camilletti:2022jms}	\\
DD2	&	1.795	&	1.527	&	1.176	&	257	&	\xmark	&	SR	&	1.589 (8.3)	&	0.146	&	0.538	&	0.156	&	\cite{Camilletti:2022jms}	\\
DD2	&	1.914	&	1.437	&	1.332	&	254	&	\xmark	&	SR	&	1.06 (2.895)	&	0.138	&	0.456	&	0.120	&	\cite{Camilletti:2022jms}	\\
DD2	&	2.149	&	1.289	&	1.667	&	248	&	\xmark	&	SR	&	0.580 (1.259)	&	0.160	&	0.437	&	0.084	&	\cite{Camilletti:2022jms}	\\
LS220	&	1.772	&	1.065	&	1.664	&	638	&	\xmark	&	SR	&	1.515 (9.503)	&	0.323	&	0.863	&	0.444	&	\cite{Bernuzzi:2020txg}	\\
LS220	&	1.772	&	1.065	&	1.664	&	638	&	\cmark	&	LR	&	1.374 (14.293)	&	0.323	&	0.539	&	0.168	&	\cite{Bernuzzi:2020txg}	\\
SFHo	&	1.772	&	1.065	&	1.664	&	386	&	\cmark	&	SR	&	1.609 (1.609)	&	0.527	&	0.493	&	0.066	&	\cite{Bernuzzi:2020txg}	\\
SFHo	&	1.795	&	1.527	&	1.176	&	103	&	\xmark	&	LR	&	0.306 (5.067)	&	0.244	&	0.233	&	0.051	&	\cite{Camilletti:2022jms}	\\
SFHo	&	1.795	&	1.527	&	1.176	&	103	&	\xmark	&	SR	&	0.412 (12.303)	&	0.180	&	0.213	&	0.040	&	\cite{Camilletti:2022jms}	\\
SFHo	&	1.654	&	1.654	&	1.000	&	102	&	\xmark	&	SR	&	0.317 (10.247)	&	0.161	&	0.192	&	0.037	&	\cite{Camilletti:2022jms}	\\
SFHo	&	1.914	&	1.437	&	1.332	&	105	&	\xmark	&	SR	&	0.307 (16.365)	&	0.196	&	0.276	&	0.052	&	\cite{Camilletti:2022jms}	\\
SLy4	&	1.772	&	1.065	&	1.664	&	358	&	\cmark	&	SR	&	0.797 (5.732)	&	0.178	&	0.476	&	0.107	&	\cite{Bernuzzi:2020txg}	\\
SLy4	&	1.856	&	1.020	&	1.820	&	357	&	\cmark	&	LR	&	0.637 (5.969)	&	0.202	&	0.551	&	0.163	&	\cite{Bernuzzi:2020txg}	\\
SLy4	&	1.654	&	1.654	&	1.000	&	89	&	\xmark	&	SR	&	0.282 (7.475)	&	0.182	&	0.170	&	0.037	&	\cite{Camilletti:2022jms}	\\
SLy4	&	1.795	&	1.527	&	1.176	&	91	&	\xmark	&	SR	&	0.27 (8.476)	&	0.186	&	0.181	&	0.040	&	\cite{Camilletti:2022jms}	\\
SLy4	&	1.914	&	1.437	&	1.332	&	93	&	\xmark	&	SR	&	0.259 (7.735)	&	0.180	&	0.207	&	0.050	&	\cite{Camilletti:2022jms}	\\\hline\hline
\multicolumn{12}{c}{\Acl{VSL}}\\\hline
SFHo	&	1.364	&	1.364	&	1.000	&	395	&	\xmark	&	SR	&	3.313 (7.634)	&	1.638	&	4.984	&	1.589	&	\cite{Bernuzzi:2020txg}	\\
SFHo	&	1.364	&	1.364	&	1.000	&	395	&	\cmark	&	SR	&	4.69 (22.715)	&	1.138	&	3.781	&	1.118	&	\cite{Bernuzzi:2020txg}	\\
SLy4	&	1.364	&	1.364	&	1.000	&	361	&	\cmark	&	SR	&	2.136 (10.128)	&	0.600	&	2.650	&	1.215	&	\cite{Bernuzzi:2020txg}	\\
SLy4	&	1.364	&	1.364	&	1.000	&	361	&	\cmark	&	SR	&	2.147 (7.14)	&	1.181	&	3.975	&	1.604	&	\cite{Bernuzzi:2020txg}	\\\hline\hline
        \end{tabular}
        \caption{List of \ac{PC} and \ac{VSL} simulations. Columns from left to
          right provide: the mass of the heaviest star; the mass of the lightest
          star; the mass ratio; the reduced dimensionless tidal deformability;
          whether the simulation employs \ac{GRLES} viscosity; the resolution;
          the time of collapse (and the time of the end of the simulation); the
          peal luminosity for the 3 neutrino flavors; the work in which the
          simulation was first presented.}
        \label{tab:sim_info_PC_VSL}
\end{table*}

\begin{table*}
\centering
        \begin{tabular}{c| c| c| c| c| c| c| c| c| c| c| c}				
            \hline\hline									
            EOS	&	$M_A$	&	 $M_B$	&	 $q$	&	 $\tilde{\Lambda}$	&	Visc	&	Res	&	  $t_\textnormal{BH}\,\left(t_\textnormal{end}\right)$	&	\multicolumn{3}{c|}{$L_{\textnormal{Peak}}\,\left[10^{53}\, \textnormal{erg}\cdot\textnormal{s}^{-1}\right]$}					&	Reference	\\
	&	$\left[M_\odot\right]$	&	$\left[M_\odot\right]$	&		&		&		&		&	$\left[\textnormal{ms}\right]$	&	$\nue$	&	$\nua$	&	$\nux$	&		\\\hline\hline
\multicolumn{12}{c}{\Acl{DC}}	\\\hline																						
LS220	&	1.350	&	1.350	&	1.000	&	684	&	\xmark	&	SR	&	22.714 (27.824)	&	0.500	&	1.274	&	0.417	&	\cite{Perego:2019adq}	\\
LS220	&	1.350	&	1.350	&	1.000	&	684	&	\cmark	&	SR	&	18.264 (35.064)	&	0.642	&	1.487	&	0.505	&	\cite{Perego:2019adq}	\\
LS220	&	1.364	&	1.364	&	1.000	&	639	&	\xmark	&	SR	&	15.475 (32.669)	&	0.668	&	1.623	&	0.457	&	\cite{Nedora:2019jhl}	\\
LS220	&	1.400	&	1.330	&	1.053	&	637	&	\xmark	&	SR	&	16.687 (23.163)	&	0.732	&	1.711	&	0.613	&	\cite{Nedora:2020pak}	\\
LS220	&	1.435	&	1.298	&	1.106	&	638	&	\xmark	&	SR	&	16.393 (24.964)	&	0.591	&	1.522	&	0.492	&	\cite{Nedora:2020pak}	\\
LS220	&	1.469	&	1.268	&	1.159	&	639	&	\cmark	&	SR	&	19.89 (33.146)	&	0.706	&	2.404	&	0.872	&	\cite{Nedora:2020pak}	\\
LS220	&	1.635	&	1.146	&	1.427	&	641	&	\cmark	&	SR	&	11.768 (11.768)	&	0.770	&	2.288	&	0.773	&	\cite{Nedora:2020pak}	\\
SFHo	&	1.350	&	1.350	&	1.000	&	422	&	\xmark	&	SR	&	7.492 (28.142)	&	1.649	&	5.409	&	1.736	&	This work \\%\cite{Radice:2018pdn}	\\
%SFHo	&	1.350	&	1.350	&	1.000	&	422	&	\xmark	&	SR	&	11.259 (22.962)	&	1.222	&	3.682	&	1.136	&	\cite{Perego:2019adq}\mc{Removed}	\\
SFHo	&	1.452	&	1.283	&	1.132	&	394	&	\xmark	&	SR	&	10.836 (11.821)	&	0.832	&	2.946	&	0.929	&	\cite{Nedora:2020pak}	\\
SFHo	&	1.452	&	1.283	&	1.132	&	394	&	\cmark	&	SR	&	5.703 (5.703)	&	0.679	&	3.037	&	0.948	&	\cite{Nedora:2020pak}	\\
SLy4	&	1.364	&	1.364	&	1.000	&	361	&	\xmark	&	SR	&	13.367 (21.903)	&	1.431	&	3.996	&	1.580	&	\cite{Nedora:2020pak}	\\
SLy4	&	1.452	&	1.283	&	1.132	&	361	&	\xmark	&	SR	&	12.461 (12.461)	&	0.976	&	3.362	&	1.090	&	\cite{Nedora:2020pak}	\\\hline\hline
\multicolumn{12}{c}{\Acl{LL}}	\\\hline									
BHB$\Lambda\phi$	&	1.364	&	1.364	&	1.000	&	808	&	\cmark	&	LR	&	$>$27.868	&	0.504	&	1.212	&	0.268	&	This work	\\
BLh	&	1.364	&	1.364	&	1.000	&	511	&	\xmark	&	HR	&	$>$51.634	&	0.930	&	2.098	&	0.511	&	This work \\
BLh	&	1.364	&	1.364	&	1.000	&	511	&	\cmark	&	SR	&	$>$91.365	&	0.769	&	1.923	&	0.578	&	\cite{Bernuzzi:2020txg}	\\
BLh	&	1.364	&	1.364	&	1.000	&	511	&	\xmark	&	LR	&	$>$36.737	&	1.045	&	2.669	&	0.697	&	\cite{Nedora:2020pak} \\
BLh	&	1.364	&	1.364	&	1.000	&	511	&	\xmark	&	SR	&	$>$97.211	&	1.075	&	3.107	&	0.850	& \cite{Nedora:2020pak}	\\
BLh	&	1.482	&	1.259	&	1.177	&	509	&	\cmark	&	LR	&	$>$69.074	&	0.678	&	2.142	&	0.804	&	\cite{Nedora:2020pak}	\\
BLh	&	1.482	&	1.259	&	1.177	&	509	&	\xmark	&	LR	&	$>$28.167	&	0.714	&	1.999	&	0.809	&	\cite{Nedora:2020pak}	\\
BLh	&	1.581	&	1.184	&	1.335	&	508	&	\cmark	&	SR	&	$>$9.683	&	0.626	&	2.130	&	0.576	&	\cite{Nedora:2020pak}	\\
BLh	&	1.581	&	1.184	&	1.335	&	508	&	\xmark	&	LR	&	$>$17.493	&	0.559	&	1.843	&	0.532	&	\cite{Nedora:2020pak}	\\
BLh	&	1.699	&	1.104	&	1.539	&	512	&	\cmark	&	LR	&	$>$45.546	&	0.685	&	1.965	&	0.646	&	\cite{Nedora:2020pak}	\\
BLh	&	1.699	&	1.104	&	1.539	&	512	&	\xmark	&	LR	&	$>$29.613	&	0.666	&	1.942	&	0.573	&	\cite{Nedora:2020pak}	\\
BLh	&	1.772	&	1.065	&	1.664	&	506	&	\cmark	&	SR	&	$>$19.987	&	0.914	&	2.118	&	0.844	&	\cite{Bernuzzi:2020txg}	\\
BLh	&	1.635	&	1.146	&	1.427	&	510	&	\cmark	&	SR	&	$>$59.32	&	1.031	&	2.598	&	0.690	&	\cite{Nedora:2020pak}	\\
DD2	&	1.300	&	1.300	&	1.000	&	1057	&	\xmark	&	LR	&	$>$70.012	&	0.515	&	1.248	&	0.284	&	This work	\\
DD2	&	1.364	&	1.364	&	1.000	&	810	&	\xmark	&	SR	&	$>$96.698	&	0.670	&	1.677	&	0.446	&	\cite{Nedora:2019jhl}	\\
DD2	&	1.364	&	1.364	&	1.000	&	810	&	\cmark	&	SR	&	$>$112.545	&	0.678	&	1.442	&	0.355	&	\cite{Nedora:2019jhl}	\\
DD2	&	1.432	&	1.300	&	1.102	&	807	&	\xmark	&	LR	&	$>$41.494	&	0.511	&	1.451	&	0.351	&	This work 	\\
DD2	&	1.435	&	1.298	&	1.106	&	806	&	\xmark	&	LR	&	$>$13.504	&	0.478	&	1.553	&	0.316	&	This work  \\
DD2	&	1.486	&	1.254	&	1.185	&	802	&	\xmark	&	HR	&	$>$58.463	&	0.614	&	1.736	&	0.281	&	This work  \\
DD2	&	1.486	&	1.254	&	1.185	&	802	&	\xmark	&	LR	&	$>$28.276	&	0.533	&	1.691	&	0.297	&	This work 	\\
DD2	&	1.497	&	1.245	&	1.202	&	801	&	\xmark	&	SR	&	$>$88.586	&	0.586	&	2.142	&	0.356	&	\cite{Nedora:2020pak}	\\
DD2	&	1.509	&	1.235	&	1.222	&	800	&	\cmark	&	SR	&	$>$85.657	&	0.592	&	1.654	&	0.332	&	\cite{Nedora:2020pak} \\
DD2	&	1.635	&	1.146	&	1.427	&	776	&	\cmark	&	LR	&	$>$37.477	&	0.551	&	1.711	&	0.411	&	\cite{Nedora:2020pak}	\\
LS220	&	1.400	&	1.200	&	1.167	&	893	&	\xmark	&	SR	&	$>$24.677	&	0.549	&	1.735	&	0.413	&	\cite{Radice:2018pdn}	\\
LS220	&	1.400	&	1.200	&	1.167	&	893	&	\cmark	&	SR	&	$>$48.494	&	0.573	&	1.707	&	0.395	&	\cite{Perego:2019adq}	\\
LS220	&	1.469	&	1.268	&	1.159	&	639	&	\xmark	&	SR	&	$>$35.937	&	0.734	&	2.310	&	0.683	&	\cite{Nedora:2020pak}	\\
SFHo	&	1.635	&	1.146	&	1.427	&	392	&	\cmark	&	SR	&	$>$42.141	&	0.604	&	2.074	&	0.510	&	\cite{Nedora:2020pak} 	\\
SLy4	&	1.635	&	1.146	&	1.427	&	361	&	\cmark	&	SR	&	$>$40.118	&	0.504	&	1.587	&	0.548	&	\cite{Nedora:2020pak} 	\\\hline\hline
        \end{tabular}
        \caption{List of \ac{DC} and \ac{LL} simulations. Columns from left to
          right provide: the mass of the heaviest star; the mass of the lightest
          star; the mass ratio; the reduced dimensionless tidal deformability;
          whether the simulation employs \ac{GRLES} viscosity; the resolution;
          the time of collapse (and the time of the end of the simulation); the
          peal luminosity for the 3 neutrino flavors; the work in which the
          simulation was first presented.}
        \label{tab:sim_info_DC_LL}
\end{table*}

\begin{table*}
        \centering
        \begin{adjustbox}{width=\textwidth}
        \begin{tabular}{c| c| c| c| c| c| c| c| c| c| c| c|c| c| c| c| c| c}
            \hline\hline
            EOS	&	$M_A$	&	 $M_B$	&	 $\tilde{\Lambda}$	&	Visc	&	Res	&	\multicolumn{3}{c|}{$\langle L \rangle\,\left[10^{53}\,\textnormal{erg}\cdot\textnormal{s}^{-1}\right]$}					&	\multicolumn{3}{c|}{$\sigma_{\langle L \rangle}\,\left[10^{53}\,\textnormal{erg}\cdot\textnormal{s}^{-1}\right]$}					&	\multicolumn{3}{c|}{$\langle E \rangle \, \left[\textnormal{MeV}\right]$}	&					\multicolumn{3}{c}{$\sigma_{\langle E \rangle} \, \left[\textnormal{MeV}\right]$}					\\
	&	$\left[M_\odot\right]$	&	$\left[M_\odot\right]$	&		&		&		&	$\nue$	&	$\nua$	&	$\nux$	&	$\nue$	&	$\nua$	&	$\nux$	&	$\nue$	&	$\nua$	&	$\nux$	&	$\nue$	&	$\nua$	&	$\nux$	\\\hline\hline
\multicolumn{18}{c}{\Acl{DC}}\\\hline
LS220	&	1.350	&	1.350	&	684	&	\xmark	&	SR	&	0.267	&	0.665	&	0.231	&	0.033	&	0.090	&	0.033	&	9.571	&	14.344	&	22.546	&	0.256	&	0.373	&	1.248	\\
LS220	&	1.350	&	1.350	&	684	&	\cmark	&	SR	&	0.460	&	0.920	&	0.272	&	0.045	&	0.101	&	0.034	&	10.672	&	14.341	&	22.393	&	0.431	&	0.573	&	1.075	\\
LS220	&	1.364	&	1.364	&	639	&	\xmark	&	SR	&	0.450	&	0.948	&	0.280	&	0.048	&	0.105	&	0.033	&	10.489	&	14.222	&	23.220	&	0.171	&	0.288	&	1.115	\\
LS220	&	1.400	&	1.330	&	637	&	\xmark	&	SR	&	0.486	&	0.997	&	0.312	&	0.057	&	0.121	&	0.049	&	10.678	&	14.250	&	22.880	&	0.313	&	0.502	&	0.936	\\
LS220	&	1.435	&	1.298	&	638	&	\xmark	&	SR	&	0.422	&	0.949	&	0.296	&	0.044	&	0.103	&	0.035	&	10.507	&	14.189	&	23.043	&	0.141	&	0.194	&	1.106	\\
LS220	&	1.469	&	1.268	&	639	&	\cmark	&	SR	&	0.465	&	1.164	&	0.390	&	0.048	&	0.166	&	0.074	&	10.263	&	13.877	&	22.621	&	0.267	&	0.264	&	1.062	\\
LS220	&	1.635	&	1.146	&	641	&	\cmark	&	SR	&	0.393	&	0.964	&	0.372	&	0.051	&	0.149	&	0.059	&	9.997	&	13.900	&	22.990	&	0.095	&	0.180	&	1.135	\\
SFHo	&	1.350	&	1.350	&	422	&	\xmark	&	SR	&	0.481	&	1.401	&	0.533	&	0.136	&	0.426	&	0.142	&	10.010	&	13.688	&	19.916	&	0.346	&	0.427	&	0.656	\\
SFHo	&	1.350	&	1.350	&	422	&	\xmark	&	SR	&	0.517	&	1.290	&	0.486	&	0.104	&	0.264	&	0.098	&	10.511	&	13.990	&	21.411	&	0.282	&	0.287	&	1.306	\\
SFHo	&	1.452	&	1.283	&	394	&	\xmark	&	SR	&	0.510	&	1.259	&	0.506	&	0.098	&	0.229	&	0.099	&	10.668	&	14.265	&	21.443	&	0.564	&	0.602	&	1.835	\\
SFHo	&	1.452	&	1.283	&	394	&	\cmark	&	SR	&	0.383	&	1.209	&	0.430	&	0.082	&	0.274	&	0.113	&	10.338	&	13.533	&	19.296	&	0.677	&	0.512	&	1.010	\\
SLy4	&	1.364	&	1.364	&	361	&	\xmark	&	SR	&	0.552	&	1.330	&	0.570	&	0.123	&	0.287	&	0.124	&	10.704	&	14.178	&	21.824	&	0.516	&	0.486	&	1.707	\\
SLy4	&	1.452	&	1.283	&	361	&	\xmark	&	SR	&	0.532	&	1.373	&	0.527	&	0.095	&	0.256	&	0.097	&	10.426	&	14.078	&	21.889	&	0.372	&	0.324	&	1.631	\\\hline\hline
\multicolumn{18}{c}{\Acl{LL}}\\\hline
BHB$\Lambda\phi$	&	1.364	&	1.364	&	808	&	\cmark	&	LR	&	0.378	&	0.926	&	0.190	&	0.043	&	0.094	&	0.027	&	9.770	&	13.509	&	22.602	&	0.205	&	0.291	&	0.711	\\
BLh	&	1.364	&	1.364	&	511	&	\xmark	&	HR	&	0.473	&	1.005	&	0.309	&	0.076	&	0.137	&	0.049	&	10.550	&	14.014	&	24.661	&	0.120	&	0.186	&	0.748	\\
BLh	&	1.364	&	1.364	&	511	&	\cmark	&	SR	&	0.495	&	1.019	&	0.340	&	0.075	&	0.135	&	0.051	&	10.386	&	13.991	&	25.587	&	0.168	&	0.181	&	1.501	\\
BLh	&	1.364	&	1.364	&	511	&	\xmark	&	LR	&	0.523	&	1.105	&	0.381	&	0.095	&	0.195	&	0.074	&	10.449	&	13.717	&	23.804	&	0.146	&	0.201	&	0.689	\\
BLh	&	1.365	&	1.365	&	508	&	\xmark	&	SR	&	0.653	&	0.937	&	0.345	&	0.017	&	0.042	&	0.030	&	10.789	&	14.324	&	26.001	&	0.132	&	0.380	&	1.798	\\
BLh	&	1.482	&	1.259	&	509	&	\cmark	&	LR	&	0.471	&	1.169	&	0.408	&	0.063	&	0.208	&	0.086	&	10.577	&	13.969	&	24.769	&	0.189	&	0.291	&	1.222	\\
BLh	&	1.482	&	1.259	&	509	&	\xmark	&	LR	&	0.462	&	1.106	&	0.399	&	0.081	&	0.168	&	0.076	&	10.155	&	13.631	&	23.094	&	0.146	&	0.225	&	1.000	\\
BLh	&	1.581	&	1.184	&	508	&	\cmark	&	SR	&	0.358	&	0.916	&	0.303	&	0.056	&	0.201	&	0.069	&	9.956	&	13.092	&	19.612	&	0.185	&	0.251	&	0.827	\\
BLh	&	1.581	&	1.184	&	508	&	\xmark	&	LR	&	0.344	&	0.908	&	0.323	&	0.045	&	0.163	&	0.061	&	10.006	&	13.412	&	22.818	&	0.135	&	0.223	&	0.887	\\
BLh	&	1.699	&	1.104	&	512	&	\cmark	&	LR	&	0.404	&	0.853	&	0.331	&	0.053	&	0.158	&	0.063	&	10.324	&	13.839	&	23.804	&	0.142	&	0.157	&	0.538	\\
BLh	&	1.699	&	1.104	&	512	&	\xmark	&	LR	&	0.348	&	0.782	&	0.280	&	0.051	&	0.162	&	0.059	&	9.881	&	13.284	&	23.301	&	0.217	&	0.217	&	1.128	\\
BLh	&	1.772	&	1.065	&	506	&	\cmark	&	SR	&	0.382	&	0.825	&	0.359	&	0.066	&	0.170	&	0.079	&	10.474	&	13.664	&	22.993	&	0.368	&	0.456	&	1.384	\\
BLh	&	1.635	&	1.146	&	510	&	\cmark	&	SR	&	0.380	&	0.954	&	0.328	&	0.074	&	0.200	&	0.075	&	10.011	&	13.597	&	24.352	&	0.236	&	0.224	&	1.166	\\
DD2	&	1.300	&	1.300	&	1057	&	\xmark	&	LR	&	0.362	&	0.895	&	0.174	&	0.045	&	0.095	&	0.022	&	9.554	&	13.336	&	23.844	&	0.284	&	0.386	&	1.166	\\
DD2	&	1.364	&	1.364	&	810	&	\xmark	&	SR	&	0.355	&	0.924	&	0.195	&	0.054	&	0.140	&	0.038	&	9.773	&	13.876	&	24.859	&	0.248	&	0.428	&	1.719	\\
DD2	&	1.364	&	1.364	&	810	&	\cmark	&	SR	&	0.419	&	1.026	&	0.216	&	0.055	&	0.107	&	0.031	&	9.790	&	13.750	&	24.456	&	0.216	&	0.399	&	1.421	\\
DD2	&	1.432	&	1.300	&	807	&	\xmark	&	LR	&	0.359	&	0.844	&	0.170	&	0.039	&	0.102	&	0.030	&	9.664	&	13.658	&	22.782	&	0.333	&	0.506	&	1.311	\\
DD2	&	1.435	&	1.298	&	806	&	\xmark	&	LR	&	0.354	&	0.927	&	0.181	&	0.037	&	0.105	&	0.025	&	9.107	&	13.434	&	21.889	&	0.477	&	0.874	&	0.749	\\
DD2	&	1.486	&	1.254	&	802	&	\xmark	&	HR	&	0.341	&	0.941	&	0.161	&	0.051	&	0.131	&	0.025	&	9.769	&	13.846	&	23.628	&	0.185	&	0.348	&	1.177	\\
DD2	&	1.486	&	1.254	&	802	&	\xmark	&	LR	&	0.336	&	0.878	&	0.174	&	0.038	&	0.141	&	0.030	&	9.585	&	13.605	&	22.659	&	0.228	&	0.335	&	1.308	\\
DD2	&	1.497	&	1.245	&	801	&	\xmark	&	SR	&	0.356	&	1.047	&	0.206	&	0.045	&	0.151	&	0.031	&	9.709	&	13.517	&	24.696	&	0.278	&	0.355	&	1.676	\\
DD2	&	1.509	&	1.235	&	800	&	\cmark	&	SR	&	0.361	&	1.021	&	0.223	&	0.043	&	0.115	&	0.028	&	10.002	&	13.842	&	24.871	&	0.372	&	0.513	&	1.708	\\
DD2	&	1.635	&	1.146	&	776	&	\cmark	&	LR	&	0.311	&	0.922	&	0.185	&	0.040	&	0.147	&	0.036	&	9.535	&	13.518	&	23.226	&	0.262	&	0.358	&	1.294	\\
LS220	&	1.400	&	1.200	&	893	&	\xmark	&	SR	&	0.341	&	0.891	&	0.214	&	0.037	&	0.125	&	0.032	&	9.463	&	13.426	&	22.881	&	0.147	&	0.354	&	1.022	\\
LS220	&	1.400	&	1.200	&	893	&	\cmark	&	SR	&	0.356	&	0.911	&	0.207	&	0.035	&	0.133	&	0.033	&	9.852	&	13.589	&	23.761	&	0.217	&	0.254	&	0.985	\\
LS220	&	1.469	&	1.268	&	639	&	\xmark	&	SR	&	0.423	&	1.116	&	0.315	&	0.054	&	0.184	&	0.059	&	10.046	&	13.729	&	23.392	&	0.322	&	0.266	&	0.999	\\
SFHo	&	1.635	&	1.146	&	392	&	\cmark	&	SR	&	0.304	&	0.819	&	0.247	&	0.039	&	0.139	&	0.046	&	10.141	&	14.091	&	24.535	&	0.174	&	0.295	&	1.176	\\
SLy4	&	1.635	&	1.146	&	361	&	\cmark	&	SR	&	0.323	&	0.821	&	0.268	&	0.035	&	0.122	&	0.048	&	10.088	&	14.323	&	23.983	&	0.141	&	0.190	&	1.299	\\\hline
        \end{tabular}
        \end{adjustbox}
        \caption{List of \ac{DC} and \ac{LL} simulations. Columns from left to
          right provide: the mass of the heaviest star; the mass of the lightest
          star; the reduced dimensionless tidal deformability; whether the
          simulation employs \ac{GRLES} viscosity; the resolution; the average
          luminosity for the 3 neutrino flavors and respective standard
          deviations; the average mean energy for the 3 neutrino flavors and
          respective standard deviations.}
        \label{tab:sim_av_info}
\end{table*}

\bibliographystyle{spphys}

\begin{thebibliography}{250}
\providecommand{\url}[1]{{#1}}
\providecommand{\urlprefix}{URL }
\expandafter\ifx\csname urlstyle\endcsname\relax
  \providecommand{\doi}[1]{DOI \discretionary{}{}{}#1}\else
  \providecommand{\doi}{DOI \discretionary{}{}{}\begingroup
  \urlstyle{rm}\Url}\fi

\bibitem{Perego:2019adq}
A.~Perego, S.~Bernuzzi, D.~Radice, Eur. Phys. J. \textbf{A55}(8), 124 (2019).
\newblock \doi{10.1140/epja/i2019-12810-7}

\bibitem{Shibata:2019wef}
M.~Shibata, K.~Hotokezaka, Ann. Rev. Nucl. Part. Sci. \textbf{69}, 41 (2019).
\newblock \doi{10.1146/annurev-nucl-101918-023625}

\bibitem{Radice:2020ddv}
D.~Radice, S.~Bernuzzi, A.~Perego, Ann. Rev. Nucl. Part. Sci. \textbf{70}
  (2020).
\newblock \doi{10.1146/annurev-nucl-013120-114541}

\bibitem{Baiotti:2016qnr}
L.~Baiotti, L.~Rezzolla, Rept. Prog. Phys. \textbf{80}(9), 096901 (2017).
\newblock \doi{10.1088/1361-6633/aa67bb}

\bibitem{LIGOScientific:2018mvr}
B.P. Abbott, et~al., Phys. Rev. \textbf{X9}(3), 031040 (2019).
\newblock \doi{10.1103/PhysRevX.9.031040}

\bibitem{LIGOScientific:2021qlt}
R.~Abbott, et~al., Astrophys. J. Lett. \textbf{915}(1), L5 (2021).
\newblock \doi{10.3847/2041-8213/ac082e}

\bibitem{TheLIGOScientific:2014jea}
J.~Aasi, et~al., Class. Quant. Grav. \textbf{32}, 074001 (2015).
\newblock \doi{10.1088/0264-9381/32/7/074001}

\bibitem{TheVirgo:2014hva}
F.~Acernese, et~al., Class. Quant. Grav. \textbf{32}(2), 024001 (2015).
\newblock \doi{10.1088/0264-9381/32/2/024001}

\bibitem{Aso:2013eba}
Y.~Aso, Y.~Michimura, K.~Somiya, M.~Ando, O.~Miyakawa, T.~Sekiguchi,
  D.~Tatsumi, H.~Yamamoto, Phys. Rev. D \textbf{88}(4), 043007 (2013).
\newblock \doi{10.1103/PhysRevD.88.043007}

\bibitem{Eichler:1989ve}
D.~Eichler, M.~Livio, T.~Piran, D.N. Schramm, Nature \textbf{340}, 126 (1989).
\newblock \doi{10.1038/340126a0}

\bibitem{Narayan:1992iy}
R.~Narayan, B.~Paczynski, T.~Piran, Astrophys. J. \textbf{395}, L83 (1992)

\bibitem{Lee:2007js}
W.H. Lee, E.~Ramirez-Ruiz, New J. Phys. \textbf{9}, 17 (2007).
\newblock \doi{10.1088/1367-2630/9/1/017}

\bibitem{Nakar:2007yr}
E.~Nakar, Phys. Rept. \textbf{442}, 166 (2007).
\newblock \doi{10.1016/j.physrep.2007.02.005}

\bibitem{Li:1998bw}
L.X. Li, B.~Paczynski, Astrophys.J. \textbf{507}, L59 (1998).
\newblock \doi{10.1086/311680}

\bibitem{Kulkarni:2005jw}
S.R. Kulkarni,   (2005)

\bibitem{Fernandez:2015use}
R.~Fernández, B.D. Metzger, Ann.\ Rev.\ Nucl.\ Part.\ Sci. \textbf{66}, 23
  (2016).
\newblock \doi{10.1146/annurev-nucl-102115-044819}

\bibitem{Metzger:2019zeh}
B.D. Metzger, Living Rev. Rel. \textbf{23}(1), 1 (2020).
\newblock \doi{10.1007/s41114-019-0024-0}

\bibitem{Cowan:2019pkx}
J.J. Cowan, C.~Sneden, J.E. Lawler, A.~Aprahamian, M.~Wiescher, K.~Langanke,
  G.~Mart\'\i{}nez-Pinedo, F.K. Thielemann, Rev. Mod. Phys. \textbf{93}(1),
  15002 (2021).
\newblock \doi{10.1103/RevModPhys.93.015002}

\bibitem{Perego:2021dpw}
A.~{Perego}, F.K. {Thielemann}, G.~{Cescutti}, in \emph{Handbook of
  Gravitational Wave Astronomy} (2021), p.~1.
\newblock \doi{10.1007/978-981-15-4702-7\_13-1}

\bibitem{Korobkin:2012uy}
O.~Korobkin, S.~Rosswog, A.~Arcones, C.~Winteler, Mon. Not. Roy. Astron. Soc.
  \textbf{426}, 1940 (2012).
\newblock \doi{10.1111/j.1365-2966.2012.21859.x}

\bibitem{Rosswog:2017sdn}
S.~Rosswog, J.~Sollerman, U.~Feindt, A.~Goobar, O.~Korobkin, R.~Wollaeger,
  C.~Fremling, M.M. Kasliwal, Astron. Astrophys. \textbf{615}, A132 (2018).
\newblock \doi{10.1051/0004-6361/201732117}

\bibitem{Kasen:2017sxr}
D.~Kasen, B.~Metzger, J.~Barnes, E.~Quataert, E.~Ramirez-Ruiz, Nature  (2017).
\newblock \doi{10.1038/nature24453}.
\newblock [Nature551,80(2017)]

\bibitem{Drout:2017ijr}
M.R. Drout, et~al., Science \textbf{358}, 1570 (2017).
\newblock \doi{10.1126/science.aaq0049}

\bibitem{TheLIGOScientific:2017qsa}
B.P. Abbott, et~al., Phys. Rev. Lett. \textbf{119}(16), 161101 (2017).
\newblock \doi{10.1103/PhysRevLett.119.161101}

\bibitem{LIGOScientific:2017ync}
B.~Abbott, et~al., Astrophys. J. Lett. \textbf{848}(2), L12 (2017).
\newblock \doi{10.3847/2041-8213/aa91c9}

\bibitem{2017ApJ...848L..33A}
I.~{Arcavi}, C.~{McCully}, G.~{Hosseinzadeh}, D.A. {Howell}, S.~{Vasylyev},
  D.~{Poznanski}, M.~{Zaltzman}, D.~{Maoz}, L.~{Singer}, S.~{Valenti},
  D.~{Kasen}, J.~{Barnes}, T.~{Piran}, W.f. {Fong}, Astrophysical Journal
  Letters \textbf{848}(2), L33 (2017).
\newblock \doi{10.3847/2041-8213/aa910f}

\bibitem{Chornock:2017sdf}
R.~Chornock, et~al., Astrophys. J. \textbf{848}(2), L19 (2017).
\newblock \doi{10.3847/2041-8213/aa905c}

\bibitem{Cowperthwaite:2017dyu}
P.S. Cowperthwaite, et~al., Astrophys. J. \textbf{848}(2), L17 (2017).
\newblock \doi{10.3847/2041-8213/aa8fc7}

\bibitem{Coulter:2017wya}
D.A. Coulter, et~al., Science  (2017).
\newblock \doi{10.1126/science.aap9811}.
\newblock [Science358,1556(2017)]

\bibitem{Evans:2017mmy}
P.A. Evans, et~al., Science \textbf{358}, 1565 (2017).
\newblock \doi{10.1126/science.aap9580}

\bibitem{Goldstein:2017mmi}
A.~Goldstein, et~al., Astrophys. J. \textbf{848}(2), L14 (2017).
\newblock \doi{10.3847/2041-8213/aa8f41}

\bibitem{Hallinan:2017woc}
G.~Hallinan, et~al., Science \textbf{358}, 1579 (2017).
\newblock \doi{10.1126/science.aap9855}

\bibitem{Kasliwal:2017ngb}
M.M. Kasliwal, et~al., Science \textbf{358}, 1559 (2017).
\newblock \doi{10.1126/science.aap9455}

\bibitem{Murguia-Berthier:2017kkn}
A.~Murguia-Berthier, et~al., Astrophys. J. Lett. \textbf{848}(2), L34 (2017).
\newblock \doi{10.3847/2041-8213/aa91b3}

\bibitem{Nicholl:2017ahq}
M.~Nicholl, et~al., Astrophys. J. \textbf{848}, L18 (2017).
\newblock \doi{10.3847/2041-8213/aa9029}

\bibitem{Smartt:2017fuw}
S.J. Smartt, et~al., Nature  (2017).
\newblock \doi{10.1038/nature24303}

\bibitem{Soares-Santos:2017lru}
M.~Soares-Santos, et~al., Astrophys. J. \textbf{848}(2), L16 (2017).
\newblock \doi{10.3847/2041-8213/aa9059}

\bibitem{Savchenko:2017ffs}
V.~Savchenko, et~al., Astrophys. J. \textbf{848}(2), L15 (2017).
\newblock \doi{10.3847/2041-8213/aa8f94}

\bibitem{Tanvir:2017pws}
N.R. Tanvir, et~al., Astrophys. J. \textbf{848}, L27 (2017).
\newblock \doi{10.3847/2041-8213/aa90b6}

\bibitem{Tanaka:2017qxj}
M.~Tanaka, et~al., Publ. Astron. Soc. Jap.  (2017).
\newblock \doi{10.1093/pasj/psx121}

\bibitem{Troja:2017nqp}
E.~Troja, et~al., Nature  (2017).
\newblock \doi{10.1038/nature24290}

\bibitem{Villar:2017wcc}
V.A. Villar, et~al., Astrophys. J. \textbf{851}(1), L21 (2017).
\newblock \doi{10.3847/2041-8213/aa9c84}

\bibitem{Waxman:2017sqv}
E.~Waxman, E.O. Ofek, D.~Kushnir, A.~Gal-Yam, Mon. Not. Roy. Astron. Soc.
  \textbf{481}(3), 3423 (2018).
\newblock \doi{10.1093/mnras/sty2441}

\bibitem{Kasliwal:2018fwk}
M.M. Kasliwal, et~al., Mon. Not. Roy. Astron. Soc. \textbf{510}(1), L7 (2022).
\newblock \doi{10.1093/mnrasl/slz007}

\bibitem{Waxman:2019png}
E.~Waxman, E.O. Ofek, D.~Kushnir, Astrophys. J. \textbf{878}(2), 93 (2019).
\newblock \doi{10.3847/1538-4357/ab1f71}

\bibitem{Abbott:2020uma}
B.~Abbott, et~al., Astrophys. J. Lett. \textbf{892}, L3 (2020).
\newblock \doi{10.3847/2041-8213/ab75f5}

\bibitem{Ruffert:1998vp}
M.~Ruffert, H.T. Janka, Astron. Astrophys. \textbf{338}, 535 (1998)

\bibitem{Rosswog:2002rt}
S.~Rosswog, E.~Ramirez-Ruiz, Mon.Not.Roy.Astron.Soc. \textbf{336}, L7 (2002).
\newblock \doi{10.1046/j.1365-8711.2002.05898.x}

\bibitem{Rosswog:2003ts}
S.~Rosswog, E.~Ramirez-Ruiz, Mon.Not.Roy.Astron.Soc. \textbf{343}, L36 (2003).
\newblock \doi{10.1046/j.1365-8711.2003.06889.x}

\bibitem{Mosta:2020hlh}
P.~M\"osta, D.~Radice, R.~Haas, E.~Schnetter, S.~Bernuzzi, Astrophys. J. Lett.
  \textbf{901}, L37 (2020).
\newblock \doi{10.3847/2041-8213/abb6ef}

\bibitem{Sun:2022vri}
L.~Sun, M.~Ruiz, S.L. Shapiro, A.~Tsokaros,   (2022)

\bibitem{Ruffert:1996by}
M.~Ruffert, H.~Janka, K.~Takahashi, G.~Sch{\"a}fer, Astron.Astrophys.
  \textbf{319}, 122 (1997)

\bibitem{Rosswog:2003tn}
S.~Rosswog, E.~Ramirez-Ruiz, M.B. Davies, Mon. Not. Roy. Astron. Soc.
  \textbf{345}, 1077 (2003).
\newblock \doi{10.1046/j.1365-2966.2003.07032.x}

\bibitem{Zalamea:2010ax}
I.~Zalamea, A.M. Beloborodov, Mon. Not. Roy. Astron. Soc. \textbf{410}, 2302
  (2011).
\newblock \doi{10.1111/j.1365-2966.2010.17600.x}

\bibitem{Dessart:2008zd}
L.~Dessart, C.~Ott, A.~Burrows, S.~Rosswog, E.~Livne, Astrophys.J.
  \textbf{690}, 1681 (2009).
\newblock \doi{10.1088/0004-637X/690/2/1681}

\bibitem{Just:2015dba}
O.~Just, M.~Obergaulinger, H.T. Janka, A.~Bauswein, N.~Schwarz, Astrophys. J.
  Lett. \textbf{816}(2), L30 (2016).
\newblock \doi{10.3847/2041-8205/816/2/L30}

\bibitem{Perego:2014fma}
A.~Perego, S.~Rosswog, R.~Cabezon, O.~Korobkin, R.~Kaeppeli, et~al.,
  Mon.Not.Roy.Astron.Soc. \textbf{443}, 3134 (2014).
\newblock \doi{10.1093/mnras/stu1352}

\bibitem{Fujibayashi:2017xsz}
S.~Fujibayashi, Y.~Sekiguchi, K.~Kiuchi, M.~Shibata, Astrophys. J.
  \textbf{846}(2), 114 (2017).
\newblock \doi{10.3847/1538-4357/aa8039}

\bibitem{Perego:2017xth}
A.~Perego, A.~Arcones, D.~Martin, H.~Yasin, JPS Conf. Proc. \textbf{14}, 020810
  (2017).
\newblock \doi{10.7566/JPSCP.14.020810}

\bibitem{Metzger:2014ila}
B.D. Metzger, R.~Fern\'{a}ndez, Mon.Not.Roy.Astron.Soc. \textbf{441}, 3444
  (2014).
\newblock \doi{10.1093/mnras/stu802}

\bibitem{Martin:2015hxa}
D.~Martin, A.~Perego, A.~Arcones, F.K. Thielemann, O.~Korobkin, S.~Rosswog,
  Astrophys. J. \textbf{813}(1), 2 (2015).
\newblock \doi{10.1088/0004-637X/813/1/2}

\bibitem{Miller:2019dpt}
J.M. Miller, B.R. Ryan, J.C. Dolence, A.~Burrows, C.J. Fontes, C.L. Fryer,
  O.~Korobkin, J.~Lippuner, M.R. Mumpower, R.T. Wollaeger, Phys. Rev.
  \textbf{D100}(2), 023008 (2019).
\newblock \doi{10.1103/PhysRevD.100.023008}

\bibitem{Foucart:2016rxm}
F.~Foucart, E.~O'Connor, L.~Roberts, L.E. Kidder, H.P. Pfeiffer, M.A. Scheel,
  Phys. Rev. \textbf{D94}(12), 123016 (2016).
\newblock \doi{10.1103/PhysRevD.94.123016}

\bibitem{Perego:2017wtu}
A.~Perego, D.~Radice, S.~Bernuzzi, Astrophys. J. \textbf{850}(2), L37 (2017).
\newblock \doi{10.3847/2041-8213/aa9ab9}

\bibitem{Nedora:2020qtd}
V.~Nedora, F.~Schianchi, S.~Bernuzzi, D.~Radice, B.~Daszuta, A.~Endrizzi,
  A.~Perego, A.~Prakash, F.~Zappa, Class. Quant. Grav. \textbf{39}(1), 015008
  (2022).
\newblock \doi{10.1088/1361-6382/ac35a8}

\bibitem{Endrizzi:2019trv}
A.~Endrizzi, A.~Perego, F.M. Fabbri, L.~Branca, D.~Radice, S.~Bernuzzi,
  B.~Giacomazzo, F.~Pederiva, A.~Lovato, Eur. Phys. J. A \textbf{56}(1), 15
  (2020).
\newblock \doi{10.1140/epja/s10050-019-00018-6}

\bibitem{Rosswog:2003rv}
S.~Rosswog, M.~Liebendoerfer, Mon.Not.Roy.Astron.Soc. \textbf{342}, 673 (2003).
\newblock \doi{10.1046/j.1365-8711.2003.06579.x}

\bibitem{Shibata:2011kx}
M.~Shibata, K.~Kiuchi, Y.i. Sekiguchi, Y.~Suwa, Prog.Theor.Phys. \textbf{125},
  1255 (2011).
\newblock \doi{10.1143/PTP.125.1255}

\bibitem{Galeazzi:2013mia}
F.~Galeazzi, W.~Kastaun, L.~Rezzolla, J.A. Font, Phys.Rev. \textbf{D88}, 064009
  (2013).
\newblock \doi{10.1103/PhysRevD.88.064009}

\bibitem{Foucart:2017mbt}
F.~Foucart, Mon. Not. Roy. Astron. Soc. \textbf{475}(3), 4186 (2018).
\newblock \doi{10.1093/mnras/sty108}

\bibitem{Ardevol-Pulpillo:2018btx}
R.~Ardevol-Pulpillo, H.T. Janka, O.~Just, A.~Bauswein, Mon. Not. Roy. Astron.
  Soc. \textbf{485}(4), 4754 (2019).
\newblock \doi{10.1093/mnras/stz613}

\bibitem{Foucart:2018gis}
F.~Foucart, M.D. Duez, L.E. Kidder, R.~Nguyen, H.P. Pfeiffer, M.A. Scheel,
  Phys. Rev. \textbf{D98}(6), 063007 (2018).
\newblock \doi{10.1103/PhysRevD.98.063007}

\bibitem{Gizzi:2019awu}
D.~Gizzi, E.~O'Connor, S.~Rosswog, A.~Perego, R.~Cabezón, L.~Nativi, Mon. Not.
  Roy. Astron. Soc. \textbf{490}(3), 4211 (2019).
\newblock \doi{10.1093/mnras/stz2911}

\bibitem{Weih:2020qyh}
L.R. Weih, A.~Gabbana, D.~Simeoni, L.~Rezzolla, S.~Succi, R.~Tripiccione, Mon.
  Not. Roy. Astron. Soc. \textbf{498}(3), 3374 (2020).
\newblock \doi{10.1093/mnras/staa2575}

\bibitem{Foucart:2020qjb}
F.~Foucart, M.D. Duez, F.~Hebert, L.E. Kidder, H.P. Pfeiffer, M.A. Scheel,
  Astrophys. J. Lett. \textbf{902}, L27 (2020).
\newblock \doi{10.3847/2041-8213/abbb87}

\bibitem{Gizzi:2021ssk}
D.~Gizzi, C.~Lundman, E.~O'Connor, S.~Rosswog, A.~Perego, Mon. Not. Roy.
  Astron. Soc. \textbf{505}(2), 2575 (2021).
\newblock \doi{10.1093/mnras/stab1432}

\bibitem{Radice.etal:2022}
D.~{Radice}, S.~{Bernuzzi}, A.~{Perego}, R.~{Haas}, "Mon.\ Not.\ Roy.\ Astron.\
  Soc."  (2022).
\newblock \doi{10.1093/mnras/stac589}

\bibitem{Janka:2017vlw}
H.T. {Janka}, in \emph{Handbook of Supernovae}, ed. by A.W. {Alsabti},
  P.~{Murdin} (2017), p. 1575.
\newblock \doi{10.1007/978-3-319-21846-5\_4}

\bibitem{Mezzacappa:2020oyq}
A.~{Mezzacappa}, E.~{Endeve}, O.E.B. {Messer}, S.W. {Bruenn}, Living Reviews in
  Computational Astrophysics \textbf{6}(1), 4 (2020).
\newblock \doi{10.1007/s41115-020-00010-8}

\bibitem{Sekiguchi:2015dma}
Y.~Sekiguchi, K.~Kiuchi, K.~Kyutoku, M.~Shibata, Phys.Rev. \textbf{D91}(6),
  064059 (2015).
\newblock \doi{10.1103/PhysRevD.91.064059}

\bibitem{Palenzuela:2015dqa}
C.~Palenzuela, S.L. Liebling, D.~Neilsen, L.~Lehner, O.L. Caballero,
  E.~O’Connor, M.~Anderson, Phys. Rev. \textbf{D92}(4), 044045 (2015).
\newblock \doi{10.1103/PhysRevD.92.044045}

\bibitem{Foucart:2015gaa}
F.~Foucart, R.~Haas, M.D. Duez, E.~O'Connor, C.D. Ott, L.~Roberts, L.E. Kidder,
  J.~Lippuner, H.P. Pfeiffer, M.A. Scheel, Phys. Rev. \textbf{D93}(4), 044019
  (2016).
\newblock \doi{10.1103/PhysRevD.93.044019}

\bibitem{Wu:2017drk}
M.R. Wu, I.~Tamborra, O.~Just, H.T. Janka, Phys. Rev. D \textbf{96}(12), 123015
  (2017).
\newblock \doi{10.1103/PhysRevD.96.123015}

\bibitem{George:2020veu}
M.~George, M.R. Wu, I.~Tamborra, R.~Ardevol-Pulpillo, H.T. Janka, Phys. Rev. D
  \textbf{102}(10), 103015 (2020).
\newblock \doi{10.1103/PhysRevD.102.103015}

\bibitem{Kullmann:2021gvo}
I.~Kullmann, S.~Goriely, O.~Just, R.~Ardevol-Pulpillo, A.~Bauswein, H.T. Janka,
  Mon. Not. Roy. Astron. Soc.  (2021).
\newblock \doi{10.1093/mnras/stab3393}

\bibitem{Yagi:2016bkt}
K.~Yagi, N.~Yunes, Phys. Rept. \textbf{681}, 1 (2017).
\newblock \doi{10.1016/j.physrep.2017.03.002}

\bibitem{Carson:2019rjx}
Z.~Carson, K.~Chatziioannou, C.J. Haster, K.~Yagi, N.~Yunes, Phys. Rev.
  \textbf{D99}(8), 083016 (2019).
\newblock \doi{10.1103/PhysRevD.99.083016}

\bibitem{Paschalidis:2017qmb}
V.~Paschalidis, K.~Yagi, D.~Alvarez-Castillo, D.B. Blaschke, A.~Sedrakian,
  Phys. Rev. \textbf{D97}(8), 084038 (2018).
\newblock \doi{10.1103/PhysRevD.97.084038}

\bibitem{Pani:2015nua}
P.~Pani, L.~Gualtieri, V.~Ferrari, Phys. Rev. D \textbf{92}(12), 124003 (2015).
\newblock \doi{10.1103/PhysRevD.92.124003}

\bibitem{Rezzolla:2016nxn}
L.~Rezzolla, K.~Takami, Phys. Rev. \textbf{D93}(12), 124051 (2016).
\newblock \doi{10.1103/PhysRevD.93.124051}

\bibitem{Godzieba:2020bbz}
D.A. Godzieba, R.~Gamba, D.~Radice, S.~Bernuzzi, Phys. Rev. D \textbf{103}(6),
  063036 (2021).
\newblock \doi{10.1103/PhysRevD.103.063036}

\bibitem{Bernuzzi:2014kca}
S.~Bernuzzi, A.~Nagar, S.~Balmelli, T.~Dietrich, M.~Ujevic, Phys.Rev.Lett.
  \textbf{112}, 201101 (2014).
\newblock \doi{10.1103/PhysRevLett.112.201101}

\bibitem{Radice:2018pdn}
D.~Radice, A.~Perego, K.~Hotokezaka, S.A. Fromm, S.~Bernuzzi, L.F. Roberts,
  Astrophys. J. \textbf{869}(2), 130 (2018).
\newblock \doi{10.3847/1538-4357/aaf054}

\bibitem{Loffler:2011ay}
F.~Loffler, et~al., Class. Quant. Grav. \textbf{29}, 115001 (2012).
\newblock \doi{10.1088/0264-9381/29/11/115001}

\bibitem{Schnetter:2003rb}
E.~Schnetter, S.H. Hawley, I.~Hawke, Class.Quant.Grav. \textbf{21}, 1465
  (2004).
\newblock \doi{10.1088/0264-9381/21/6/014}

\bibitem{Reisswig:2012nc}
C.~Reisswig, R.~Haas, C.D. Ott, E.~Abdikamalov, P.~Mösta, D.~Pollney,
  E.~Schnetter, Phys. Rev. \textbf{D87}(6), 064023 (2013).
\newblock \doi{10.1103/PhysRevD.87.064023}

\bibitem{Bernuzzi:2009ex}
S.~Bernuzzi, D.~Hilditch, Phys. Rev. \textbf{D81}, 084003 (2010).
\newblock \doi{10.1103/PhysRevD.81.084003}

\bibitem{Pollney:2009yz}
D.~Pollney, C.~Reisswig, E.~Schnetter, N.~Dorband, P.~Diener, Phys. Rev.
  \textbf{D83}, 044045 (2011).
\newblock \doi{10.1103/PhysRevD.83.044045}

\bibitem{Reisswig:2013sqa}
C.~Reisswig, C.~Ott, E.~Abdikamalov, R.~Haas, P.~M{\"o}sta, et~al.,
  Phys.Rev.Lett. \textbf{111}, 151101 (2013).
\newblock \doi{10.1103/PhysRevLett.111.151101}

\bibitem{Radice:2012cu}
D.~Radice, L.~Rezzolla, Astron. Astrophys. \textbf{547}, A26 (2012).
\newblock \doi{10.1051/0004-6361/201219735}

\bibitem{Radice:2013hxh}
D.~Radice, L.~Rezzolla, F.~Galeazzi, Mon.Not.Roy.Astron.Soc. \textbf{437}, L46
  (2014).
\newblock \doi{10.1093/mnrasl/slt137}

\bibitem{Radice:2013xpa}
D.~Radice, L.~Rezzolla, F.~Galeazzi, Class.Quant.Grav. \textbf{31}, 075012
  (2014).
\newblock \doi{10.1088/0264-9381/31/7/075012}

\bibitem{Radice:2016dwd}
D.~Radice, F.~Galeazzi, J.~Lippuner, L.F. Roberts, C.D. Ott, L.~Rezzolla, Mon.
  Not. Roy. Astron. Soc. \textbf{460}(3), 3255 (2016).
\newblock \doi{10.1093/mnras/stw1227}

\bibitem{Berger:1984zza}
M.J. Berger, J.~Oliger, J.Comput.Phys. \textbf{53}, 484 (1984)

\bibitem{Berger:1989a}
M.J. {Berger}, P.~{Colella}, Journal of Computational Physics \textbf{82}, 64
  (1989).
\newblock \doi{10.1016/0021-9991(89)90035-1}

\bibitem{Bernuzzi:2020txg}
S.~Bernuzzi, et~al., Mon. Not. Roy. Astron. Soc.  (2020).
\newblock \doi{10.1093/mnras/staa1860}

\bibitem{Favata:2004wz}
M.~Favata, S.A. Hughes, D.E. Holz, Astrophys. J. \textbf{607}, L5 (2004).
\newblock \doi{10.1086/421552}

\bibitem{Hinderer:2007mb}
T.~Hinderer, Astrophys.J. \textbf{677}, 1216 (2008).
\newblock \doi{10.1086/533487}

\bibitem{Gourgoulhon:2001ec}
E.~Gourgoulhon, P.~Grandclement, S.~Bonazzola, Phys. Rev. \textbf{D65}, 044020
  (2002).
\newblock \doi{10.1103/PhysRevD.65.044020}

\bibitem{Ruffert:1996qu}
M.~Ruffert, M.~Rampp, H.T. Janka, Astron. Astrophys. \textbf{321}, 991 (1997)

\bibitem{Thompson:2002mw}
T.A. Thompson, A.~Burrows, P.A. Pinto, Astrophys. J. \textbf{592}, 434 (2003).
\newblock \doi{10.1086/375701}

\bibitem{Burrows:2004vq}
A.~Burrows, S.~Reddy, T.A. Thompson, Nucl. Phys. \textbf{A777}, 356 (2006).
\newblock \doi{10.1016/j.nuclphysa.2004.06.012}

\bibitem{Neilsen:2014hha}
D.~Neilsen, S.L. Liebling, M.~Anderson, L.~Lehner, E.~O’Connor, et~al.,
  Phys.Rev. \textbf{D89}(10), 104029 (2014).
\newblock \doi{10.1103/PhysRevD.89.104029}

\bibitem{Ruffert:1995fs}
M.H. Ruffert, H.T. Janka, G.~Sch{\"a}fer, Astron. Astrophys. \textbf{311}, 532
  (1996)

\bibitem{OConnor:2009iuz}
E.~O'Connor, C.D. Ott, Class. Quant. Grav. \textbf{27}, 114103 (2010).
\newblock \doi{10.1088/0264-9381/27/11/114103}

\bibitem{Sekiguchi:2011zd}
Y.~Sekiguchi, K.~Kiuchi, K.~Kyutoku, M.~Shibata, Phys.Rev.Lett. \textbf{107},
  051102 (2011).
\newblock \doi{10.1103/PhysRevLett.107.051102}

\bibitem{Lehner:2016lxy}
L.~Lehner, S.L. Liebling, C.~Palenzuela, O.L. Caballero, E.~O'Connor,
  M.~Anderson, D.~Neilsen, Class. Quant. Grav. \textbf{33}(18), 184002 (2016).
\newblock \doi{10.1088/0264-9381/33/18/184002}

\bibitem{Bruenn:1985en}
S.W. Bruenn, Astrophys. J. Suppl. \textbf{58}, 771 (1985).
\newblock \doi{10.1086/191056}

\bibitem{Shapiro:1983du}
S.L. Shapiro, S.A. Teukolsky, \emph{{Black holes, white dwarfs, and neutron
  stars: The physics of compact objects}} (Wiley, New York, USA, 1983)

\bibitem{Lattimer:1991nc}
J.M. Lattimer, F.D. Swesty, Nucl. Phys. \textbf{A535}, 331 (1991).
\newblock \doi{10.1016/0375-9474(91)90452-C}

\bibitem{Douchin:2001sv}
F.~Douchin, P.~Haensel, Astron. Astrophys. \textbf{380}, 151 (2001)

\bibitem{daSilvaSchneider:2017jpg}
A.S. Schneider, L.F. Roberts, C.D. Ott, Phys. Rev. \textbf{C96}(6), 065802
  (2017).
\newblock \doi{10.1103/PhysRevC.96.065802}

\bibitem{Typel:2009sy}
S.~Typel, G.~Ropke, T.~Klahn, D.~Blaschke, H.H. Wolter, Phys. Rev.
  \textbf{C81}, 015803 (2010).
\newblock \doi{10.1103/PhysRevC.81.015803}

\bibitem{Hempel:2009mc}
M.~Hempel, J.~Schaffner-Bielich, Nucl. Phys. \textbf{A837}, 210 (2010).
\newblock \doi{10.1016/j.nuclphysa.2010.02.010}

\bibitem{Steiner:2012xt}
A.W. Steiner, J.M. Lattimer, E.F. Brown, Astrophys. J. \textbf{765}, L5 (2013).
\newblock \doi{10.1088/2041-8205/765/1/L5}

\bibitem{Banik:2014qja}
S.~Banik, M.~Hempel, D.~Bandyopadhyay, Astrophys. J. Suppl. \textbf{214}(2), 22
  (2014).
\newblock \doi{10.1088/0067-0049/214/2/22}

\bibitem{Logoteta:2020yxf}
D.~Logoteta, A.~Perego, I.~Bombaci, Astron. Astrophys. \textbf{646}, A55
  (2021).
\newblock \doi{10.1051/0004-6361/202039457}

\bibitem{Abbott:2018exr}
B.P. Abbott, et~al., Phys. Rev. Lett. \textbf{121}(16), 161101 (2018).
\newblock \doi{10.1103/PhysRevLett.121.161101}

\bibitem{LIGOScientific:2019fpa}
B.P. Abbott, et~al., Phys. Rev. D \textbf{100}(10), 104036 (2019).
\newblock \doi{10.1103/PhysRevD.100.104036}

\bibitem{De:2018uhw}
S.~De, D.~Finstad, J.M. Lattimer, D.A. Brown, E.~Berger, C.M. Biwer, Phys. Rev.
  Lett. \textbf{121}(9), 091102 (2018).
\newblock \doi{10.1103/PhysRevLett.121.259902, 10.1103/PhysRevLett.121.091102}.
\newblock [Erratum: Phys. Rev. Lett.121,no.25,259902(2018)]

\bibitem{Tsang:2012se}
M.B. Tsang, et~al., Phys. Rev. C \textbf{86}, 015803 (2012).
\newblock \doi{10.1103/PhysRevC.86.015803}

\bibitem{Lattimer:2012xj}
J.M. Lattimer, Y.~Lim, Astrophys. J. \textbf{771}, 51 (2013).
\newblock \doi{10.1088/0004-637X/771/1/51}

\bibitem{PREX:2021umo}
D.~Adhikari, et~al., Phys. Rev. Lett. \textbf{126}(17), 172502 (2021).
\newblock \doi{10.1103/PhysRevLett.126.172502}

\bibitem{Bombaci:2018ksa}
I.~Bombaci, D.~Logoteta, Astron. Astrophys. \textbf{609}, A128 (2018).
\newblock \doi{10.1051/0004-6361/201731604}

\bibitem{Machleidt:2011zz}
R.~Machleidt, D.R. Entem, Phys. Rept. \textbf{503}, 1 (2011).
\newblock \doi{10.1016/j.physrep.2011.02.001}

\bibitem{Piarulli:2016vel}
M.~Piarulli, L.~Girlanda, R.~Schiavilla, A.~Kievsky, A.~Lovato, L.E. Marcucci,
  S.C. Pieper, M.~Viviani, R.B. Wiringa, Phys. Rev. \textbf{C94}(5), 054007
  (2016).
\newblock \doi{10.1103/PhysRevC.94.054007}

\bibitem{Logoteta:2016nzc}
D.~Logoteta, I.~Bombaci, A.~Kievsky, Phys. Rev. \textbf{C94}(6), 064001 (2016).
\newblock \doi{10.1103/PhysRevC.94.064001}

\bibitem{Radice:2017zta}
D.~Radice, Astrophys. J. \textbf{838}(1), L2 (2017).
\newblock \doi{10.3847/2041-8213/aa6483}

\bibitem{Duez:2020lgq}
M.D. Duez, A.~Knight, F.~Foucart, M.~Haddadi, J.~Jesse, F.~Hebert, L.E. Kidder,
  H.P. Pfeiffer, M.A. Scheel, Phys. Rev. D \textbf{102}(10), 104050 (2020).
\newblock \doi{10.1103/PhysRevD.102.104050}

\bibitem{Radice:2020ids}
D.~Radice, Symmetry \textbf{12}(8), 1249 (2020).
\newblock \doi{10.3390/sym12081249}

\bibitem{Kiuchi:2017zzg}
K.~Kiuchi, K.~Kyutoku, Y.~Sekiguchi, M.~Shibata, Phys. Rev. \textbf{D97}(12),
  124039 (2018).
\newblock \doi{10.1103/PhysRevD.97.124039}

\bibitem{Nedora:2019jhl}
V.~Nedora, S.~Bernuzzi, D.~Radice, A.~Perego, A.~Endrizzi, N.~Ortiz, Astrophys.
  J. \textbf{886}(2), L30 (2019).
\newblock \doi{10.3847/2041-8213/ab5794}

\bibitem{Nedora:2020pak}
V.~Nedora, S.~Bernuzzi, D.~Radice, B.~Daszuta, A.~Endrizzi, A.~Perego,
  A.~Prakash, M.~Safarzadeh, F.~Schianchi, D.~Logoteta, Astrophys. J.
  \textbf{906}(2), 98 (2021).
\newblock \doi{10.3847/1538-4357/abc9be}

\bibitem{Camilletti:2022jms}
A.~Camilletti, L.~Chiesa, G.~Ricigliano, A.~Perego, L.C. Lippold, S.~Padamata,
  S.~Bernuzzi, D.~Radice, D.~Logoteta, F.M. Guercilena,   (2022)

\bibitem{Chatziioannou:2020pqz}
K.~Chatziioannou, Gen. Rel. Grav. \textbf{52}(11), 109 (2020).
\newblock \doi{10.1007/s10714-020-02754-3}

\bibitem{Lai:1993di}
D.~Lai, Mon. Not. Roy. Astron. Soc. \textbf{270}, 611 (1994).
\newblock \doi{10.1093/mnras/270.3.611}

\bibitem{Radice:2018xqa}
D.~Radice, A.~Perego, S.~Bernuzzi, B.~Zhang, Mon. Not. Roy. Astron. Soc.
  \textbf{481}(3), 3670 (2018).
\newblock \doi{10.1093/mnras/sty2531}

\bibitem{Fernandez:2013tya}
R.~Fernández, B.D. Metzger, Mon.\ Not.\ Roy.\ Astron.\ Soc. \textbf{435}, 502
  (2013).
\newblock \doi{10.1093/mnras/stt1312}

\bibitem{Bernuzzi:2015opx}
S.~Bernuzzi, D.~Radice, C.D. Ott, L.F. Roberts, P.~Moesta, F.~Galeazzi, Phys.
  Rev. \textbf{D94}(2), 024023 (2016).
\newblock \doi{10.1103/PhysRevD.94.024023}

\bibitem{Zappa:2017xba}
F.~Zappa, S.~Bernuzzi, D.~Radice, A.~Perego, T.~Dietrich, Phys. Rev. Lett.
  \textbf{120}(11), 111101 (2018).
\newblock \doi{10.1103/PhysRevLett.120.111101}

\bibitem{Martin:2017dhc}
D.~Martin, A.~Perego, W.~Kastaun, A.~Arcones, Class. Quant. Grav.
  \textbf{35}(3), 034001 (2018).
\newblock \doi{10.1088/1361-6382/aa9f5a}

\bibitem{Qian:1996xt}
Y.~Qian, S.~Woosley, Astrophys. J. \textbf{471}, 331 (1996).
\newblock \doi{10.1086/177973}

\bibitem{Rosswog:1998hy}
S.~Rosswog, M.~Liebendoerfer, F.~Thielemann, M.~Davies, W.~Benz, et~al.,
  Astron.Astrophys. \textbf{341}, 499 (1999)

\bibitem{Rosswog:2001fh}
S.~Rosswog, M.B. Davies, Mon. Not. Roy. Astron. Soc. \textbf{345}, 1077 (2003).
\newblock \doi{10.1046/j.1365-2966.2003.07032.x}

\bibitem{Oechslin:2006uk}
R.~Oechslin, H.T. Janka, A.~Marek, Astron. Astrophys. \textbf{467}, 395 (2007).
\newblock \doi{10.1051/0004-6361:20066682}

\bibitem{Rosswog:2012wb}
S.~Rosswog, T.~Piran, E.~Nakar, Mon. Not. Roy. Astron. Soc. \textbf{430}, 2585
  (2013).
\newblock \doi{10.1093/mnras/sts708}

\bibitem{Bauswein:2013yna}
A.~Bauswein, S.~Goriely, H.T. Janka, Astrophys.J. \textbf{773}, 78 (2013).
\newblock \doi{10.1088/0004-637X/773/1/78}

\bibitem{Sekiguchi:2016bjd}
Y.~Sekiguchi, K.~Kiuchi, K.~Kyutoku, M.~Shibata, K.~Taniguchi, Phys. Rev.
  \textbf{D93}(12), 124046 (2016).
\newblock \doi{10.1103/PhysRevD.93.124046}

\bibitem{Bovard:2017mvn}
L.~Bovard, D.~Martin, F.~Guercilena, A.~Arcones, L.~Rezzolla, O.~Korobkin,
  Phys. Rev. \textbf{D96}(12), 124005 (2017).
\newblock \doi{10.1103/PhysRevD.96.124005}

\bibitem{Vincent:2019kor}
T.~Vincent, F.~Foucart, M.D. Duez, R.~Haas, L.E. Kidder, H.P. Pfeiffer, M.A.
  Scheel, Phys. Rev. \textbf{D101}(4), 044053 (2020).
\newblock \doi{10.1103/PhysRevD.101.044053}

\bibitem{Perego:2020evn}
A.~Perego, et~al., Astrophys. J. \textbf{925}(1), 22 (2022).
\newblock \doi{10.3847/1538-4357/ac3751}

\bibitem{Metzger:2008av}
B.~Metzger, A.~Piro, E.~Quataert, Mon.Not.Roy.Astron.Soc. \textbf{390}, 781
  (2008).
\newblock \doi{10.1111/j.1365-2966.2008.13789.x}

\bibitem{Metzger:2008jt}
B.D. Metzger, A.L. Piro, E.~Quataert, Mon. Not. Roy. Astron. Soc. \textbf{396},
  304 (2009).
\newblock \doi{10.1111/j.1365-2966.2008.14380.x}

\bibitem{Siegel:2014ita}
D.M. Siegel, R.~Ciolfi, L.~Rezzolla, Astrophys. J. \textbf{785}, L6 (2014).
\newblock \doi{10.1088/2041-8205/785/1/L6}

\bibitem{Just:2014fka}
O.~Just, A.~Bauswein, R.A. Pulpillo, S.~Goriely, H.T. Janka, Mon. Not. Roy.
  Astron. Soc. \textbf{448}(1), 541 (2015).
\newblock \doi{10.1093/mnras/stv009}

\bibitem{Fujibayashi:2017puw}
S.~Fujibayashi, K.~Kiuchi, N.~Nishimura, Y.~Sekiguchi, M.~Shibata, Astrophys.
  J. \textbf{860}(1), 64 (2018).
\newblock \doi{10.3847/1538-4357/aabafd}

\bibitem{Siegel:2017jug}
D.M. Siegel, B.D. Metzger, Astrophys. J. \textbf{858}(1), 52 (2018).
\newblock \doi{10.3847/1538-4357/aabaec}

\bibitem{Metzger:2018uni}
B.D. Metzger, T.A. Thompson, E.~Quataert, Astrophys. J. \textbf{856}(2), 101
  (2018).
\newblock \doi{10.3847/1538-4357/aab095}

\bibitem{Fernandez:2018kax}
R.~Fern{\'a}ndez, A.~Tchekhovskoy, E.~Quataert, F.~Foucart, D.~Kasen, Mon. Not.
  Roy. Astron. Soc. \textbf{482}(3), 3373 (2019).
\newblock \doi{10.1093/mnras/sty2932}

\bibitem{Fujibayashi:2020dvr}
S.~{Fujibayashi}, S.~{Wanajo}, K.~{Kiuchi}, K.~{Kyutoku}, Y.~{Sekiguchi},
  M.~{Shibata}, Astrophys.\ J. \textbf{901}(2), 122 (2020).
\newblock \doi{10.3847/1538-4357/abafc2}

\bibitem{Ciolfi:2020wfx}
R.~Ciolfi, J.V. Kalinani, Astrophys. J. Lett. \textbf{900}(2), L35 (2020).
\newblock \doi{10.3847/2041-8213/abb240}

\bibitem{Just:2021cls}
O.~Just, S.~Goriely, H.T. Janka, S.~Nagataki, A.~Bauswein, Mon. Not. Roy.
  Astron. Soc. \textbf{509}(1), 1377 (2021).
\newblock \doi{10.1093/mnras/stab2861}

\bibitem{Shibata:2021bbj}
M.~Shibata, S.~Fujibayashi, Y.~Sekiguchi, Phys. Rev. D \textbf{103}(4), 043022
  (2021).
\newblock \doi{10.1103/PhysRevD.103.043022}

\bibitem{Wu:2016pnw}
M.R. Wu, R.~Fern\'{a}ndez, G.~Martínez-Pinedo, B.D. Metzger, Mon. Not. Roy.
  Astron. Soc. \textbf{463}(3), 2323 (2016).
\newblock \doi{10.1093/mnras/stw2156}

\bibitem{Lippuner:2015gwa}
J.~Lippuner, L.F. Roberts, Astrophys. J. \textbf{815}(2), 82 (2015).
\newblock \doi{10.1088/0004-637X/815/2/82}

\bibitem{Perego:2015agy}
A.~Perego, R.~Cabezon, R.~Kaeppeli, Astrophys. J. Suppl. \textbf{223}(2), 22
  (2016).
\newblock \doi{10.3847/0067-0049/223/2/22}

\bibitem{Richers:2017awc}
S.~Richers, H.~Nagakura, C.D. Ott, J.~Dolence, K.~Sumiyoshi, S.~Yamada,
  Astrophys. J. \textbf{847}(2), 133 (2017).
\newblock \doi{10.3847/1538-4357/aa8bb2}

\bibitem{OConnor:2018sti}
E.~O'Connor, et~al., J. Phys. \textbf{G45}(10), 104001 (2018).
\newblock \doi{10.1088/1361-6471/aadeae}

\bibitem{Pan:2018vkx}
K.C. Pan, C.~Mattes, E.P. O'Connor, S.M. Couch, A.~Perego, A.~Arcones, J. Phys.
  \textbf{G46}(1), 014001 (2019).
\newblock \doi{10.1088/1361-6471/aaed51}

\bibitem{Cabezon:2018lpr}
R.M. Cabezón, K.C. Pan, M.~Liebendörfer, T.~Kuroda, K.~Ebinger, O.~Heinimann,
  F.K. Thielemann, A.~Perego, Astron. Astrophys. \textbf{619}, A118 (2018).
\newblock \doi{10.1051/0004-6361/201833705}

\bibitem{Horowitz:2001xf}
C.J. Horowitz, Phys. Rev. \textbf{D65}, 043001 (2002).
\newblock \doi{10.1103/PhysRevD.65.043001}

\bibitem{Roberts:2016mwj}
L.F. Roberts, S.~Reddy, Phys. Rev. \textbf{C95}(4), 045807 (2017).
\newblock \doi{10.1103/PhysRevC.95.045807}

\bibitem{Oertel:2020pcg_pub}
M.~Oertel, A.~Pascal, M.~Mancini, J.~Novak, Phys. Rev. C \textbf{102}(3),
  035802 (2020).
\newblock \doi{10.1103/PhysRevC.102.035802}

\bibitem{Malkus:2012ts}
A.~Malkus, J.P. Kneller, G.C. McLaughlin, R.~Surman, Phys. Rev. D \textbf{86},
  085015 (2012).
\newblock \doi{10.1103/PhysRevD.86.085015}

\bibitem{Malkus:2015mda}
A.~Malkus, G.C. McLaughlin, R.~Surman, Phys. Rev. D \textbf{93}(4), 045021
  (2016).
\newblock \doi{10.1103/PhysRevD.93.045021}

\bibitem{Zhu:2016mwa}
Y.L. Zhu, A.~Perego, G.C. McLaughlin, Phys. Rev. D \textbf{94}(10), 105006
  (2016).
\newblock \doi{10.1103/PhysRevD.94.105006}

\bibitem{Wu:2017qpc}
M.R. Wu, I.~Tamborra, Phys. Rev. D \textbf{95}(10), 103007 (2017).
\newblock \doi{10.1103/PhysRevD.95.103007}

\bibitem{Richers:2019grc}
S.A. Richers, G.C. McLaughlin, J.P. Kneller, A.~Vlasenko, Phys. Rev. D
  \textbf{99}(12), 123014 (2019).
\newblock \doi{10.1103/PhysRevD.99.123014}

\bibitem{Frensel:2016fge}
M.~Frensel, M.R. Wu, C.~Volpe, A.~Perego, Phys. Rev. D \textbf{95}(2), 023011
  (2017).
\newblock \doi{10.1103/PhysRevD.95.023011}

\bibitem{2018arXiv180504163H}
K.~{Abe}, et~al., arXiv e-prints arXiv:1805.04163 (2018)

\bibitem{Shakura:1972te}
N.I. Shakura, R.A. Sunyaev, Astron. Astrophys. \textbf{24}, 337 (1973)

\end{thebibliography}

\end{document}